\documentstyle[preprint,prd,aps,epsf]{revtex}

\newcommand{\Journal}[4]{{#1}{\bf #2}, {#4} {(#3)}}

\newcommand{\epj}{Eur.\ Phys\ J.\ }
\newcommand{\epjc}{\epj C }
\newcommand{\jh}{J.\ High Energy Phys. }
\newcommand{\np}{Nucl.\ Phys.\ }
\newcommand{\npb}{\np {\bf B}}
\newcommand{\plb}{\pl B }
\newcommand{\ptp}{Prog.\ Theor.\ Phys.\ }
\newcommand{\zp}{Z.\ Phys.\ }
\newcommand{\zpc}{\zp C }

\newcommand{\ibid}{\it ibid.\ }
\newcommand{\etal}{{\it et al.\/}}

\newcommand{\bb}{$B^0-\bar{B}^0$}
\newcommand{\kk}{$K^0-\bar{K}^0$}
\newcommand{\ek}{$\varepsilon_K$}
\newcommand{\bsbs}{$B_s-\bar{B}_s$}
\newcommand{\bdbd}{$B_d-\bar{B}_d$}
\newcommand{\dmbs}{$\Delta m_{B_s}$}
\newcommand{\dmbd}{$\Delta m_{B_d}$}
\newcommand{\dmbsd}{$\Delta m_{B_s}/\Delta m_{B_d}$}
\newcommand{\bsg}{$b \to s\,\gamma$}
\newcommand{\meg}{$\mu \to e\,\gamma$}
\newcommand{\tmg}{$\tau \to \mu\,\gamma$}
\newcommand{\teg}{$\tau \to e\,\gamma$}
\newcommand{\Bbsg}{$\text{B}(b \to s\,\gamma)$}
\newcommand{\Bmeg}{$\text{B}(\mu \to e\,\gamma)$}
\newcommand{\Btmg}{$\text{B}(\tau \to \mu\,\gamma)$}
\newcommand{\Bteg}{$\text{B}(\tau \to e\,\gamma)$}
\newcommand{\dlt}{$\delta_{13}$}
\newcommand{\amu}{$a_\mu^{\rm SUSY}$}

\newcommand{\BJKS}{$B\to J/\psi\,K_S$}
\newcommand{\Bsg}{$B\to M_s\,\gamma$}
\newcommand{\rek}{$\varepsilon_K/(\varepsilon_K)_{\rm SM}$}
\newcommand{\Ameg}{$A(\mu \to e\,\gamma)$}
\newcommand{\AtBsg}{$A_{CP}(B\to M_s\,\gamma)$}
\newcommand{\AtBJKS}{$A_{CP}(B\to J/\psi\,K_S)$}

\newcommand{\lsim}{\lesssim}
\newcommand{\gsim}{\gtrsim}

\begin{document}

\preprint{
\begin{tabular}{l}
KEK-TH-760 \\
ICRR-Report-476-2001-6 \\
\rule{0em}{6ex}
\end{tabular}
}

\draft

\title{
Muon anomalous magnetic moment, lepton flavor violation, and
 flavor changing neutral current processes 
 in SUSY GUT with right-handed neutrino 
}
\author{
Seungwon Baek,$^1$
Toru Goto,$^2$
Yasuhiro Okada,$^{2,3}$
and
Ken-ichi Okumura$^{4}$
}

\address{
$^1$Department of Physics, National Taiwan University, Taipei 106, Taiwan
\\
$^2$Theory Group, KEK, Tsukuba, Ibaraki, 305-0801 Japan
\\
$^3$Department of Particle and Nuclear Physics,\\
 The Graduate University of Advanced Studies, Tsukuba, Ibaraki,
 305-0801 Japan\\
$^4$Institute for Cosmic Ray Research, University of Tokyo,\\
Kashiwanoha 5-1-5, Kashiwa, 277-8582 Japan
}

\date{April 16, 2001}

\maketitle

\begin{abstract}
Motivated by the large mixing angle solutions for the atmospheric
 and solar neutrino anomalies, 
flavor changing neutral current processes
 and lepton flavor violating processes as well as
the muon anomalous magnetic moment are analyzed
in the framework of SU(5) SUSY GUT with right-handed neutrino.
In order to explain realistic mass relations for quarks and leptons,
we take into account effects of   
higher dimensional operators above the GUT scale.
It is shown that the supersymmetric (SUSY) contributions to the 
CP violation parameter in \kk\ mixing, \ek,
the \meg\ branching ratio, and the muon anomalous
magnetic moment become large in a wide range of parameter space.
We also investigate correlations among these quantities. 
Within the current experimental bound of \Bmeg,
large SUSY contributions are possible either in the muon anomalous 
magnetic moment
or in \ek. In the former case, the favorable value of the recent 
muon anomalous magnetic moment measurement at the BNL E821 experiment can 
be accommodated.
 In the latter case,
 the allowed region of the Kobayashi-Maskawa phase
can be different from the prediction within the Standard Model (SM)
and therefore  
the measurements of the CP asymmetry of \BJKS\ mode 
and \dmbs\
could discriminate this case from the SM.
We also show that the \tmg\ branching ratio 
can be close to the current experimental upper bound and the 
mixing induced CP asymmetry of the radiative $B$ decay can be enhanced 
in the case where the neutrino parameters correspond to the 
Mikheyev-Smirnov-Wolfenstein small mixing angle solution. 
\end{abstract}

\section{
  Introduction
}

In order to explore physics beyond the Standard Model (SM),
indirect searches play an important role complimentary
to direct searches of new particles at high energy frontiers. 
The indirect searches include flavor changing neutral current (FCNC)
processes, lepton flavor violation (LFV) and the muon anomalous 
magnetic moment. In the minimal SM,  
lepton flavor is conserved and FCNC is forbidden at tree level, 
so that $B$, $K$ and $\mu$ decay experiments 
have supplied severe constraints on models beyond the SM. 
At the recent BNL E821 experiment, it was reported that the muon 
anomalous magnetic moment had 2.6 $\sigma$ deviation
 from the SM prediction \cite{E821}. 
If the deviation is confirmed by improvement in both statistics and 
understanding theoretical uncertainty of the SM prediction, the 
muon anomalous magnetic moment becomes a clear signal of physics 
beyond the SM.

Among candidates of the physics beyond the SM, supersymmetry (SUSY) is
the most attractive one.
Because of the cancellation of quadratic divergence in the
renormalization of the Higgs field, SUSY models do not have the
hierarchy problem of the SM.
Furthermore, the gauge coupling unification is
realized in the SUSY grand unified theories (SUSY GUT) based on SU(5)
gauge group or its extensions.

In view of the flavor physics, it is important that the scalar partners
of quarks and leptons, namely squarks and sleptons, have a new source
of flavor mixing.
Due to the new flavor mixing, LFV and FCNC such as \meg, \bsg\ and
\kk/\bb\ mixing could be induced through SUSY loop diagrams.
Because these processes receive too large contributions for generic
flavor mixing in squark and slepton sector, the structure of SUSY
breaking sector of the Lagrangian is required to have a special form,
unless the masses of the SUSY particles are beyond multi-TeV region
\cite{SUSY-Flavor}.
The simplest possibility to avoid this problem is that the SUSY breaking
mechanism is assumed to be flavor blind. 
However, even in such a case, the squark and slepton mass matrices
receive radiative corrections from interactions below the scale where
the SUSY breaking is originated, and flavor blindness is broken
\cite{gluino-FCNC,86HaKoRa}.
In particular, effects of the large top Yukawa coupling constant can not
be neglected.
A number of analyses have been done in the context of minimal
supergravity (minimal SUGRA) anzats where SUSY breaking parameters are
assumed to be flavor blind at the Planck scale
\cite{Krimoto-Cho,91BeBoMaRi,Goto,99GoOkSh}.
It was shown that the flavor mixing is controlled by the
Cabibbo-Kobayashi-Maskawa (CKM) matrix element.
As a result the maximal deviation from the SM in the CP violating
parameter of \kk\ mixing, \ek, and \bdbd/\bsbs\ mixing is of the order
of 10~\%, while the SUSY contribution to \bsg\ process can be much
important \cite{Goto,99GoOkSh}.
In the GUT scenario, there are additional contributions to FCNC/LFV processes
 from GUT interactions \cite{86HaKoRa,GUT-classic,LFV,FCNC-GUT,GUT}.
As for LFV processes the \meg\ branching ratio is close to the current
experimental bound, especially for SO(10) model
\cite{LFV}.

Recent experimental evidences of neutrino oscillation indicate existence
of small neutrino mass and large flavor mixings in the lepton sector
\cite{neutrino-exp}.
A natural explanation for small neutrino mass is the seesaw mechanism
\cite{seesaw}.
In this mechanism, heavy right-handed neutrinos are introduced, and
these neutrinos have Majorana mass term and new Yukawa interactions.
Because the neutrino Yukawa coupling constants can be as large as the
top Yukawa coupling constant if the right-handed neutrinos are
$O(10^{14})$ GeV, radiative corrections from these interactions
contribute to the renormalization of slepton mass matrix above the mass
scale of the right-handed neutrinos.
Within the minimal SUGRA scenario, it was shown that the branching
ratios of LFV processes becomes large enough to be measured in
near-future experiments \cite{ynu-LFV,98HiNoYa,99HiNo}.
Some GUT models which have predictable neutrino mass and mixing are
already constrained \cite{ynu-LFV-GUT}.
In the context of SUSY GUT, these new interactions in the lepton sector
also contribute to the quark sector, because radiative corrections on
the squark from neutrino  interactions can become a new source of quark
FCNC processes as well as the LFV processes.
Recently, these processes are analyzed in the minimal SU(5) SUSY GUT
with right-handed neutrino and large deviations from the SM are
predicted \cite{01BaGoOkOk,00Moroi}.
However, in these analyses, simple flavor structure was assumed so that
the correct mass relations between the down-type quarks and charged
leptons in the first and second generations can not be realized.

Very recently, the BNL E821 experiment reported a new result on the muon
anomalous magnetic moment \cite{E821}.
The measured value of $a_{\mu}=(g_{\mu}-2)/2$ is
$a_{\mu}({\rm exp})=11659202(14)(6) \times 10^{-10}$, which is compared
to the SM prediction
$a_{\mu}({\rm SM})=11659159.6(6.7) \times 10^{-10}$.
It was concluded that the theory and experiment had 2.6 $\sigma$ difference
$a_{\mu}({\rm exp}) - a_{\mu}({\rm SM})= 43(16) \times 10^{-10}$.
The deviation can be explained in the context of the SUSY model
\cite{g-2new}. 
In contrast to the LFV and FCNC processes which are very sensitive to
the origin on the flavor mixing at high energy scale, the muon anomalous
magnetic moment can provide us  information on the slepton masses,
rather independent of the flavor structure of the slepton mass
matrices. 

In this paper, we discuss FCNC/LFV processes in the SU(5) supersymmetric
 grand unified theory with
right-handed neutrino (SU(5)RN SUSY GUT) taking account of realistic mass
relations. 
The seesaw mechanism generates small neutrino masses and large mixing
angles which incorporate atmospheric and solar neutrino anomalies.
In order to reproduce realistic mass relations, we introduce a higher
dimensional operator including {\bf 24} superfield which gives
contributions to the Yukawa coupling matrices for the down-type quarks
and the charged leptons in a different  manner.
Moreover, new degrees of freedom arise in the choice of the bases when
the MSSM multiplets are embedded in the SU(5) multiplets.
We show that the main effect of these new mixings is described by two
mixing angles which parameterize rotations of the bases between the
down-type quarks and charged leptons in the first and second
generations.
We perform numerical analysis on FCNC/LFV processes taking account of
various sources of flavor mixing.
We also calculate the muon anomalous magnetic moment and investigate the 
correlations among these quantities.
Solving renormalization group equations for the Yukawa coupling matrices
and the SUSY breaking parameters, the flavor mixing in the squark and
slepton sectors is evaluated at the electroweak (EW) scale.
In addition to the muon anomalous magnetic moment, we calculate
following FCNC/LFV observables: the branching ratios of \meg, \tmg\ and
\bsg, \ek, the mass differences in \bdbd\ mixing and \bsbs\ mixing, and
the time-dependent CP asymmetries of \BJKS\ and \Bsg\ where $M_s$ is a CP
eigenstate including a strange quark.
We find that the SUSY contributions to \ek, the \meg\ branching ratio,
and the muon anomalous magnetic moment become large in a wide range of
parameter space.
Within the current experimental bound on \Bmeg, large SUSY contributions
are possible either in the muon anomalous magnetic moment or in \ek.
In the former case, the favorable value of the recent muon anomalous
magnetic moment measurement at the BNL experiment can be accommodated.
In the latter case, the new contribution \ek\ modifies the
constraint for the CKM matrix elements and affects $B$ decay observables
because allowed region of \dmbsd\ and the time-dependent CP asymmetry of
the \BJKS\ mode can be quite different from that of the SM or MSSM
without the new flavor mixing source.
We also show that \Btmg\ and the indirect CP asymmetry of the radiative $B$
decay can be large in the case where the neutrino parameters correspond
to the small mixing angle Mikheyev-Smirnov-Wolfenstein (MSW) solution.
We also notice that the branching ratio of \meg\ can be close to the
present experimental upper limit both in the large and small mixing
angle MSW solutions.

The rest of this paper is organized as follows.
In Sec.~\ref{sec:model}, the SU(5)RN SUSY GUT is introduced.
The higher dimensional operators are included to incorporate the
realistic fermion mass relation.
Two new mixing angles are defined to parameterize the effect of these
operators.
In Sec.~\ref{sec:mSUGRA}, the minimal SUGRA model is introduced for the
SUSY breaking sector.
Radiative corrections for the SUSY breaking parameters and FCNC/LFV
processes are discussed qualitatively using approximated
formulas. 
In Sec.~\ref{sec:numerical-results}, the numerical results for the muon
anomalous magnetic moment and FCNC/LFV processes are presented.
Sec.~\ref{sec:conclusion} is devoted for conclusion and discussions.
In Appendices, useful formulas are collected.

\section{
  SU(5) SUSY GUT with right-handed neutrino
}
\label{sec:model}

In this section we discuss quark and lepton Yukawa couplings in the
SU(5)RN SUSY GUT.
Before introducing higher dimensional operators, we first discuss the
case without them.
Later, we introduce those operators to accommodate realistic mass
relation.
Without higher dimensional operators, the Yukawa coupling and the
Majorana mass term of the superpotential for this model are given by
\begin{eqnarray}
{\cal W}_{\rm SU(5)RN} &=&
 \frac{1}{8}\epsilon_{abcde}(\lambda_u)_{ij}(T^i)^{ab}(T^j)^{cd}H^e
 +(\lambda_d)_{ij}(\overline{F}^i)_a(T^j)^{ab}\overline{H}_b
\nonumber\\&&
 +(\lambda_{\nu})_{ij}\overline{N}^i(\overline{F}^j)_aH^a
 +\frac{1}{2}(M_{\nu})_{ij}\overline{N}^i\overline{N}^j,
\label{eq:SU5RN}
\end{eqnarray}
where $T^i$, $\overline{F}^i$ and $\overline{N}^i$ are ${\bf 10}$,
${\bf \overline{5}}$ and ${\bf 1}$ representation of SU(5) gauge group,
 respectively.
$i,\,j$ are generation indices and $a,\,b,\,c,\,d$ and $e$ are SU(5)
indices.
$\epsilon_{abcde}$ is the totally antisymmetric tensor of SU(5) gauge
group.
$H$ and $\overline{H}$ are Higgs superfields with ${\bf 5}$ and
${\bf \overline{5}}$ representations.
In terms of SU(3)$\times$SU(2)$_L\times$U(1)$_Y$, $T^i$ contains
$Q^i({\bf 3,2}, \frac{1}{6})$,
$\overline{U}^i({\bf \overline{3}, 1}, -\frac{2}{3})$
and $\overline{E}^i({\bf 1,1},1)$ superfields.
Here the representations for SU(3) and SU(2) groups and the $U(1)_Y$ charge
are indicated in the parentheses.
$\overline{F}^i$ includes $\overline{D}^i({\bf 3,1},\frac{1}{3})$
and $L^i({\bf 1,2},-\frac{1}{2})$,
and $\overline{N}^i$ is a singlet of
SU(3)$\times$SU(2)$_L\times$U(1)$_Y$.
$H$ consists of $H_C({\bf 3,1},0)$ and $H_2({\bf 1,2},\frac{1}{2})$ and
$\overline{H}$ contains $\overline{H}_C({\bf \overline{3},1},0)$
and $H_1({\bf 1,2},-\frac{1}{2})$.
$(\lambda_u)_{ij}$, $(\lambda_d)_{ij}$ and $(\lambda_{\nu})_{ij}$ are
Yukawa coupling matrices and $(M_{\nu})_{ij}$ is a Majorana mass matrix.
In addition to the above formula, we also need a superpotential for
Higgs superfields, ${\cal W}_H(H,\overline{H},\Sigma)$ where
$\Sigma^a_{~b}$ is a {\bf 24} representation of SU(5) group.
It is assumed to develop vacuum expectation values as
$\langle \Sigma^a_{~b} \rangle=
\text{diag}(\frac{1}{3},\frac{1}{3},\frac{1}{3},
-\frac{1}{2},-\frac{1}{2})v_G$ at the GUT scale
($M_G\approx2\times10^{16}$~GeV) and breaks the SU(5) symmetry
to SU(3)$\times$SU(2)$_L \times$U(1)$_Y$.

Below the GUT scale, the heavy superfields such as $H_C$,
$\overline{H}_C$ and $\Sigma$ are integrated out and the superpotential
of the MSSM with right-handed neutrino (MSSMRN) is given by
\begin{eqnarray}
{\cal W}_{\rm MSSMRN} &=&
   (y_u)_{ij}\overline{U}^iQ^jH_2
 + (y_d)_{ij}\overline{D}^iQ^jH_1 
 + (y_e)_{ij}\overline{E}^iL^jH_1
 + \mu H_1 H_2
\nonumber\\&&
 + (y_{\nu})_{ij}\overline{N}^iL^jH_2
 + \frac{1}{2}(M_{\nu})_{ij}\overline{N}^i\overline{N}^j,
\label{eq:MSSMRN}
\end{eqnarray}
where Yukawa coupling matrices are related to those of the SU(5)RN as
$(y_u)_{ij}=(\lambda_u)_{ij}$,
$(y_d)_{ij}=(y_l^T)_{ij}=(\lambda_d)_{ij}$ and
$(y_{\nu})_{ij}=(\lambda_{\nu})_{ij}$.
Below the Majorana mass scale ($\equiv M_R$), the singlet fields are
also integrated out from the superpotential  and a dimension five
operator is generated as follows:
\begin{equation}
\Delta{\cal W}_{\nu} = -\frac{1}{2}(K_{\nu})_{ij}(L_i H_2)(L_j H_2),
~~~~~
K_{\nu} = (y_{\nu}^T)_{ik}(\frac{1}{M_{\nu}})^{kl}(y_{\nu})_{lj}.
\label{eq:seesaw1}
\end{equation}
After the EW symmetry breaking, this operator induces by the
seesaw mechanism the following neutrino mass matrix,
\begin{eqnarray}
(m_{\nu})_{ij} &=& (K_{\nu})_{ij}\langle H_2 \rangle^2.
\label{eq:seesaw2}
\end{eqnarray}
In this model, the naive GUT relation is predicted at the GUT scale,
\begin{eqnarray}
(y_e)_{ij} &=& (y_d)_{ji}.
\label{eq:naive GUT relation}
\end{eqnarray}
Although this relation gives a reasonable agreement for $m_b$ and
$m_{\tau}$, it is well known that the mass ratio of down-type quarks and
charged leptons in the first and the second generations can not be
explained in this way.
One possibility to remedy this defect is to introduce higher dimensional
operators above the GUT scale because they can give different 
contributions to the Yukawa coupling  matrices
 of down-type quarks and charged leptons
  after the SU(5) symmetry breaking.

We consider higher dimensional operators including the {\bf 24} Higgs
superfield up to dimension five terms.
Relevant parts of the superpotential is parameterized as follows:
\begin{eqnarray}
\Delta{\cal W}_{\rm SU(5)RN} &=& \frac{1}{M_X}
   \left[
    \frac{1}{4}\epsilon_{abcde}(\kappa_u^{+})_{ij}
    \{\Sigma^a_{~f}(T^i)^{fb}(T^j)^{cd}
  + (T^i)^{ab}\Sigma^c_{~f}(T^j)^{fd}\}H^e
   \right.
\nonumber\\&&
\phantom{\frac{1}{M_X}}
   \left.
  + \frac{1}{4}\epsilon_{abcde}(\kappa_u^{-})_{ij}
               \{\Sigma^a_{~f}(T^i)^{fb}(T^j)^{cd}
              -(T^i)^{ab}\Sigma^c_{~f}(T^j)^{fd}\}H^e
   \right.
\nonumber\\&&
\phantom{\frac{1}{M_X}\frac{1}{4}}
   \left.
  + (\kappa_d)_{ij}
    (\overline{F}^i)_a\Sigma^a_{~b}(T^j)^{bc}\overline{H}_c
   \right.
\nonumber\\&&
\phantom{\frac{1}{M_X}\frac{1}{4}}
   \left.
  + (\overline{\kappa}_d)_{ij}
    (\overline{F}^j)_a(T^j)^{ab}\Sigma^c_{~b}\overline{H}_c
   \right.
\nonumber\\&&
   \left.
\phantom{\frac{1}{M_X}\frac{1}{4}}
  + (\kappa_{\nu})_{ij}
    \overline{N}^i(\overline{F}^j)_a\Sigma^a_{~b} H^b
   \right],
\label{eq:dim5 superpotential}
\end{eqnarray}
where $M_X$ is the cut-off scale which we take as the Planck mass $M_P$.
We also assume that the elements of coupling matrices $\kappa_u^{\pm}$,
$\kappa_d$, $\overline{\kappa}_d$ and $\kappa_{\nu}$ are smaller than
$O(1)$.
After SU(5) symmetry is broken, they give contributions of the order of
$\xi=v_G/M_X\approx 0.01$ to the Yukawa coupling constants of the MSSMRN
as follows:
\begin{mathletters}
\label{eq:yukawa-matching}
\begin{eqnarray}
(y_u)_{ij} &=& (\lambda_u)_{ij}
  +\xi\left\{
     \frac{1}{2}(\kappa^{+}_u)_{ij}
    +\frac{5}{6}(\kappa^{-}_u)_{ij}
  \right\},
\\
(y_d)_{ij} &=& (\lambda_d)_{ij}
  +\xi\left\{
     \frac{1}{3}(\kappa_d)_{ij}
    -\frac{1}{2}(\overline{\kappa}_d)_{ij}
  \right\},
\\
(y_e)_{ij} &=& (\lambda_d^T)_{ij}
  +\xi\left\{
    -\frac{1}{2}(\kappa_d^T)_{ij}
    -\frac{1}{2}(\overline{\kappa}_d^T)_{ij}
  \right\},
\\
(y_{\nu})_{ij} &=& (\lambda_{\nu})_{ij}
    -\frac{\xi}{2}(\kappa_{\nu})_{ij}.
\end{eqnarray}
\end{mathletters}
The naive GUT relation between the lepton and the down-type quark Yukawa
coupling matrices in Eq.~(\ref{eq:naive GUT relation}) is modified to
\begin{eqnarray}
  (y_e)_{ij} &=& (y_d)_{ji} + \frac{5}{6}\xi(\kappa_d)_{ji}.
\end{eqnarray}
With this small contribution from the higher dimensional operator,
realistic mass relations between the down-type quarks and charged
leptons can be incorporated in the model.
In the following analysis we take
$(\kappa_u^+)_{ij}=(\kappa_u^-)_{ij}=(\overline{\kappa}_d)_{ij}
=(\kappa_{\nu})_{ij}=0$
because they are not necessarily required to reproduce the realistic
mass relations.

In the following, we show that new mixing angles are introduced at
the GUT scale because of $\kappa_d$.
Using SU(5) symmetry, we can rotate the generation indices of
superfields in Eq.~(\ref{eq:MSSMRN}) so that the Yukawa coupling
constants and the Majorana mass matrix are parameterized as follows:
\begin{mathletters}
\begin{eqnarray}
  (y_u)_{ij} &=&
  (V_{\rm CKM}^TV_U^*)_i^{~k} {y_u}_k (V_{\rm CKM})^k_{~j},
\\
  (y_d)_{ij} &=& (V_D^*)_i^{~j} {y_d}_j,
\\
  (y_e)_{ij} &=& (V_E^*)_i^{~j} {y_e}_j,
\\
  (y_{\nu})_{ij} &=& {y_{\nu}}_i (V_L)^i_{~j},
\\
  (M_{\nu})_{ij} &=& (V_N^T)_{i}^{~k}{M_{\nu}}_k(V_N)^k_{~j}, 
\end{eqnarray}
\end{mathletters}
where $V_U$, $V_E$, $V_D$ and $V_N$ are unitary matrices and
$V_{\rm CKM}$ is the CKM matrix at the GUT scale.
${y_u}_i$, ${y_d}_i$, ${y_e}_i$, ${y_{\nu}}_i$,
and ${M_{\nu}}_i$ represent the eigenvalues of the Yukawa
coupling matrices and the Majorana mass matrix. 
The GUT relation between the two Yukawa coupling constants is then
given by
\begin{equation}
  (V_D^*)_i^{~j}{y_d}_j - {y_e}_i(V_E^{\dag})^i_{~j} =
  \frac{5}{6}\xi (\kappa_d)_{ij}.
\label{eq:GUT relation}
\end{equation}
From this formula we can derive the following approximate relations for
the $1$-$3$ and $2$-$3$ ($3$-$1$ and $3$-$2$) elements
 of the mixing matrices because the
Yukawa coupling constants of the first and second generations are much
smaller than that of the third generation,
\begin{mathletters}
\begin{eqnarray}
  (V_D)^i_{~3} &\approx&
  \frac{5}{6}\frac{\xi}{y_{b}}(\kappa_d^*)_{i3},
\\
  (V_E)^i_{~3} &\approx&
  -\frac{5}{6}\frac{\xi}{y_{\tau}}(\kappa_d^{\dag})_{i3},
\\
  (V_E)^3_{~i} &\approx&
  \frac{5}{6}\frac{\xi}{y_{b}}(V_D^T\kappa_d V_E)_{3i},
\\
  (V_D)^3_{~i} &\approx&
  -\frac{5}{6}\frac{\xi}{y_{\tau}}(V_E^T\kappa_d^T V_D)_{3i},
\end{eqnarray}
\end{mathletters}
for $i=1,2$.
We can estimate $\xi/y_b$ and $\xi/y_{\tau}$ as
\begin{equation}
  \frac{\xi}{y_b} \approx
  \frac{\xi}{y_{\tau}} \approx
  -\frac{\xi v \cos\beta}{\sqrt{2}m_{\tau}},
~~~~~
  \frac{v}{\sqrt{2}} =
  \sqrt{\langle H_1 \rangle^2+\langle H_2 \rangle^2},
\end{equation}
where $\beta$ is a vacuum angle of two Higgs vacuum expectation values
($\tan\beta = \langle H_2^0 \rangle/\langle H_1^0 \rangle$).
Assuming the condition $(\kappa_d)_{ij}\lsim O(1)$,
we can conclude that the magnitude of these elements are constrained to
be smaller than $(\tan\beta)^{-1}$ because lower $\tan\beta$ region is
excluded from Higgs boson search.
On the other hand, $1$-$2$ ($2$-$1$) element is not constrained from such a
consideration. 
Motivated by this observation we assume the following form:
\begin{mathletters}
\begin{eqnarray}
  (V_D)^i_{~j} &=& e^{i \gamma_D}
  \left( 
    \begin{array}{ccc}
      e^{i\alpha_D}\cos\theta_D & -e^{-i\beta_D}\sin\theta_D  & 0 \\
      e^{i\beta_D}\sin\theta_D  &  e^{-i\alpha_D}\cos\theta_D & 0\\
      0             & 0              & e^{i(- \gamma_D+\delta)}
    \end{array}
  \right),
\\
  (V_E)^i_{~j} &=& e^{i \gamma_E}
  \left(
    \begin{array}{ccc}
      e^{i\alpha_E}\cos\theta_E & -e^{-i\beta_E}\sin\theta_E  & 0\\
      e^{i\beta_E}\sin\theta_E  &  e^{-i\alpha_E}\cos\theta_E & 0\\
      0             & 0             & e^{i(- \gamma_E+\delta)}
    \end{array}
  \right).
\end{eqnarray}
\label{eq:GUT mixing angle}
\end{mathletters}
The antisymmetric part of the Yukawa matrix for the up-type quarks is
also written by the coefficients of the dimension five operator as
follows:
\begin{eqnarray}
  (V_U^*)_i^{~j}{y_u}_j - {y_u}_i(V_U^{\dag})^i_{~j} &=&
  \frac{5}{6}\xi(V_{\rm CKM}^*\kappa_u^-V_{\rm CKM}^{\dag})_{ij}.
\end{eqnarray}
Because we set $(\kappa_u^-)_{ij}=0$ for simplicity, $(V_U)^i_{~j}$
becomes $e^{i{\phi_U}_i}\delta^i_{~j}$ in our analysis.

The neutrino Yukawa coupling matrix and the Majorana mass matrix are
constrained from the oscillation solutions of the atmospheric and solar
neutrino anomalies.
In the basis where the charged lepton mass matrix is diagonal, the
neutrino mass matrix is written as follows:
\begin{equation}
  (m_{\nu})_{ij} =
  (V_{\rm MNS}^*)_i^{~k}{m_{\nu}}_k(V_{\rm MNS}^{\dag})^k_{~j},
\label{eq:MNS}
\end{equation} 
where $V_{\rm MNS}$ is the Maki-Nakagawa-Sakata (MNS) matrix
\cite{62MaNaSa}.
At the Majorana mass scale Eqs.~(\ref{eq:seesaw1}) and
(\ref{eq:seesaw2}) are solved as follows:
\begin{equation}
  (V_N^*)_i^{~k}{y_{\nu}}_k(V_L)^k_{~j} =
  \frac{1}{\langle H_2 \rangle}\sqrt{{M_{\nu}}_i}
  (O_{\nu}^T)_i^{~k}\sqrt{{m_{\nu}}_k}(V_{\rm MNS}^{\dag})^k_{~j},
\label{eq:ynu}
\end{equation}
where $O_{\nu}$ is a complex orthogonal matrix which can not be determined
from the low energy experiments.
Although we neglect a running effect of the neutrino mass matrix
  between the low energy scale and the GUT scale in
 Eqs.~(\ref{eq:MNS}) and (\ref{eq:ynu}), later
  we fully take account of this effect in the numerical
  calculation in Sec.~\ref{sec:numerical-results}.

\section{
  Minimal SUGRA and the muon anomalous magnetic moment and the
  FCNC/LFV processes
}
\label{sec:mSUGRA}

In Subsection \ref{sec:radiative-corrections} we first discuss the
flavor mixing of squark and slepton mass matrices 
induced by the radiative
correction due to the Yukawa coupling constants.
In order to explain qualitative features we show the one-loop
logarithmic terms for SUSY breaking parameters.
In the numerical calculation in Sec.~\ref{sec:numerical-results},
however, we use the full renormalization group equation (RGE) and solve them
numerically.
In Subsection \ref{sec:fcnc-lfv-processes}, we give a brief
description on the SUSY contribution to the muon anomalous magnetic
moment and various FCNC and LFV processes.

\subsection{
  Minimal SUGRA and radiative corrections to the SUSY breaking
  parameters
}
\label{sec:radiative-corrections}

The soft SUSY breaking terms of the MSSM are given by  
\begin{eqnarray}
  {\cal L}_{\rm soft} &=&
  -(m^{2}_{Q})^i_{~j}\widetilde{Q}_i^{\dag} \widetilde{Q}^{j}
  -(m^{2}_{U})_i^{~j}\widetilde{U}^{i*} \widetilde{U}_{j}  
  -(m^{2}_{D})_i^{~j}\widetilde{D}^{i*} \widetilde{D}_{j}
\nonumber \\&&
  -(m^{2}_{L})^i_{~j}\widetilde{L}_i^{\dag} \widetilde{L}^{j}
  -(m^{2}_{E})_i^{~j}\widetilde{E}^{i*} \widetilde{E}_{j}
  - m^{2}_{H_{2}}H_{2}^{\dag}H_{2}
  - m^{2}_{H_{1}}H_{1}^{\dag}H_{1}
\nonumber \\&&
  -\left\{(\widetilde{y}_{u})_{ij}  \widetilde{U}^{i*}\widetilde{Q}^{j} H_2
   +(\widetilde{y}_{d})_{ij}  \widetilde{D}^{i*}\widetilde{Q}^{j} H_1
   \phantom{\text{H.c.}}\right.
\nonumber \\&&
   \left.
   +(\widetilde{y}_{e})_{ij}  \widetilde{E}^{i*}\widetilde{L}^{j} H_1
   +\mu B H_1 H_2 + \text{H.c.}
   \right\}
\nonumber \\&&
   +\frac{1}{2}M_{1}\overline{\widetilde{B}} \widetilde{B}
   +\frac{1}{2}M_{2}\overline{\widetilde{W}} \widetilde{W}
   +\frac{1}{2}M_{3}\overline{\widetilde{G}} \widetilde{G},
\label{eq:Soft Breaking MSSM}
\end{eqnarray}
where $\widetilde{Q}^i$, $\widetilde{U}^{i*}$, $\widetilde{D}^{i*}$,
$\widetilde{L}^i$ and $\widetilde{E}^{i*}$ are scalar components of
$Q^i$, $\overline{U}^i$, $\overline{D}^i$, $L^i$ and $\overline{E}^i$,
respectively.
We use the same symbols as superfields for scalar components of the
Higgs supermultiplets.
$\widetilde{B}$, $\widetilde{W}$ and $\widetilde{G}$ are U(1), SU(2) and
 SU(3) gauginos, respectively.
At the GUT scale, these soft SUSY breaking parameters are determined by
the following SUSY breaking terms of the SU(5)RN SUSY GUT,
\begin{eqnarray}
  {\cal L} &=&
  -(m^2_T)^i_{~j}(\widetilde{T}_i^*)_{ab}(\widetilde{T}^j)^{ab}
  -(m^2_{\overline{F}})^i_{~j}
   (\widetilde{\overline{F}}_i^*)^a (\widetilde{\overline{F}}^j)_a
  -(m^2_{\overline{N}})^i_{~j}\widetilde{\overline{N}}_i^*\widetilde{\overline{N}}^j
\nonumber\\&&
  -(m^2_{H}) H^*_{~a} H^a
  -(m^2_{\overline{H}}) \overline{H}^{*a} \overline{H}_a
\nonumber\\&&
  -\left\{
      \frac{1}{8}\epsilon_{abcde}(\widetilde{\lambda}_u)_{ij}
      (\widetilde{T}^i)^{ab}(\widetilde{T}^j)^{cd} H^e
    + (\widetilde{\lambda}_d)_{ij}(\widetilde{\overline{F}}^i)_a
      (\widetilde{T}^j)^{ab} \overline{H}_b
    \right.
\nonumber\\&&\phantom{+}
    \left.
    + (\widetilde{\lambda}_{\nu})_{ij}
      \widetilde{\overline{N}}^i(\widetilde{\overline{F}}^j)_a H^a
    + \frac{1}{2}(\widetilde{M}_{\nu})_{ij}
      \widetilde{\overline{N}}^i\widetilde{\overline{N}}^j
      + \text{H.c.} \right\}
\nonumber\\&&
-\frac{1}{M_X}\left[
     \frac{1}{4}\epsilon_{abcde}(\widetilde{\kappa}_u^{+})_{ij}
      \left\{ \Sigma^a_{~f}(\widetilde{T}^i)^{fb}(\widetilde{T}^j)^{cd}
         +(\widetilde{T}^i)^{ab}\Sigma^c_{~f}
          (\widetilde{T}^j)^{fd} \right\} H^e
    \right.
\nonumber\\&&\phantom{+\frac{1}{M_X}}
    +\frac{1}{4}\epsilon_{abcde}
     (\widetilde{\kappa}_u^{-})_{ij}
     \left\{ \Sigma^a_{~f}(\widetilde{T}^i)^{fb}(\widetilde{T}^j)^{cd}
        -(\widetilde{T}^i)^{ab}\Sigma^c_{~f}
         (\widetilde{T}^j)^{fd} \right\} H^e
\nonumber\\&&\phantom{+\frac{1}{M_X}}
    +(\widetilde{\kappa}_d)_{ij}
     (\widetilde{\overline{F}}^i)_a\Sigma^a_{~b}(\widetilde{T}^j)^{bc}
     \overline{H}_c
\nonumber\\&&\phantom{+\frac{1}{M_X}}
    +(\widetilde{\overline{\kappa}}_d)_{ij}
     (\widetilde{\overline{F}}^j)_a (\widetilde{T}^j)^b
     \Sigma^c_{~b} \overline{H}_c
\nonumber\\&&\phantom{+\frac{1}{M_X}}
  \left.
    +(\widetilde{\kappa}_{\nu})_{ij}
     \widetilde{\overline{N}}^i
     (\widetilde{\overline{F}}^j)_a\Sigma^a_{~b} H^b + \text{H.c.}
   \right]
\nonumber\\&&
    +\frac{1}{2}M_5\overline{\widetilde{G}_5}\widetilde{G}_5,
\label{eq:SUSY-Breaking-SU5RN}
\end{eqnarray}
where $\widetilde{T}^i$, ${\widetilde{\overline{F}}}^i$ and
${\widetilde{\overline{N}}}^i$ are scalar components of $T^i$,
$\overline{F}^i$ and $\overline{N}^i$ and $\widetilde{G}_5$ represents
SU(5) gaugino.
We assume the minimal supergravity scenario for the origin of SUSY
breaking and set the following boundary conditions for the SUSY breaking
parameters at the Planck scale,
\begin{mathletters}
\begin{eqnarray}
  (m^2_T)^i_{~j} &=&
  (m^2_{\overline{F}})^i_{~j} =
  (m^2_{\overline{N}})^i_{~j} =
  m_0^2\delta^i_{j},
\\
  (\widetilde{\lambda})_{ij} &=& m_0 A_0 (\lambda)_{ij},
~~~( \lambda = \lambda_u, \lambda_d, \lambda_\nu ), 
\\
  (\widetilde{\kappa})_{ij} &=&
  m_0(A_0+\Delta A_0) (\kappa)_{ij},
~~~(\kappa = \kappa^\pm_u, \kappa_d, \overline{\kappa}_d, \kappa_\nu),
\label{eq:minimal-SUGRA-dA0}
\\
  M_5 &=& M_0.
\end{eqnarray}
\label{eq:minimal-SUGRA}
\end{mathletters}
If we ignore radiative corrections from the gauge and Yukawa coupling
constants and assume $\Delta A_0 = 0$, the soft SUSY breaking terms are
given by
\begin{mathletters}
\begin{eqnarray}
  (m_Q^2)^i_{~j} &=& (m_L^2)^i_{~j} =
  (m_U^2)_i^{~j} = (m_E^2)_i^{~j} = (m_D^2)_i^{~j} = m_0^2\delta_i^{j},
\\
  (\widetilde{y})_{ij} &=& m_0 A_0 (y)_{ij},~~~( y = y_u, y_d, y_e ).
\end{eqnarray}
\end{mathletters}
Then LFV processes are forbidden and SUSY contributions to FCNC
processes are suppressed.
We consider $\Delta A_0 \neq 0$ case later.

Radiative corrections between the Planck scale and the EW scale modify
the above structure of the soft SUSY breaking terms.
In particular, the corrections from the Yukawa coupling constants
associated with the colored Higgs supermultiplets and right-handed
neutrino supermultiplets are important because they have different
flavor structure from the Yukawa coupling constants of the MSSM.

Let us estimate these corrections using approximate formulas only
considering logarithmic terms to see qualitative features of FCNC/LFV
processes in the model.
The Yukawa couplings including colored Higgs supermultiplets are
parameterized as follows:
\begin{eqnarray}
  {\cal W}_C &=&
  -(y_{CR})_{ij}H_C\overline{U}^i\overline{E}^j
  -\frac{1}{2}(y_{CL})_{ij}H_C Q^i Q^j
\nonumber\\&&
  -(y_{\overline{C}R})_{ij}\overline{H}_C \overline{D}^i\overline{U}^j
  -(y_{\overline{C}L})_{ij}\overline{H}_C L^i Q^j
  +(y_{CN})_{ij}H_C \overline{N}^i \overline{D}^j.
\label{eq:colored Higgs}
\end{eqnarray}
It is convenient to work in the basis where the down-type quark and
charged lepton mass matrices are diagonal,
\begin{mathletters}
\begin{eqnarray}
  (y_u)_{ij} &=& {y_u}_i (V_{\rm CKM})^i_{~j},
\\
  (y_d)_{ij} &=& {y_d}_i \delta^i_{j},
\\
  (y_e)_{ij} &=& {y_e}_i \delta^i_{j},
\\
  (y_{\nu})_{ij} &=& {y_{\nu}}_i (V_L)^i_{~j}.
\end{eqnarray}
\label{eq:MSSM Yukawa}
\end{mathletters}
In this basis, Yukawa coupling matrices in Eq.~(\ref{eq:colored Higgs})
are given by
\begin{mathletters}
\begin{eqnarray}
  (y_{CR})_{ij} &=& {y_u}_i(V_{\rm CKM} V_E)^i_{~j},
\\
  (y_{CL})_{ij} &=&
   \frac{1}{2}\{(V_{\rm CKM}^T)_i^{~j}e^{i{\phi_U}_j}{y_u}_j
  +e^{i{\phi_U}_i}{y_u}_i(V_{\rm CKM})^i_{~j}\}, 
\\
  (y_{\overline{C}R})_{ij} &=& {y_d}_ie^{i{\phi_U}_i}\delta^i_{j},
\\
  (y_{\overline{C}L})_{ij} &=& {y_e}_i (V_E^\dag)^i_{~j},
\\
  (y_{CN})_{ij} &=& {y_{\nu}}_i (V_L V_D)^i_{~j}.
\end{eqnarray}
\label{eq:colored-Higgs-Yukawa}
\end{mathletters}
The radiative corrections to squark and slepton mass matrices from these
Yukawa coupling constants are approximated as follows:
\begin{mathletters}
\begin{eqnarray}
  \Delta m_Q^2 &\approx&
  -2\left(
      y_u^{\dag} y_u
    +2 y_{CL}^{\dag} y_{CL}
    + y_d^{\dag}  y_d
    + y_{\overline{C}L}^{\dag} y_{\overline{C}L}
  \right)
  (3+|A_0|^2)m_0^2 t_G
\nonumber\\&&
  -2\left(
       y_u^{\dag} y_u
    +  y_d^{\dag} y_d
    \right)
    (3+|A_0|^2)m_0^2 t_W,
\\
  \Delta m_U^2 &\approx&
  -2\left(
     2 y_u y_u^\dag
    +  y_{CR} y_{CR}^\dag
    +2 y_{\overline{C}R}^T y_{\overline{C}R}^*
  \right)
  (3+|A_0|^2)m_0^2 t_G
\nonumber\\&&
  -4 y_u y_u^\dag (3+|A_0|^2)m_0^2 t_W,
\\
  \Delta m_E^2 &\approx&
  -2\left(
    2 y_e y_e^\dag+3 y_{CR}^T y_{CR}^*
  \right)
  (3+|A_0|^2)m_0^2 t_G
\nonumber\\&&
  -4 y_e y_e^\dag (3+|A_0|^2)m_0^2 t_W,
\\
  \Delta m_D^2 &\approx&
  -2\left(
    2 y_d  y_d^\dag
    +2 y_{\overline{C}R} y_{\overline{C}R}^\dag
    + y_{CN}^T y_{CN}^*
  \right)
  (3+|A_0|^2)m_0^2 t_G
\nonumber\\&&
  -4 y_d y_d^\dag (3+|A_0|^2)m_0^2 t_W,
\\
  \Delta m_L^2 &\approx&
  -2\left(
       y_e^{\dag} y_e
    +3 y_{\overline{C}L}^* y_{\overline{C}L}^T
    +  y_{\nu}^{\dag} y_{\nu}
  \right)
  (3+|A_0|^2)m_0^2 t_G
\nonumber\\&&
  -2 y_e^{\dag} y_e (3+|A_0|^2)m_0^2 t_W
  -2 y_{\nu}^{\dag} y_{\nu} (3+|A_0|^2)m_0^2 t_R,
\end{eqnarray}
\end{mathletters}
where we only take account
 of logarithmic terms so that 
$t_G = \frac{1}{(4\pi)^2}\ln(\frac{M_P}{M_G})$,
$t_R = \frac{1}{(4\pi)^2}\ln(\frac{M_G}{M_R})$ and
$t_W = \frac{1}{(4\pi)^2}\ln(\frac{M_G}{M_{\rm SUSY}})$.
$M_{\rm SUSY}$ is a characteristic mass scale of the SUSY particles
 and identified to the EW scale.
The off-diagonal elements of the above formulas are sources of the LFV
and FCNC processes.
Keeping only possible large Yukawa coupling constants, the off-diagonal
elements of the mass matrices are approximated as follows:
\begin{mathletters}
\label{eq:mass-correction}
\begin{eqnarray}
  (m_Q^2)^i_{~j} &\approx&
  -2(V_{\rm CKM}^{\dag})^i_{~3} {y_t}^2(V_{\rm CKM})^3_{~j}
  (3+|A_0|^2)m_0^2(t_G+t_W)
\nonumber\\&&
  -\left\{
      (V_{\rm CKM}^{\dag})^i_{~3}{y_t}^2\delta^3_{~j}
    + \delta^i_{~3}{y_t}^2(V_{\rm CKM})^3_{~j}
  \right.
\nonumber \\&&\phantom{-}
  \left.
    +(V_{\rm CKM}^{\dag})^i_{~3}{y_t}^2(V_{\rm CKM})^3_{~j}
  \right\}
  (3+|A_0|^2)m_0^2 t_G,~~~~~~~~~~( i \neq j ),
\label{eq:mass correction 1}
\\
  (m_E^2)_i^{~j} &\approx&
  -6(V_E^T V_{\rm CKM}^T)_i^{~3} {y_t}^2 (V_{\rm CKM}^*V_E^*)_3^{~j}
  (3+|A_0|^2)m_0^2t_G,~~~~~~( i \neq j ),
\label{eq:mass correction 2}
\\
  (m_D^2)_i^{~j} &\approx&
  -2 (V_D^T V_L^T)_i^{~k}{{y_{\nu}}_k}^2(V_L^* V_D^*)_k^{~j}
  (3+|A_0|^2)m_0^2t_G,~~~~~~~~~~( i \neq j ),
\label{eq:mass correction 3}
\\
  (m_L^2)^i_{~j} &\approx&
  -2(V_L^{\dag})^i_{~k} {{y_{\nu}}_k}^2 (V_L)^k_{~j} 
  (3+|A_0|^2)m_0^2(t_G+t_R),~~~~~~~~~~( i \neq j ).
\label{eq:mass correction 4}
\end{eqnarray}
\end{mathletters}
$(m_Q^2)^i_{~j}$ corresponds to the flavor mixing due to the large top
Yukawa coupling constant which already exists within
 the MSSM based on the minimal SUGRA.
$(m_E^2)_i^{~j}$ receives radiative correction from the up-type Yukawa
coupling constant between the Planck scale and the GUT scale.
This is a well-known mechanism to induce LFV processes in the SUSY
GUT \cite{86HaKoRa}.
We notice that the following important features.
\begin{itemize}
\item
  There are flavor mixings in the right-handed down-type squark and the
  left-handed slepton sectors.
  These mixings are absent in the minimal SUGRA model without
  right-handed neutrino supermultiplet.
\item
  Because $V_L$ can be related to the MNS matrix, large mixing is possible.
\item
  The main effect of the higher dimensional operator under assumption of
  Eq.~(\ref{eq:GUT mixing angle}) is only rotating the basis of light
  fermions between $d_R$ and $s_R$, $e_R$ and $\mu_R$.
  For each of these mixings the rotation is described by one parameter
  $\theta_D$ in $V_D$ or $\theta_E$ in $V_E$.
\end{itemize}

In the above discussion, we only considered the radiative correction to
squarks and slepton mass matrices, however, the trilinear scalar
coupling constants, $\widetilde{y}_u$, $\widetilde{y}_d$ and
$\widetilde{y}_e$ also receive corrections as follows:
\begin{mathletters}
\label{eq:A-term-correction}
\begin{eqnarray}
  (\widetilde{y}_u)_{ij} &\approx&
  m_0 A_u {y_u}_i(V_{\rm CKM})^i_{~j} 
\nonumber\\&&
  -\frac{m_0 A_0}{3+|A_0|^2}
   \frac{
      (\Delta m^2_U)_i^{~k} {y_u}_k(V_{\rm CKM})^k_{~j} 
     +{y_u}_i(V_{\rm CKM})^i_{~k} (\Delta m^2_Q)^k_{~j}
     }{m_0^2},
\label{eq:A-term correction 1}
\\
  (\widetilde{y}_d)_{ij} &\approx&
  m_0 A_d {y_d}_i \delta^i_{j}
\nonumber\\&&
  -\frac{m_0 A_0}{3+|A_0|^2}
   \frac{
      (\Delta m^2_D)_i^{~j} {y_d}_j
     +{y_d}_i(\Delta m^2_Q)^i_{~j}
     }{m_0^2}
\nonumber\\&&
  -\frac{2}{5}m_0 \Delta A
   \left\{
      {y_d}_i\delta^i_{j}
     -(V_D^T)_i^{~k}{y_e}_k(V_E^{\dag})^k_{~j}
   \right\},
\label{eq:A-term correction 2}
\\
  (\widetilde{y}_e)_{ij} &\approx&
  m_0 A_e {y_e}_i \delta^i_{j}
\nonumber\\&&
  -\frac{m_0 A_0}{3+|A_0|^2}
   \frac{
      (\Delta m^2_E)_i^{~j}{y_e}_j
     +{y_e}_i(\Delta m^2_L)^i_{~j}
     }{m_0^2}
\nonumber\\&&
  -\frac{3}{5}m_0 \Delta A
   \left\{
     {y_e}_i\delta^i_{j}
     -(V_E^T)_i^{~k}{y_d}_k(V_D^{\dag})^k_{~j}
   \right\},
\label{eq:A-term correction 3}
\end{eqnarray}
\end{mathletters}
where $A_u$, $A_d$, $A_e$ and $\Delta A$ are given by
\begin{mathletters}
\begin{eqnarray}
  A_u &\approx& A_0
  -\frac{192}{5}g_5^2\frac{M_0}{m_0}t_G
  -\frac{276}{15}g_5^2\frac{M_0}{m_0}t_W
  +3y_t^2(t_G+t_W)+\sum^3_{i=1} {y_{\nu}}_i^2(t_G+t_R),
\\
  A_d &\approx& A_0
  -\frac{168}{5}g_5^2\frac{M_0}{m_0}t_G
  -\frac{88}{5}g_5^2\frac{M_0}{m_0}t_W,
\\
  A_e &\approx& A_0
  -\frac{168}{5}g_5^2\frac{M_0}{m_0}t_G
  -\frac{48}{5}g_5^2\frac{M_0}{m_0}t_W,
\\
  \Delta A &\approx&
  -20 g_5^2 \frac{M_0}{m_0} t_G.
\label{eq:delta-a}
\end{eqnarray}
\end{mathletters}
The second terms in the right-hand side of
Eqs.~(\ref{eq:A-term-correction}) are
induced by the radiative corrections from the Yukawa interactions.
The third terms in Eqs.~(\ref{eq:A-term correction 2}) and
 (\ref{eq:A-term correction 3}) come
 from the higher dimensional operators.
They break proportionality between the trilinear scalar coupling
matrix and the corresponding Yukawa coupling matrix and generate
flavor mixings in the left-right mixing mass matrices of squarks and
sleptons after the EW symmetry breaking.
If $\Delta A_0 \neq 0$ in Eq.~(\ref{eq:minimal-SUGRA-dA0}),  we have an
extra contribution, $\Delta A_0$ to Eq.~(\ref{eq:delta-a}) and there are
corrections of order $m_0^2 A_0\Delta A_0\xi t_G$ to
Eqs.~(\ref{eq:mass-correction}) and
corrections of order $m_0 \Delta A_0\xi t_X$ ($X=G,R,W$) to
Eqs.~(\ref{eq:A-term-correction}).
Notice that even if we assume $\Delta A_0 = 0$ at the Planck scale
$\Delta A$ is induced by  the gauge interaction
 as shown in Eq.~(\ref{eq:delta-a}) because
the renormalization of $\widetilde{\lambda}_d$ and
$\widetilde{\kappa}_d$ are different due to the wave-function
renormalization of $\Sigma$.
We therefore expect analysis with $\Delta A_0 =0$
 gives us a qualitative feature for general cases.

\subsection{
  The muon anomalous magnetic moment,
  FCNC and LFV processes
}
\label{sec:fcnc-lfv-processes}

Let us discuss the muon anomalous magnetic
moment, FCNC and LFV processes in the model according to the
approximations of the previous section.
In the following we make a simplification in the neutrino sector to
estimate possible deviations from the SM in the FCNC/LFV processes.
We assume the Majorana masses of the right-handed neutrinos are
universal at the Majorana mass scale $M_R$ as
$(M_{\nu})_{ij}=\delta_{ij}M_R$.
We also neglect any CP violating phase in the model except for the
Kobayashi-Maskawa phase and assume that $V_N$, $V_L$, $O_{\nu}$ and
$V_{\rm MNS}$ are real matrices and
$\alpha_{D,E}=\beta_{D,E}=\gamma_{D,E}=\delta=0$ in
Eq.~(\ref{eq:GUT mixing angle}) and $\phi_{Ui}=0$ in
 Eq.~(\ref{eq:colored-Higgs-Yukawa}).
If we include these phases, new contributions to the electron
and neutron electric dipole moments (EDMs) are induced 
so that we have to take into
account constraints to SUSY parameters from the upper bound of the EDMs.
In the numerical calculation we evaluate these EDMs and check that these
constraints are satisfied in the case that new CP phases are set to
vanish.
With the above simplification, the mixing matrix $V_L$ and the neutrino
Yukawa coupling can be related to the low energy observables according to
 Eq.~(\ref{eq:ynu}),
\begin{equation}
  V_L = V_{\rm MNS}^{\dag},
~~~~~
  {y_{\nu}}_i = \sqrt{M_R {m_{\nu}}_i}/\langle H_2 \rangle.
\label{eq:seesaw}
\end{equation}
We parameterize the MNS matrix assuming maximal mixing for the
atmospheric neutrino oscillation as follows: 
\begin{eqnarray}
  V_{\rm MNS} &=&
  \left(
    \begin{array}{ccc}
      \cos\theta_{\rm sun} & \sin\theta_{\rm sun} & 0 \\
      -\frac{\sin\theta_{\rm sun}}{\sqrt{2}} &
       \frac{\cos\theta_{\rm sun}}{\sqrt{2}} &
       \frac{1}{\sqrt{2}} \\
       \frac{\sin\theta_{\rm sun}}{\sqrt{2}} &
      -\frac{\cos\theta_{\rm sun}}{\sqrt{2}} &
       \frac{1}{\sqrt{2}}
    \end{array}
  \right),
\label{eq:MNS matrix}
\end{eqnarray}
where $\theta_{\rm sun}$ is the mixing angle for the solar neutrino
oscillation.
We assume the $1$-$3$ element of the MNS matrix is zero in our analysis
because it is known to be small from the result of the CHOOZ experiment
\cite{Apollonio:1999ae}.
We  will comment on the nonzero case later.
We assume the hierarchical pattern of neutrino mass, namely
${m_{\nu}}_1 < {m_{\nu}}_2 \ll {m_{\nu}}_3$.
From the following relation,
\begin{equation}
  {m_{\nu}}_2^2 = \Delta m_{\rm sun}^2+{m_{\nu}}_1^2,
~~~~~ 
  {m_{\nu}}_3^2 = \Delta m_{\rm atm}^2+{m_{\nu}}_2^2,
\end{equation}
 where $\Delta m^2_{\rm sun}$ and
$\Delta m^2_{\rm atm}$ are the mass differences of
 solar and atmospheric neutrino
oscillation, 
${m_{\nu}}_3$ and ${m_{\nu}}_2$ are determined once we fix ${m_{\nu}}_1$.
Then using Eq.~(\ref{eq:seesaw}) we can calculate ${y_{\nu}}_i$ for a fixed value of
$M_R$.

\subsubsection{
  The muon anomalous magnetic moment
}

We consider the SUSY contribution to the muon anomalous magnetic moment
\cite{99GoOkSh,g-2new,g-2,tanb-enhancement}.
The muon anomalous magnetic moment $a_{\mu}$ is defined by the following
effective Lagrangian:
\begin{equation}
  {\cal L}^{g-2} = \frac{1}{2}\left(\frac{e}{2 m_{\mu}}\right) a_{\mu}
  \overline{\mu} \sigma^{\alpha \beta} \mu F_{\alpha \beta},
\end{equation}
where $e$ is the positron charge, $m_{\mu}$ is the muon mass 
 , $F_{\alpha \beta}$ is the electromagnetic field tensor and 
 $\sigma_{\alpha \beta}=i[\gamma_{\alpha},\gamma_{\beta}]/2$.
The SUSY contribution to $a_{\mu}$ ($\equiv a^{\rm SUSY}_{\mu}$) is
obtained from the flavor diagonal parts of photon-penguin diagrams
including smuon-neutralino and sneutrino-chargino.
\amu\ depends on the slepton mass and the neutralino/chargino mass
and mixing, but it is rather insensitive to the flavor mixing of the
slepton sector.
Therefore we expect \amu\ is almost the same as a result in the minimal
SUGRA model.
In the MSSM based on the minimal SUGRA, it is known that the main
contribution comes from the sneutrino-chargino diagram which contains a
component proportional to $\mu \tan\beta$ \cite{tanb-enhancement}.
Then \amu\ preferred by the recent results from BNL E821 experiment is
achieved in the large $\tan\beta$ region of the parameter space.
In this region the sign of \amu\ is correlated to the branching ratio of
\bsg\ through the sign of the Higgsino mass parameter $\mu$ so that
\amu\ is positive when \bsg\ is suppressed and negative when \bsg\
is enhanced.

\subsubsection{\meg, \tmg}

We consider LFV decays of charged leptons.
Radiative decays of charged leptons occur through photon-penguin
diagrams including sleptons, neutralinos and charginos.
The effective Lagrangian for these processes is described as follows:
\begin{eqnarray}
  {\cal L}^{\rm LFV} &=& -\frac{4G_F}{\sqrt{2}}
  \left\{
      {m_e}_i A_R^{ij}(\overline{l_{Ri}}\sigma^{\mu \nu}l_{Lj})F_{\mu \nu}
    + {m_e}_i A_L^{ij}(\overline{l_{Li}}\sigma^{\mu \nu}l_{Rj})F_{\mu \nu}
  \right\}
  +{\rm H.c.} ~~~~~ (i>j),
\label{eq:mueg amplitudes}
\end{eqnarray}
where $G_F$ is the Fermi constant and $i,j$ denote generation indices.
$A^{ij}_R$ corresponds to the amplitude for
$l_{i}^+ \to l_{j}^+ \gamma_R$ and $A^{ij}_L$ for
$l_{i}^+ \to l_{j}^+ \gamma_L$.
The branching ratios are calculated from these amplitudes as
$\text{B}(l_i^+ \to l_j^+ \gamma) =
384\pi^2(|A_R^{ij}|^2 + |A_L^{ij}|^2)$.

\begin{figure} 
\begin{center}
\makebox[0em]{
\def\epsfsize#1#2{0.4#1}
\epsfbox{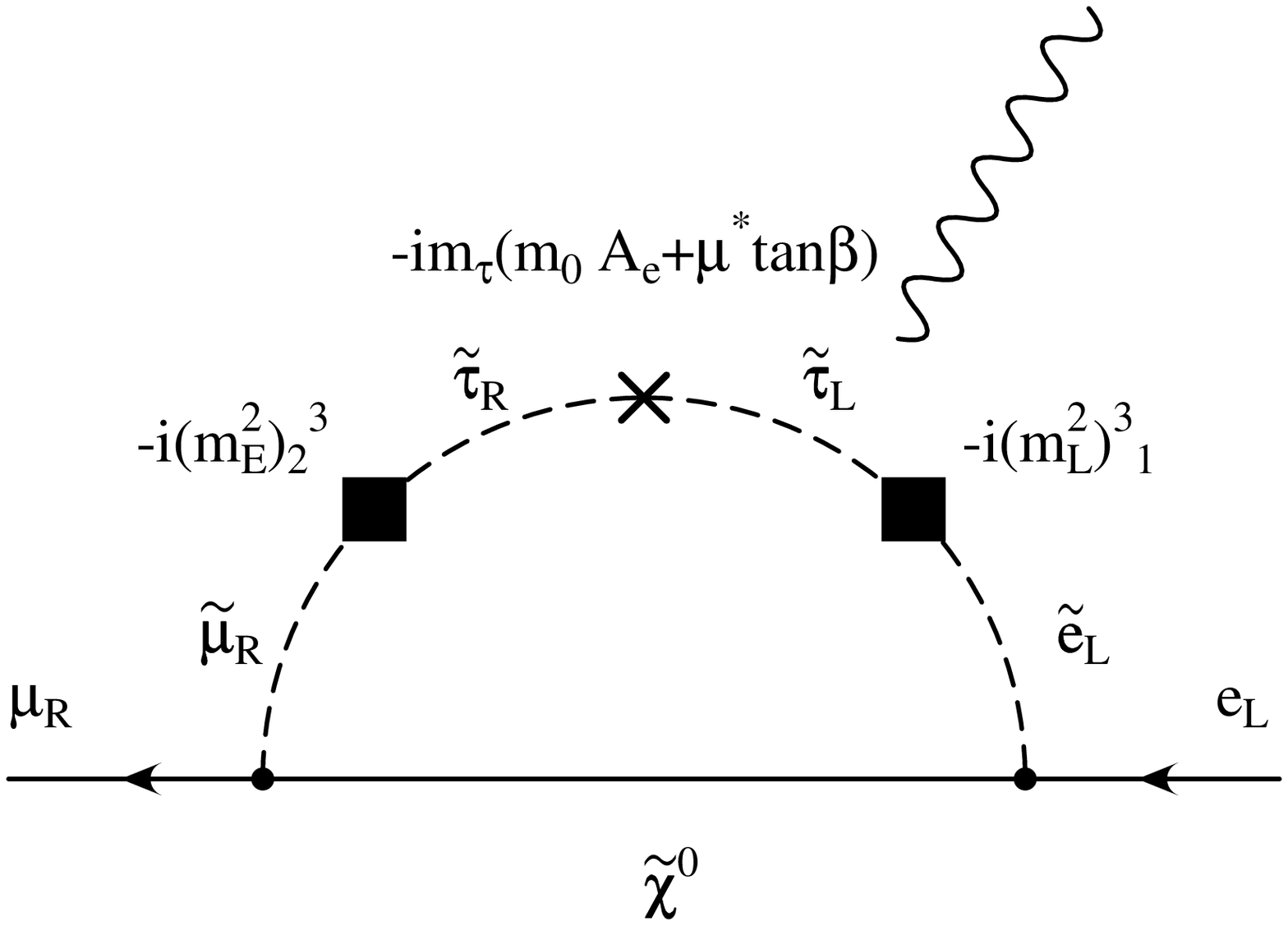}
\def\epsfsize#1#2{0.4#1}
\epsfbox{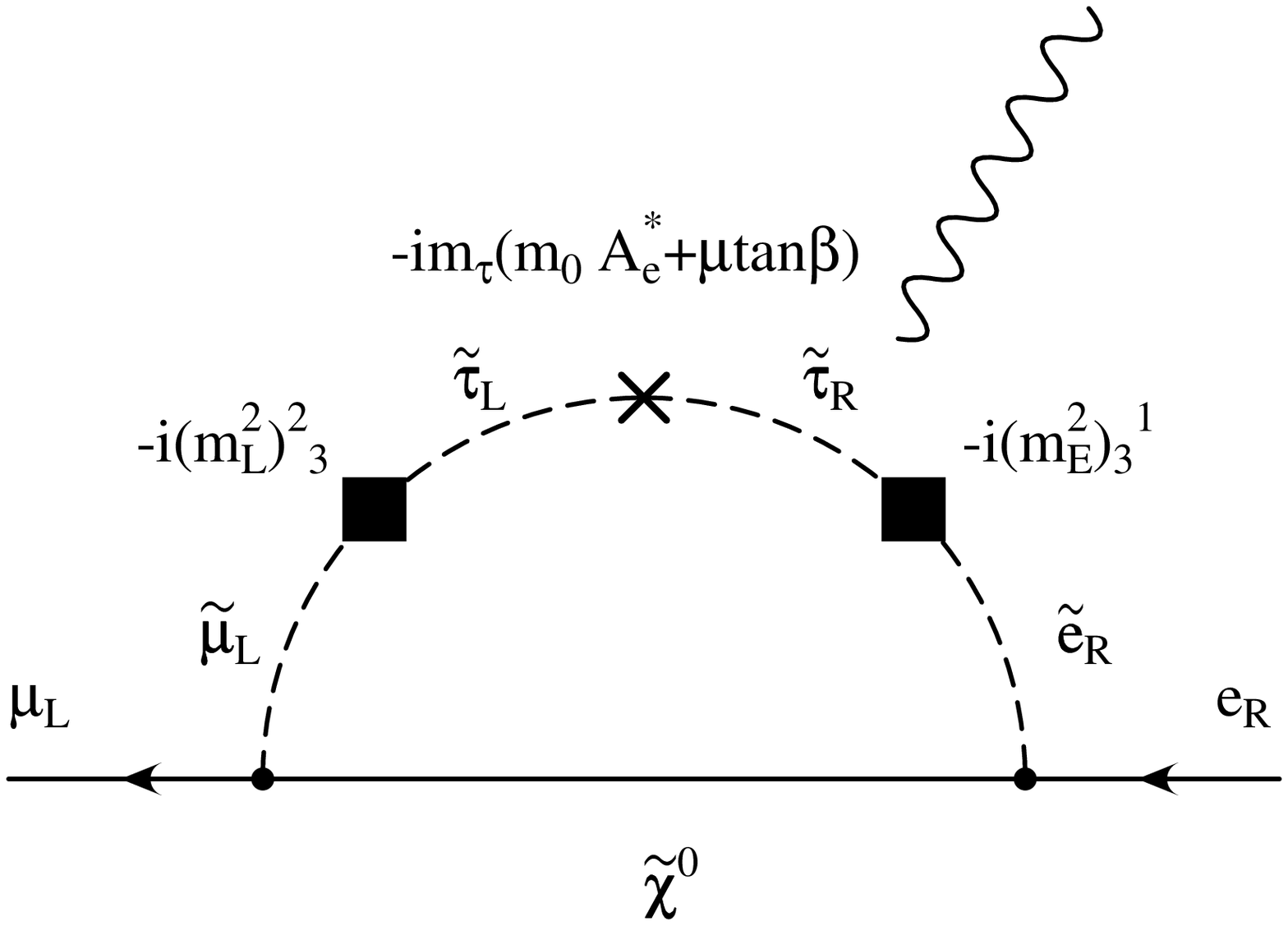}
}
\end{center}
\caption{
Possible large contributions to \meg\ amplitudes, $A_R^{21}$ and
$A_L^{21}$ in the present model. They are enhanced with a factor
$m_{\tau}/m_{\mu}$ compared to the other
contributions.
}
\label{fig:mueg1}
\end{figure}

\begin{figure} 
\begin{center}
\makebox[0em]{
\def\epsfsize#1#2{0.4#1}
\epsfbox{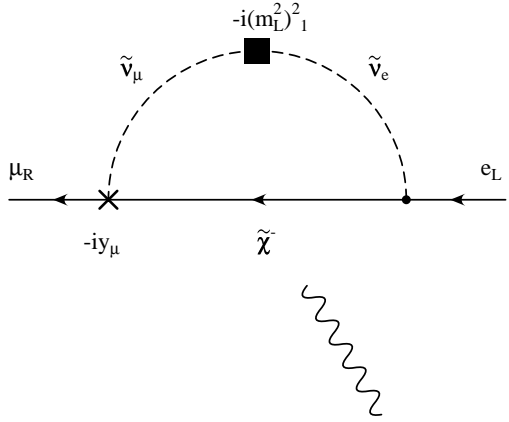}
}
\end{center}
\caption{
A possible large contribution to \meg\  amplitude $A_R^{21}$ in the
present model.
}
\label{fig:mueg2}
\end{figure}

\begin{figure} 
\begin{center}
\makebox[0em]{
\def\epsfsize#1#2{0.4#1}
\epsfbox{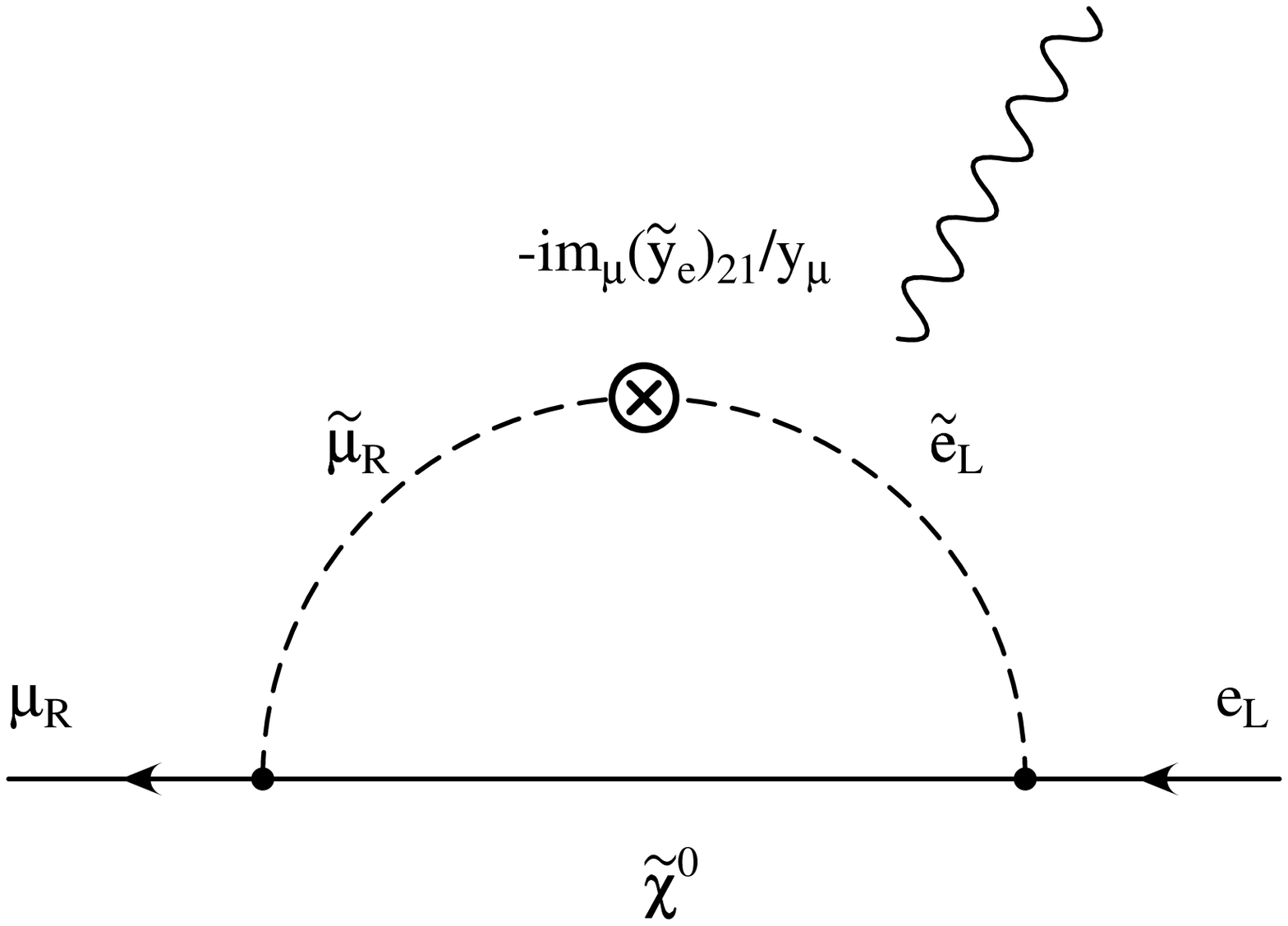}
\def\epsfsize#1#2{0.4#1}
\epsfbox{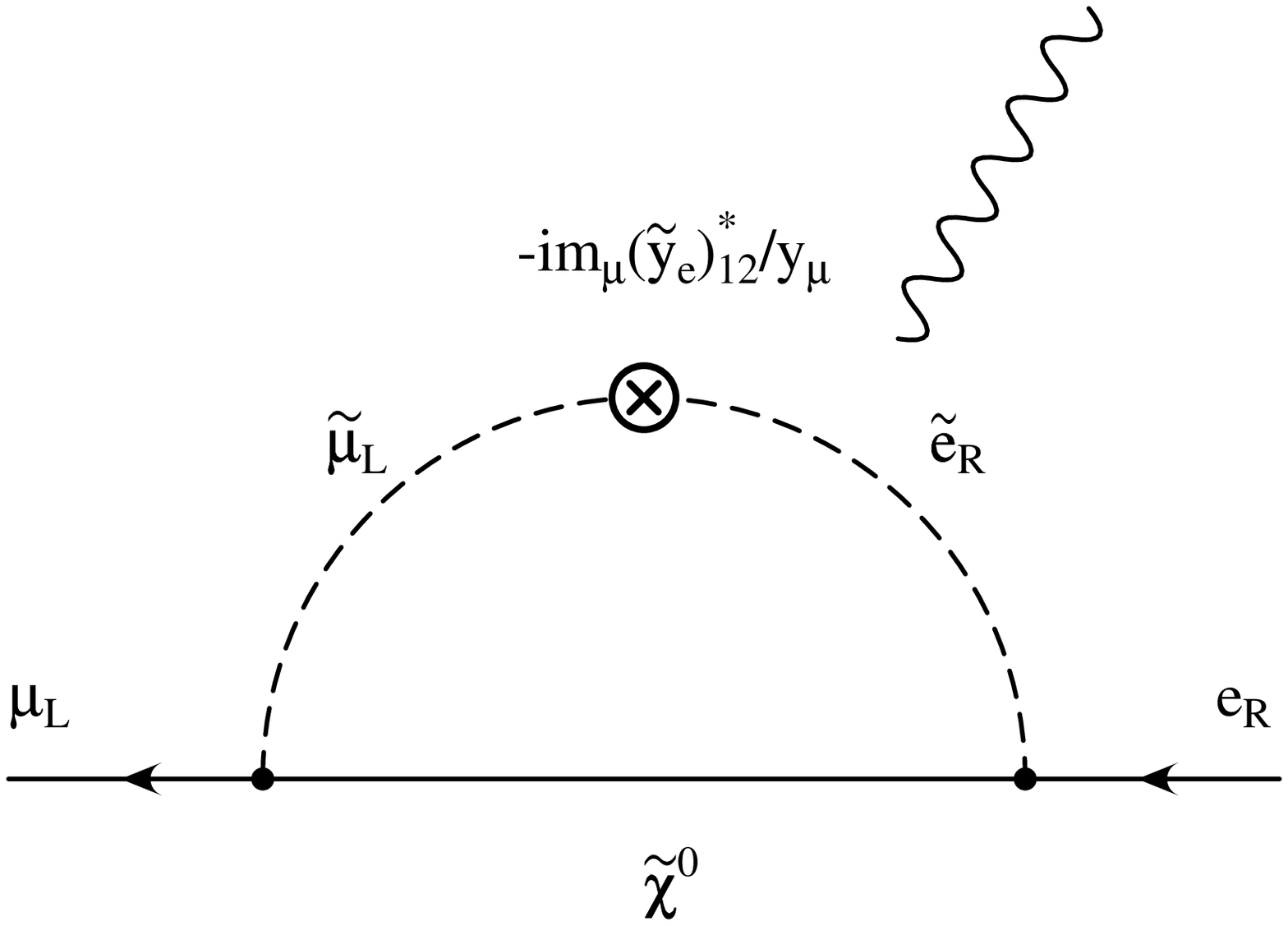}
}
\end{center}
\caption{
Possible large contributions to \meg\  amplitudes $A_R^{21}$ and
$A_L^{21}$ in the present model.
The flavor mixing in the left-right mixing is induced by the gauge
interaction between the Planck scale and the GUT scale through the
wavefunction renormalization of {\bf 24} Higgs supermultiplet.
}
\label{fig:mueg3}
\end{figure}

For \meg, it is known that if both left-handed and right-handed sectors
have flavor mixing, there are the diagrams which have an enhancement
factor $m_{\tau}$ as shown in Fig.~\ref{fig:mueg1}
\cite{LFV,98HiNoYa,99HiNo}.
In our model we find that diagrams corresponding to
Figs.~\ref{fig:mueg2} and \ref{fig:mueg3} also give large contributions.
The flavor mixing in the left-right mixing term in Fig.~\ref{fig:mueg3}
is induced by renormalization between the Planck and GUT scale as shown
in Eq.~(\ref{eq:A-term correction 3}).
Approximate formulas for $A_R^{21}$ and $A_L^{21}$ from these
contributions are given by
\begin{mathletters}
\begin{eqnarray}
  A_R^{21} &\approx&
  \frac{1}{\sqrt{2}}\sin 2\theta_{\rm sun}
    ({{y_{\nu}}_2}^2-{{y_{\nu}}_1}^2)
    y_t^2
    \left[
      \left\{
        (V_{\rm CKM})^3_{~1}\sin\theta_E +(V_{\rm CKM})^3_{~2}\cos\theta_E
      \right\} \frac{m_{\tau}}{m_{\mu}}a^n_2
      - a^c
    \right]
\nonumber\\&&
  - \cos\theta_E\sin\theta_D~a^n_1,
\\
  A_L^{21} &\approx&
  - ({{y_{\nu}}_3}^2-{{y_{\nu}}_2}^2) 
    y_t^2
    \left\{
      (V_{\rm CKM})^3_{~1}\cos\theta_E -(V_{\rm CKM})^3_{~2}\sin\theta_E
    \right\} \frac{m_{\tau}}{m_{\mu}}a^n_2
    +\sin\theta_E\cos\theta_D~a^n_1,
\end{eqnarray}
\label{eq:mueg-approximate-formulas}
\end{mathletters}
where we explicitly show $\theta_{\rm sun}$, ${y_{\nu}}_i$, $y_t$,
$\theta_E$ and $\theta_D$ dependence.
$a^n_2$, $a^c$ and $a^n_1$ are functions of the slepton masses and the
chargino and neutralino masses and mixings.
These contributions correspond to 
Fig.~\ref{fig:mueg1}, Fig.~\ref{fig:mueg2} and Fig.~\ref{fig:mueg3}, respectively.
The explicit forms of the functions are given in Appendix
\ref{sec:approximation-mueg}.
Because $(V_{\rm CKM})^3_{~2} \gg (V_{\rm CKM})^3_{~1}$, the mixing angle
$\theta_E$ can enhance the amplitude $A_L^{21}$ compared to the case
$\theta_E=0$.

The ratio of the magnitudes of $A^{21}_L$ and $A^{21}_R$ can be measured
by the P-odd asymmetry of \meg\ process, \Ameg\ \cite{polarization}.
With the help of initial muon polarization, we define \Ameg\ as follows:
\begin{mathletters}
\begin{eqnarray}
  \frac{d\text{B}(\mu^+ \to e^+\,\gamma)}{d\cos\theta} &=& 
  \frac{1}{2}\text{B}(\mu^+ \to e^+\,\gamma)
  \left\{
    1+A(\mu^+ \to e^+\,\gamma)P\cos\theta
  \right\},
\\
  A(\mu^+ \to e^+\,\gamma) &=&
  \frac{|A_L^{21}|^2-|A_R^{21}|^2}{|A_L^{21}|^2+|A_R^{21}|^2},
\end{eqnarray}
\end{mathletters}
where $P$ is the polarization of initial $\mu^+$ and $\theta$ is the angle
between the polarization and the momentum of the decay positron.
For \tmg, a similar P-odd asymmetry can be measured in the $e^+e^- \to
\tau^+\tau^-$ process using spin correlation of the $\tau$ pair
\cite{00KiOk}.

\subsubsection{\bsg}

\begin{figure}
\begin{center}
\makebox[0em]{
\def\epsfsize#1#2{0.4#1}
\epsfbox{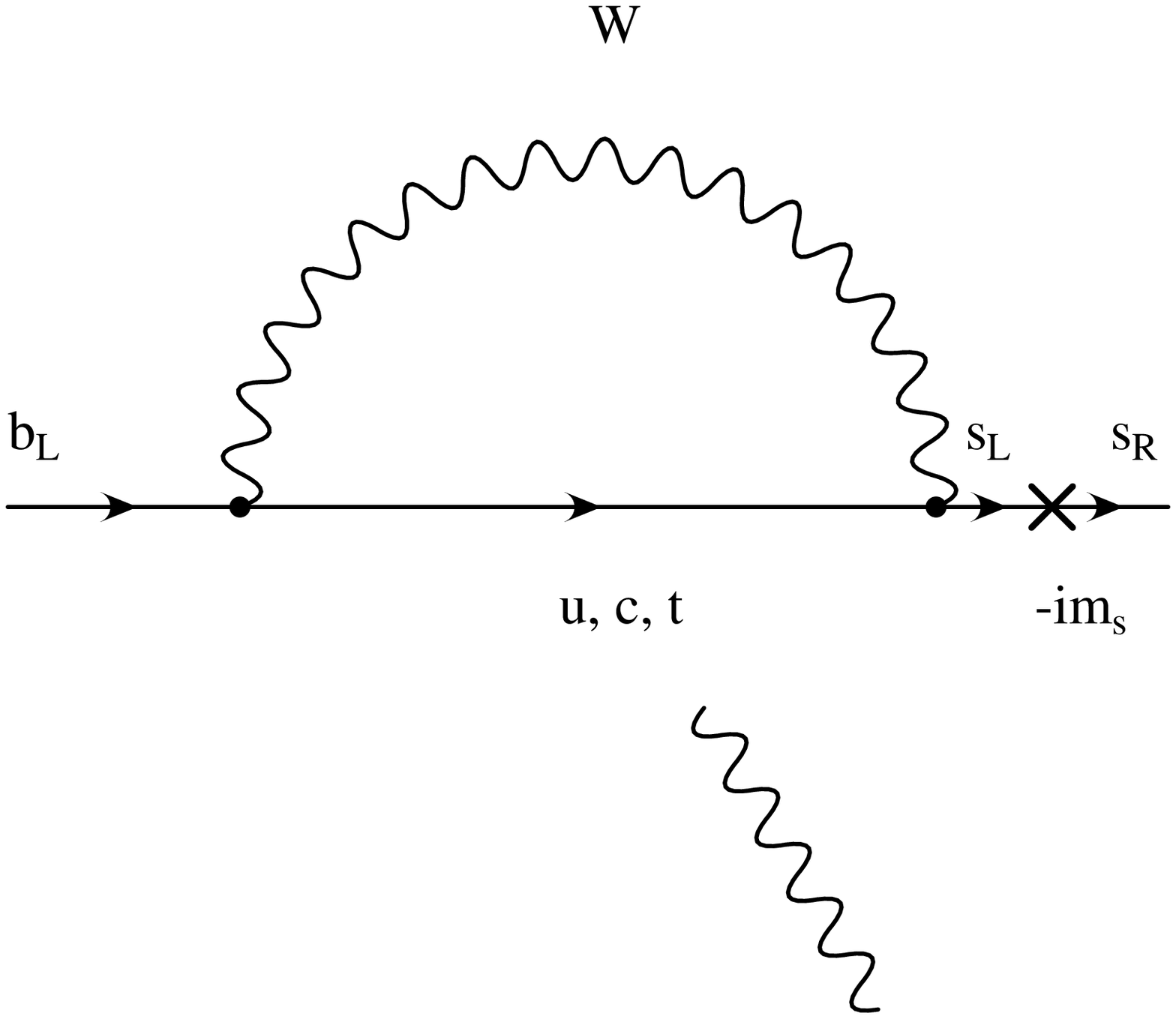}
\def\epsfsize#1#2{0.4#1}
\epsfbox{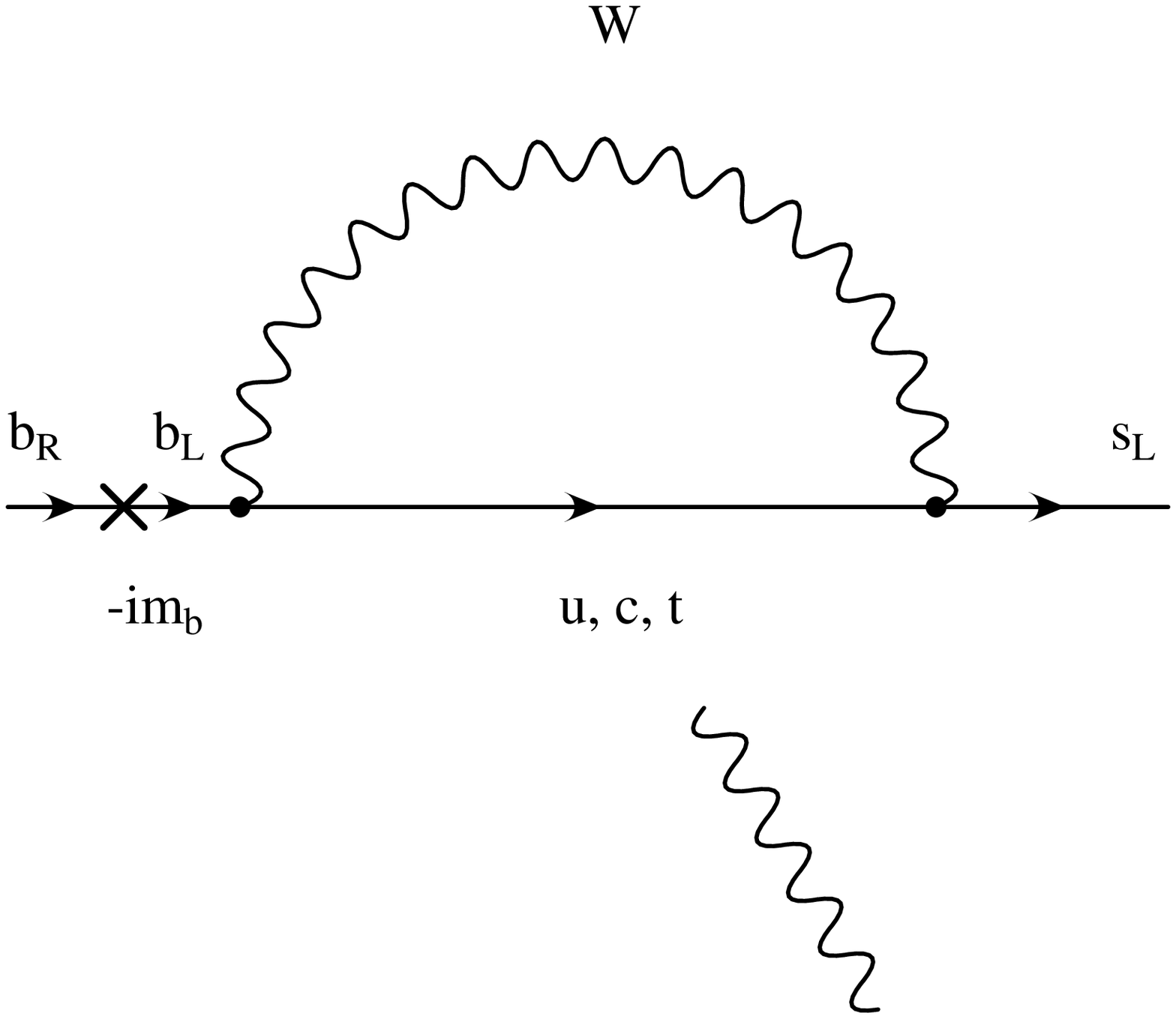}
}
\end{center}
\caption{
One-loop diagrams which contribute to \bsg\ in the SM.
The contribution to $C_7'$ (left) is suppressed by $m_s/m_b$
compared to the contribution to $C_7$ (right).
}
\label{fig:bsg-SM}
\end{figure}

\begin{figure}
\begin{center}
\makebox[0em]{
\def\epsfsize#1#2{0.4#1}
\epsfbox{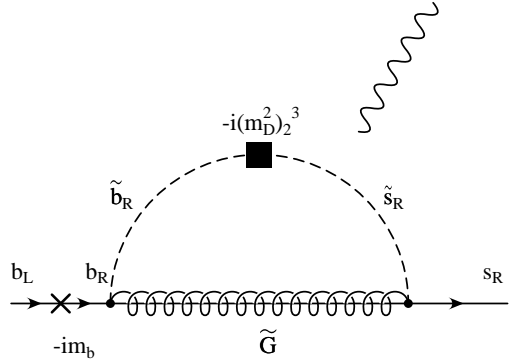}
}
\end{center}
\caption{
A possible large contribution to \bsg\ amplitude $C_7'$
in the SU(5)RN SUSY GUT which is not suppressed by $m_s/m_b$. 
}
\label{fig:bsg-gluino}
\end{figure}

The $\Delta B =1$ FCNC effective Lagrangian for the radiative $B$ decay
is written as follows:
\begin{eqnarray}
  {\cal L}^{\Delta S =1} &=& -\frac{4G_F}{\sqrt{2}}
  \left\{
      C_7'(\overline{s_R}\sigma^{\mu \nu} b_L) F_{\mu\nu}
    + C_7 (\overline{s_L}\sigma^{\mu \nu} b_R) F_{\mu\nu}
  \right\}
  +{\rm H.c.}.
\end{eqnarray}
In the SM case, the process occurs through photon penguin diagrams which
exchange a $W$ boson as Fig.~\ref{fig:bsg-SM} and $C_7'$ is suppressed by
a factor of $m_s/m_b$ compared to $C_7$.
In the minimal SUGRA model without the right-handed neutrino
supermultiplets, the flavor mixing in the squark mass matrices only
appears in that of left-handed squark and the same argument can be
applied.
In the present model, however, a gluino exchanging diagram
(Fig.~\ref{fig:bsg-gluino}) can give a large contribution to $C_7'$
because of the new flavor mixing in the right-handed down-type squarks
$(m_D^2)_2^{~3}$.

If $C_7'$ has a similar magnitude as $C_7$, the time-dependent CP
asymmetry of \Bsg\ may be observed where $M_s$ is a CP eigenstate which
includes a strange quark such as $K_1$ ($\to K_S\rho^0$) or $K^*$
($\to K_S\pi^0$) \cite{01BaGoOkOk,atbsg}.
The asymmetry is defined as follows:
\begin{equation}
  \frac{\Gamma(t)-\overline{\Gamma}(t)}
       {\Gamma(t)+\overline{\Gamma}(t)} =
  \eta A_{CP}(B\to M_s\,\gamma)\sin\Delta m_{B_d} t,
~~~~~
  A_{CP}(B\to M_s\,\gamma) =
  \frac{2\text{Im}(e^{-i\theta_B}C_7 C_7')}{|C_7|^2+|C_7'|^2},
\end{equation}
where $\Gamma(t)$ ($\overline{\Gamma}(t)$) is the decay width of
$B^0(t)\to M_s\,\gamma$ ($\overline{B}^0(t)\to M_s\,\gamma$). 
$\eta$ is $+1$ if $M_s$ is a CP even state and $-1$ if $M_s$ is a CP odd state.
$\theta_B$ is the phase of \bdbd\ mixing amplitude $M_{12}(B_d)$
which defined below in Eq.~(\ref{eq:dmbd}).
In the SM case, this asymmetry is only a few percent, however, it may be
considerably enhanced by the new SUSY contribution to $C_7'$. 

\subsubsection{\ek}

\begin{figure}
\begin{center}
\makebox[0em]{
\def\epsfsize#1#2{0.4#1}
\epsfbox{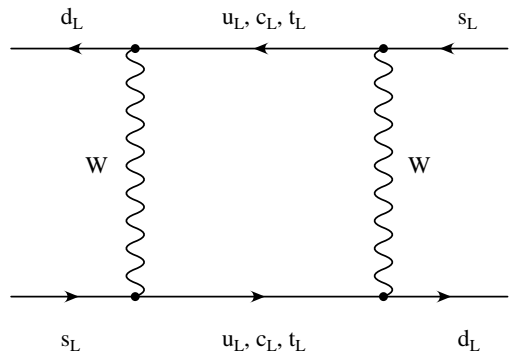}
}
\end{center}
\caption{
A one-loop diagram which contributes to $\varepsilon_K$ in the SM.
A similar diagram which contribute to \bdbd\ (\bsbs)  mixing in the SM
can be obtained by replacing quarks in the external lines so that
$s \to b$ ($s \to b$ and $d \to s$).
}
\label{fig:epsilon_SM}
\end{figure}

\begin{figure}
\begin{center}
\makebox[0em]{
\def\epsfsize#1#2{0.4#1}
\epsfbox{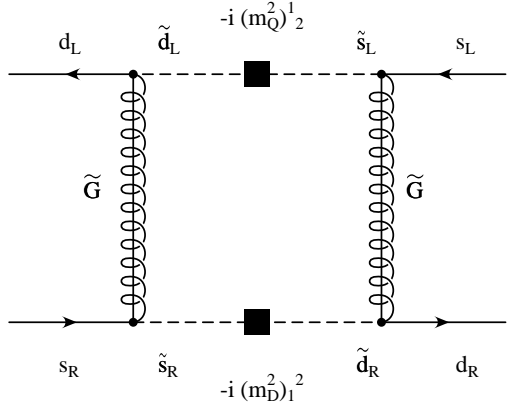}
}
\end{center}
\caption{
A possible large contribution to $\varepsilon_K$ in the 
 SU(5)RN SUSY GUT.
There is also a crossed diagram because of Majorana nature of gluino.
}
\label{fig:epsilon_gluino}
\end{figure}

\kk\ mixing is described by the $\Delta S=2$ FCNC effective Lagrangian.
The general form is given by
\begin{eqnarray}
  {\cal L}^{\Delta S=2} &=&
  -\frac{8 G_F}{\sqrt{2}}
  \Biggl\{
    \frac{1}{2} g^V_R
    (\overline{d}_R^{\alpha}\gamma^{\mu} s_{R\alpha})
    (\overline{d}_R^{\beta}\gamma_{\mu} s_{R\beta})
  + \frac{1}{2} g^V_L
    (\overline{d}_L^{\alpha}\gamma^{\mu} s_{L\alpha})
    (\overline{d}_L^{\beta}\gamma_{\mu} s_{L\beta})
\nonumber\\&&\phantom{-\frac{8 G_F}{\sqrt{2}}}
  + \frac{1}{2} g^S_{RR}
    (\overline{d}_L^{\alpha} s_{R\alpha})
    (\overline{d}_L^{\beta} s_{R\beta})
  + \frac{1}{2} g^S_{LL}
    (\overline{d}_R^{\alpha} s_{L\alpha})
    (\overline{d}_R^{\beta} s_{L\beta})
\nonumber\\&&\phantom{-\frac{8 G_F}{\sqrt{2}}}
  + \frac{1}{2} {g^S_{RR}}'
    (\overline{d}_L^{\alpha} s_{R\beta})
    (\overline{d}_L^{\beta} s_{R\alpha})
  + \frac{1}{2} {g^S_{LL}}'
    (\overline{d}_R^{\alpha} s_{L\beta})
    (\overline{d}_R^{\beta} s_{L\alpha}) 
\nonumber\\&&\phantom{-\frac{8 G_F}{\sqrt{2}}}
  + g^S_{RL}
    (\overline{d}_L^{\alpha} s_{R\alpha})
    (\overline{d}_R^{\beta} s_{L\beta})
  + {g^S_{RL}}'
    (\overline{d}_L^{\alpha} s_{R\beta})
    (\overline{d}_R^{\beta} s_{L\alpha})
  \Biggr\}
  + {\rm H.c.},
\label{eq:epsilon_K}
\end{eqnarray}
where $\alpha$ and $\beta$ denote color indices.
The explicit forms of effective coupling constants in the above formula
 are given in Appendix~\ref{sec:FCNC-effective-couplings}.
The CP violation parameter in \kk\ mixing, \ek\ is calculated from the
above formula as follows:
\begin{equation}
  \varepsilon_K =
  \frac{e^{\frac{\pi}{4}i}}{\sqrt{2}}
  \frac{\text{Im}\{M_{12}(K)\}}{\Delta m_K}, 
~~~~~
  M_{12}(K) =
  -\frac{\langle K^0|{\cal L}^{\Delta S=2}|\bar{K}^0\rangle}{2 m_K}.
\end{equation}
In the case of the SM, the process occurs through a box diagram in which
two $W$ bosons are exchanged between the down-type quarks
(Fig.~\ref{fig:epsilon_SM}) so that it is dominated by the $g^V_L$-term.
However, we have a new flavor mining, $(m_D^2)_1^{~2}$ in the
right-handed down-type squark sector.
We can draw gluino exchanging diagrams as Fig.~\ref{fig:epsilon_gluino}
which include the CP violating phase of CKM matrix in $(m_Q^2)^1_{~2}$
on one of the squark lines and large flavor mixing in $(m_D^2)_1^{~2}$ on
the other.
These diagrams contribute to the coupling constants $g^S_{RL}$ and
${g^S_{RL}}'$.
In the actual numerical calculation we first derive the effective Lagrangian
at the energy scale $M_{\rm SUSY}$.
According to the reference \cite{QCD-BBKK}, we include QCD corrections and derive
the effective Lagrangian at the hadronic scale.
The matrix elements for the dominant operators are parameterized as follows:
\begin{mathletters}
\begin{eqnarray}
  \langle K^0|
  (\overline{d}_L^{\alpha}\gamma^{\mu} s_{L\alpha})
  (\overline{d}_L^{\beta}\gamma_{\mu} s_{L\beta})
  |\bar{K}^0 \rangle
  &=&
  \frac{2}{3}m_K^2 f_K^2 B_K,
\\
  \langle K^0|
  (\overline{d}_R^{\alpha} s_{L\alpha})
  (\overline{d}_L^{\beta} s_{R\beta})
  |\bar{K}^0 \rangle
  &=&
  \frac{1}{2}\left(\frac{m_K}{m_s + m_d}\right)^2m_K^2 f_K^2 (B_K)_{RL}^S,
\\
  \langle K^0|
  (\overline{d}_R^{\alpha} s_{L\beta})
  (\overline{d}_L^{\beta} s_{R\alpha})
  |\bar{K}^0 \rangle
  &=&
  \frac{1}{6}\left(\frac{m_K}{m_s + m_d}\right)^2m_K^2 f_K^2 {(B_K)_{RL}^S}',
\end{eqnarray}
\end{mathletters}
where $B_K$, $(B_K)^S_{RL}$ and ${(B_K)^S_{RL}}'$ are bag parameters
calculated by the lattice QCD method.
Because there is a large enhancement factor of order $(m_K/m_s)^2$ in
the matrix elements and large mixngs originate from the MNS matrix, SUSY
contribution to $\varepsilon_K$ is expected to be large in this model.

\subsubsection{\bdbd/\bsbs\ mixing }

\begin{figure}
\begin{center}
\makebox[0em]{
\def\epsfsize#1#2{0.4#1}
\epsfbox{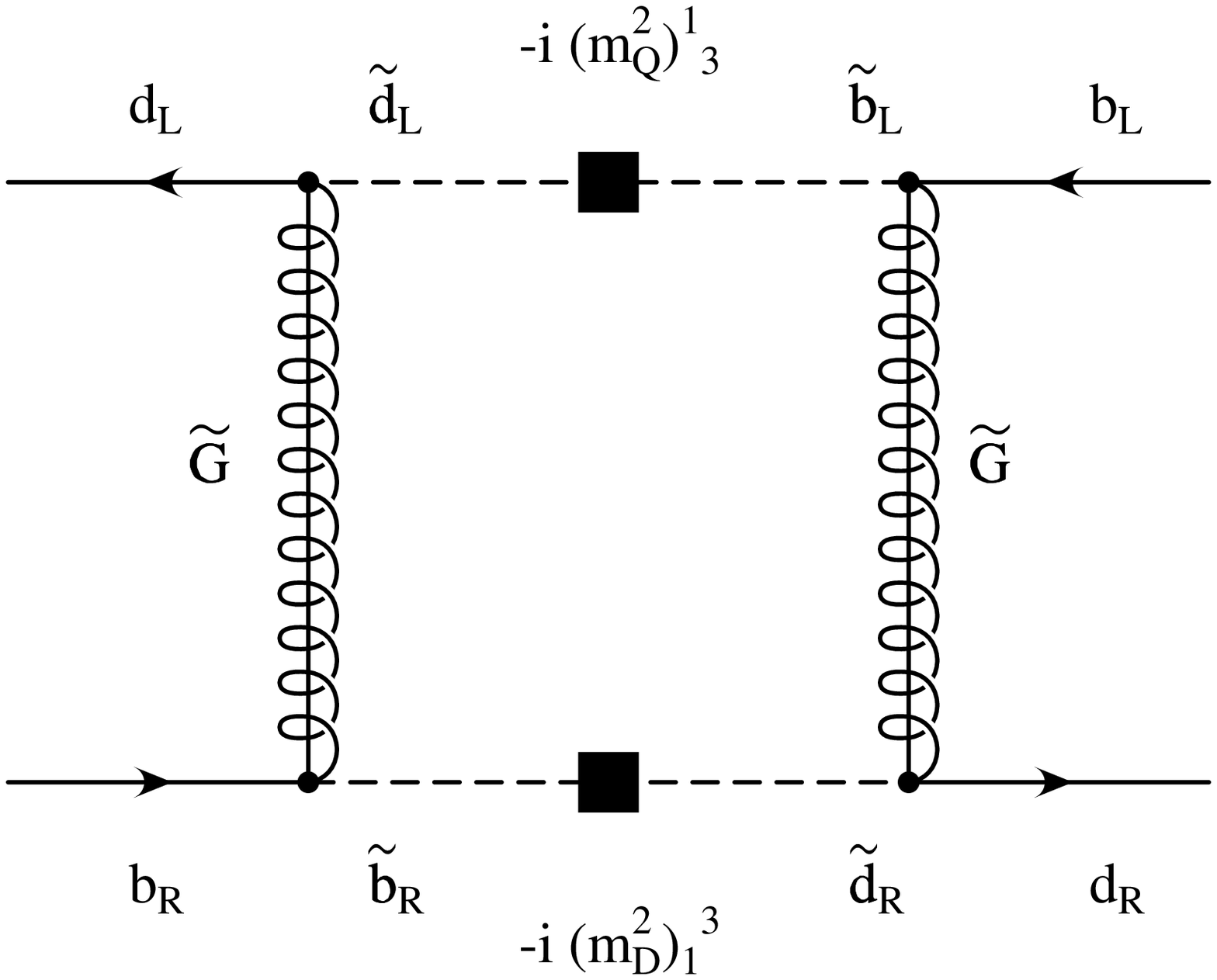}
\def\epsfsize#1#2{0.4#1}
\epsfbox{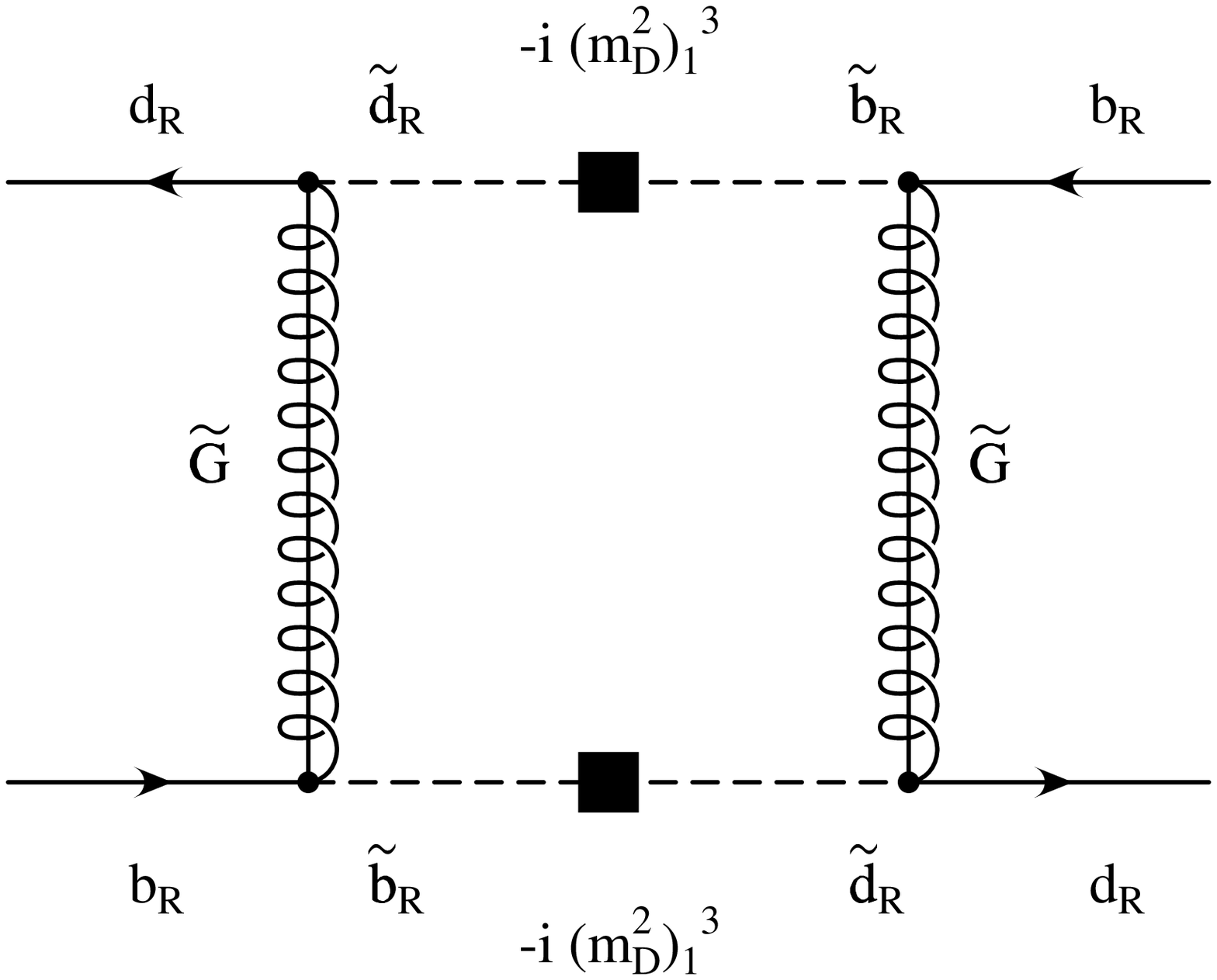}
}
\end{center}
\caption{
Possible large contributions to \bdbd\ mixing in the present
model. Similar diagrams which contribute to \bsbs\  mixing can be
obtained by replacing the down quark/squark in the diagrams to the
strange quark/squark.
There are also crossed diagrams.
}
\label{fig:dmbd_gluino}
\end{figure}

The $\Delta B=2$ effective Lagrangians for \bdbd\ and \bsbs\ mixings are
parameterized in the same manner as \kk\ mixing.
For \bdbd\ mixing, it is obtained by replacing the strange quark with
the bottom quark in Eq.~(\ref{eq:epsilon_K}).
For \bsbs\ mixing, we further replace the down quark by the strange quark.
The mass difference of \bdbd\ mixing, \dmbd\ is calculated from the
effective Lagrangian as follows:
\begin{equation}
  \Delta m_{B_d} = 2|M_{12}(B_{d})|,
~~~~~
  M_{12}(B_{d}) =
  -\frac{\langle B^0_d|{\cal L}^{\Delta B=2}|\bar{B}^0_d \rangle}
        {2 m_{B_{d}}}.
\label{eq:dmbd}
\end{equation}
In the SM case, this process occurs through the $W$ boson exchanging
diagram and the $g^V_L$-term gives a dominant contribution.
In the present case, there are new diagrams as shown in
Fig.~\ref{fig:dmbd_gluino} which contain the new flavor mixing
 in $(m_D^2)_1^{~3}$ on one of the down-type squark lines or on the both of
them.
The former contributes to $g^S_{RL}$ and ${g^S_{RL}}'$ and the latter to
$g^V_R$.
Unlike \kk\ mixing, the scalar-scalar matrix elements do not
have an enhancement factor for \bdbd\ mixing case because $m_{B_d}/m_b$
is $O(1)$.
Similar argument holds for \bsbs\ mixing.
In the numerical calculation we use next leading order QCD correction
 for $g^V_R$ and
$g^V_L$ and leading order QCD formulas
 for other contributions \cite{QCD-BBKK} and bag parameters
are calculated by the lattice QCD.
Numerical values are shown later.

\section{
  Results of numerical calculations
}
\label{sec:numerical-results}

In this section we present our numerical results on the FCNC/LFV
processes in the SU(5)RN SUSY GUT.

In the present analysis we assume that the SUSY breaking terms have
the minimal SUGRA type boundary condition at the Planck scale and that
the K\"{a}hler potential is flat.
Adopting the simplifications discussed in the previous section, we have
the following input parameters.
\begin{itemize}
\item Parameters at the Planck scale:
the universal scalar mass $m_0$, the universal gaugino mass $M_0$, and
the universal coefficient for the scalar couplings $A_0$.
\item Parameters at the GUT scale:
mixing angles $\theta_D$ and $\theta_E$.
\item Parameter at the right-handed neutrino mass scale:
Majorana mass of the right-handed neutrino $M_R$, which is also used as
the matching scale.
\item Parameters at the EW scale:
quark, lepton and neutrino masses, mixing matrices $V_{\rm CKM}$ and
$V_{\rm MNS}$, $\tan\beta$ and the sign of the
Higgsino mass parameter $\mu$ in Eq.~(\ref{eq:MSSMRN}).
\end{itemize}
Throughout the following calculation, we fix some of the parameters as
shown in Table~\ref{tab:fixedparameters}.
We consider two cases for the neutrino parameters, corresponding to the
large mixing angle (LMA) and the small mixing angle (SMA) MSW solutions
of the solar neutrino anomaly.
The parameters we used in the neutrino sector for each case are given in
Table~\ref{tab:nuparam}.
For $M_R$ and $\tan\beta$, we take several cases to see the dependences
(see Table~\ref{tab:paramcontours}).
SUSY breaking parameters $m_0$, $M_0$ and $A_0$ are varied
 and $\Delta A_0$ is fixed to zero.

With these parameters, we solve the RGEs of the mass parameters and the
coupling constants between the
Planck and the EW scale taking all the flavor mixings into account.
Detail of our method is explained in Appendix~\ref{sec:RGE}.
The magnitude of $\mu$ is determined by the radiative EW symmetry
breaking condition, in which the minimum of the one-loop effective
potential for the Higgs fields is evaluated.
Then we obtain all the masses and mixings of the SUSY particles at the
EW scale and calculate the FCNC/LFV observables as functions of above
parameters.
We calculate the following quantities:
\begin{itemize}
\item The SUSY contribution to the muon anomalous magnetic moment \amu;
\item Branching ratios of \bsg, \meg\ and \tmg;
\item P-odd asymmetry of \meg;
\item \bb\ mass splittings \dmbd\ and \dmbs;
\item CP violation parameter \ek;
\item Time-dependent CP asymmetries of \Bsg\ and \BJKS.
\end{itemize}
In order to find the allowed region in the parameter space, we impose
the constraints from the experimental results of the direct searches of
SUSY particles \cite{susy-search}
and Higgs bosons \cite{higgs-search}
and the measurements of \Bbsg\ \cite{bsg-exp}.
Also it turns out that, in some parameter region, the branching ratio of
\meg\ exceeds the present upper limit and hence this process already
gives an important constraint on the parameter space.
We discuss the constraints from the measured values of \ek\ and \dmbd\ and
the lower bound of \dmbs\ later, since it depends on the CKM
parameters, namely $|V_{ub}|$ and \dlt\ \cite{CK-PDG}.

\subsection{
  $\theta_E$ and $\theta_D$ dependence of \meg\ and \ek
}

Let us first discuss the $\theta_E$ and $\theta_D$ dependence of the
\meg\ decay and \ek.

\begin{figure}
\begin{center}
\makebox[0em]{
\def\epsfsize#1#2{0.5#1}
\epsfbox{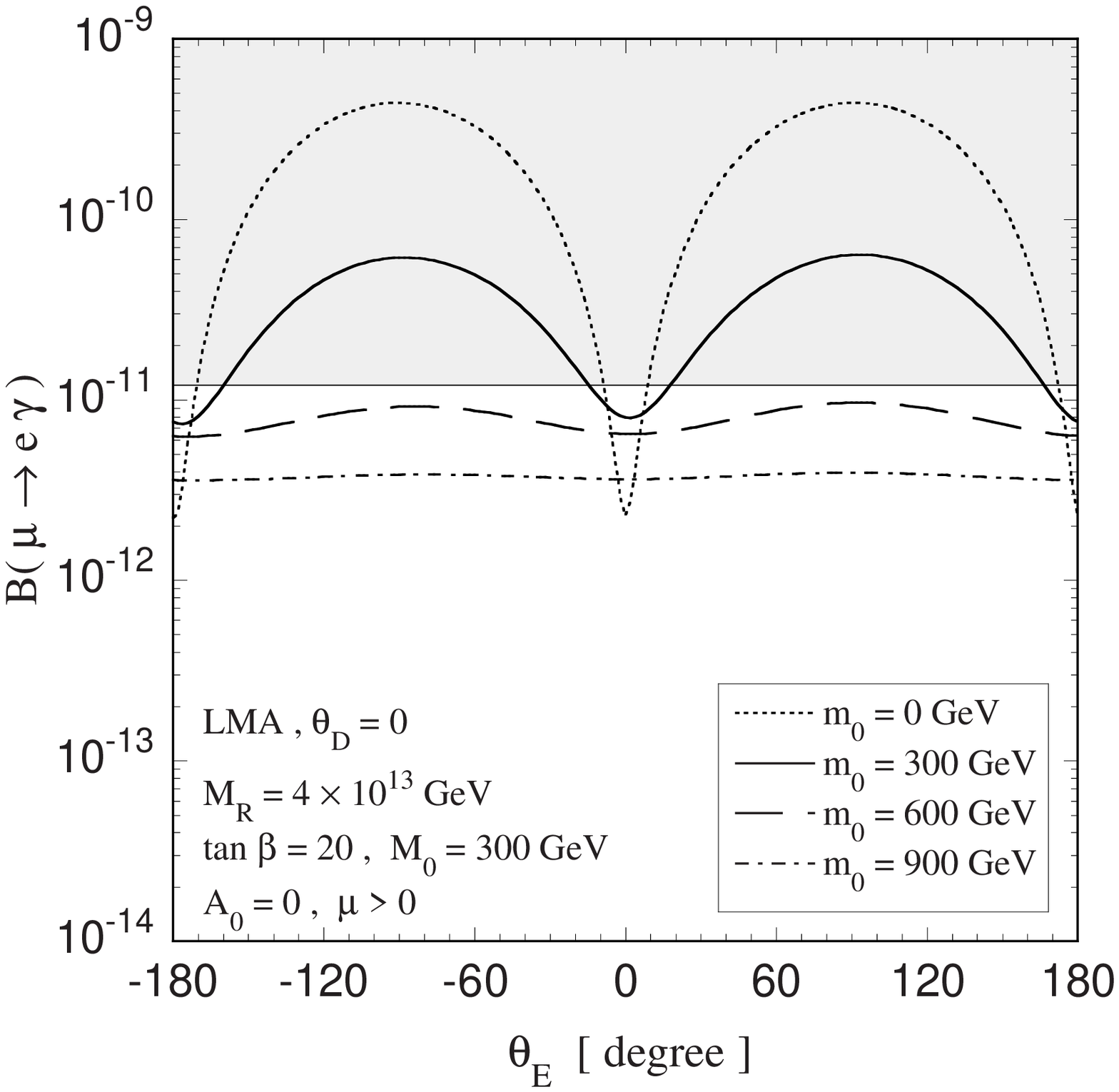}
\makebox[0em][r]{\raisebox{1ex}{(a)~~}}
\def\epsfsize#1#2{0.5#1}
\epsfbox{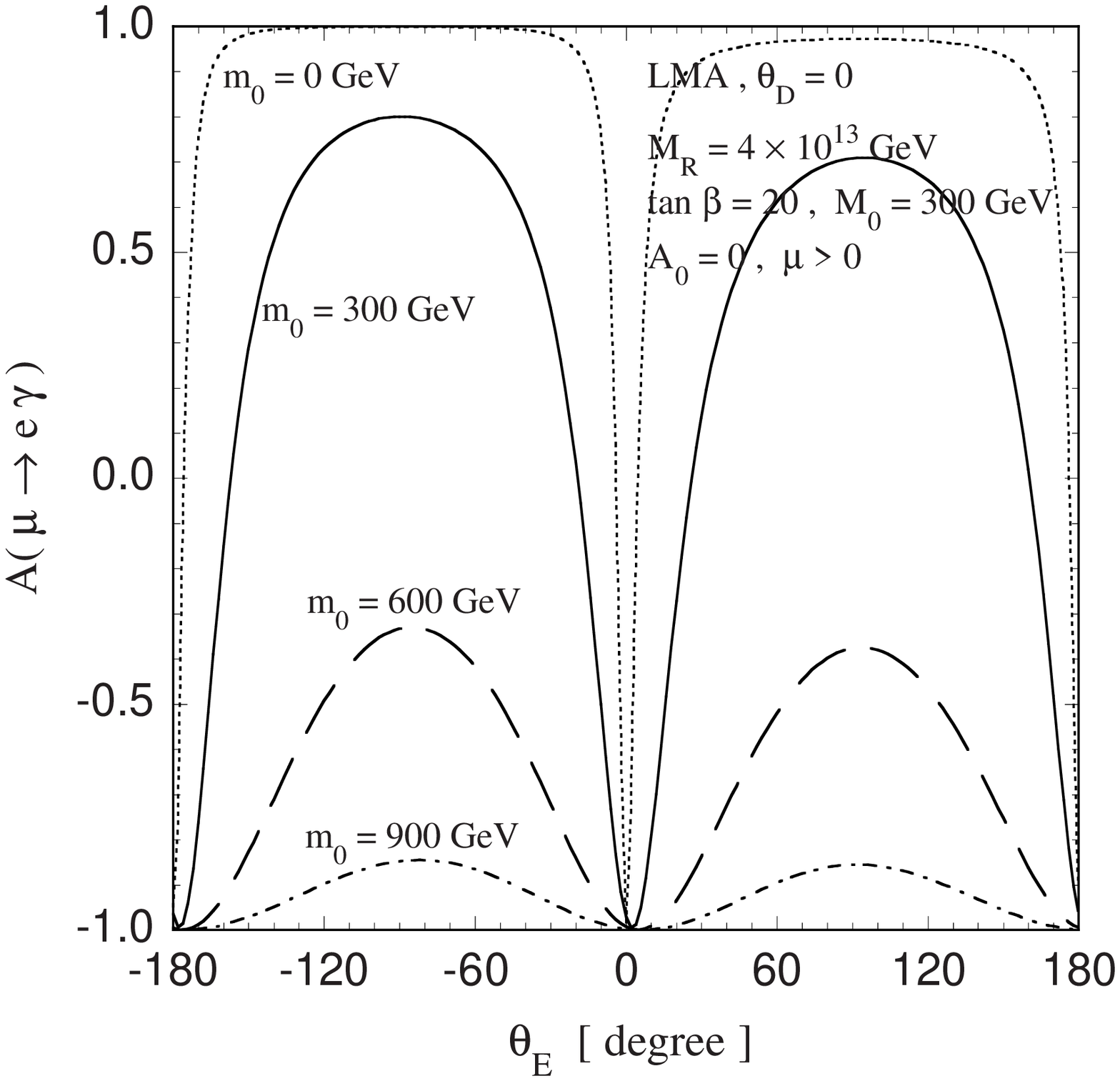}
\makebox[0em][r]{\raisebox{1ex}{(b)~~}}
}
\end{center}
\begin{center}
\makebox[0em]{
\def\epsfsize#1#2{0.5#1}
\epsfbox{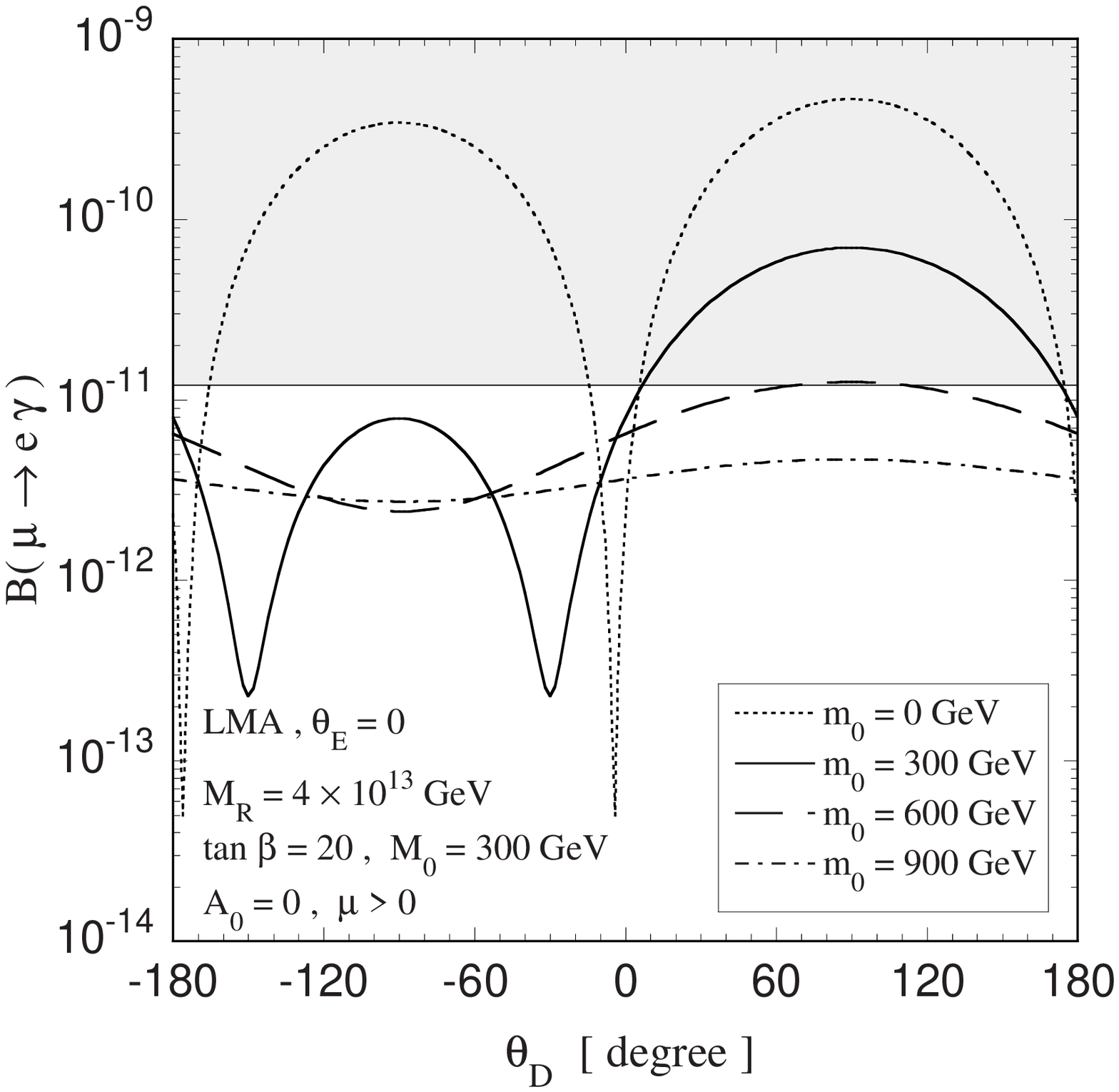}
\makebox[0em][r]{\raisebox{1ex}{(c)~~}}
\def\epsfsize#1#2{0.5#1}
\epsfbox{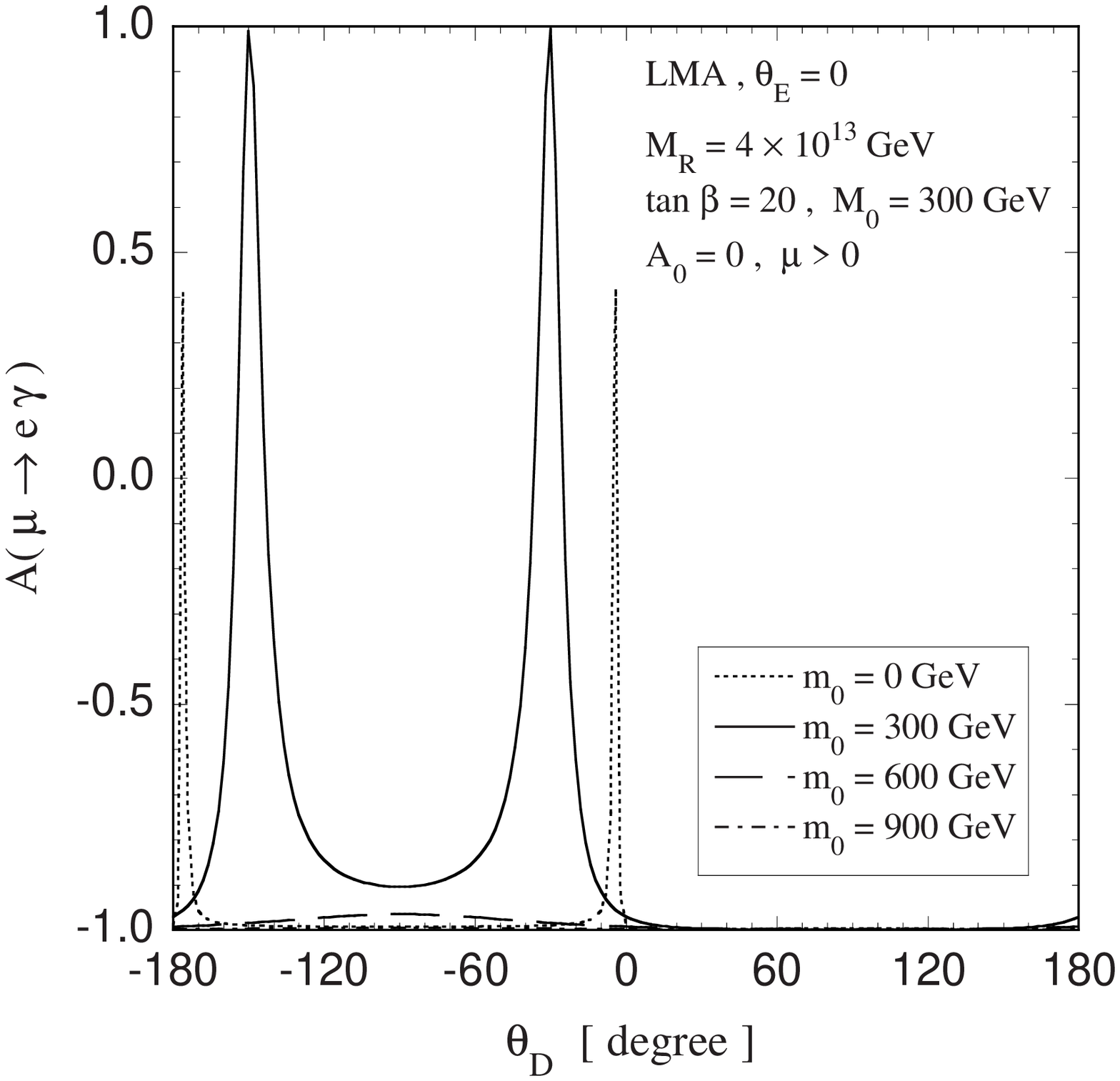}
\makebox[0em][r]{\raisebox{1ex}{(d)~~}}
}
\end{center}
\caption{
Branching ratio and P-odd asymmetry of \meg\ as functions of $\theta_E$
and $\theta_D$.
}  
\label{fig:meg-thDthE}
\end{figure}

As given in Eq.~(\ref{eq:mueg-approximate-formulas}), the decay
amplitudes $A_{R}^{21}$ and $A_{L}^{21}$ depend on $\theta_E$ and
$\theta_D$ differently, so that both of the branching ratio and P-odd
asymmetry are affected.
In Fig.~\ref{fig:meg-thDthE} we show \Bmeg\ and \Ameg\ as functions of
$\theta_E$ and $\theta_D$.
The shaded regions are excluded by the upper bound of
 \Bmeg \cite{Brooks:1999pu}.
Here we take the LMA case for the neutrino parameters,
$M_R=4\times10^{13}$~GeV, $\tan\beta=20$, $\mu>0$.
SUSY breaking parameters are also fixed as $M_0=300$~GeV, $A_0=0$ and
$m_0=0$, $300$, $600$, $900$~GeV.
For the fixed $\theta_D=0$ case ((a) and (b)), we can see that the
amplitude $A_{L}^{21}$ is enhanced for a nonvanishing $\theta_E$ and
relatively small $m_0$.
In the parameter region $\theta_E\sim90^\circ$, \Bmeg\ becomes larger
than that for $\theta_E\sim0$ and \Ameg\ approaches to $+1$, reflecting
that $A_{L}^{21}$ is enhanced and dominates over $A_{R}^{21}$.
For $\theta_E=0$ case ((c) and (d)), $A_{R}^{21}$ dominates in the most
of the range of $\theta_D$ and hence \Ameg\ is close to $-1$.
In some special case, $\theta_D=-30^\circ$ and $m_0=300$~GeV for
example, a cancellation among contributions to $A_{R}^{21}$ occurs and
the branching ratio is suppressed.
In such a case the P-odd asymmetry approaches to $+1$.

\begin{figure}
\begin{center}
\makebox[0em]{
\def\epsfsize#1#2{0.5#1}
\epsfbox{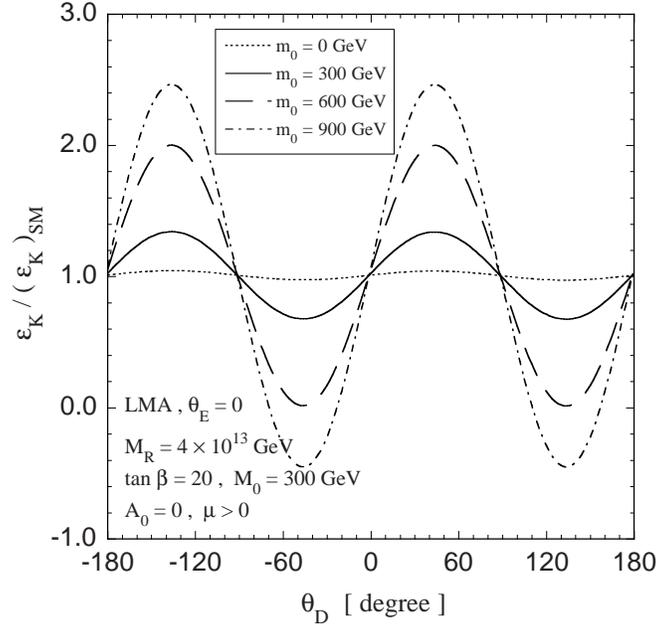}
}
\end{center}
\caption{
\ek\ normalized to the SM value as a function of $\theta_D$.
Parameters are the same as those in Fig.~\protect\ref{fig:meg-thDthE}(c)
and (d).
}
\label{fig:ek-thD}
\end{figure}

Fig.~\ref{fig:ek-thD} shows the $\theta_D$ dependence of \ek\ for the
same parameter set as Fig.~\ref{fig:meg-thDthE}(c) and (d).
This dependence comes from $g^{S}_{RL}$ and ${g^{S}_{RL}}'$ in
Eq.~(\ref{eq:epsilon_K}) since $\theta_D$ directly affects the mixing
between the right-handed down-type squarks of the first and the second
generations.
We have checked that $\theta_E$ dependence is negligible for \ek.

Hereafter we fix $\theta_E$ as $\theta_E=0$ and in most cases we also
fix $\theta_D=0$.

\subsection{
  \amu, FCNC and LFV observables for different sets of the neutrino and
  the SUSY parameters
}

\begin{figure}
\begin{center}
\makebox[0em]{
\def\epsfsize#1#2{#1}
\epsfbox{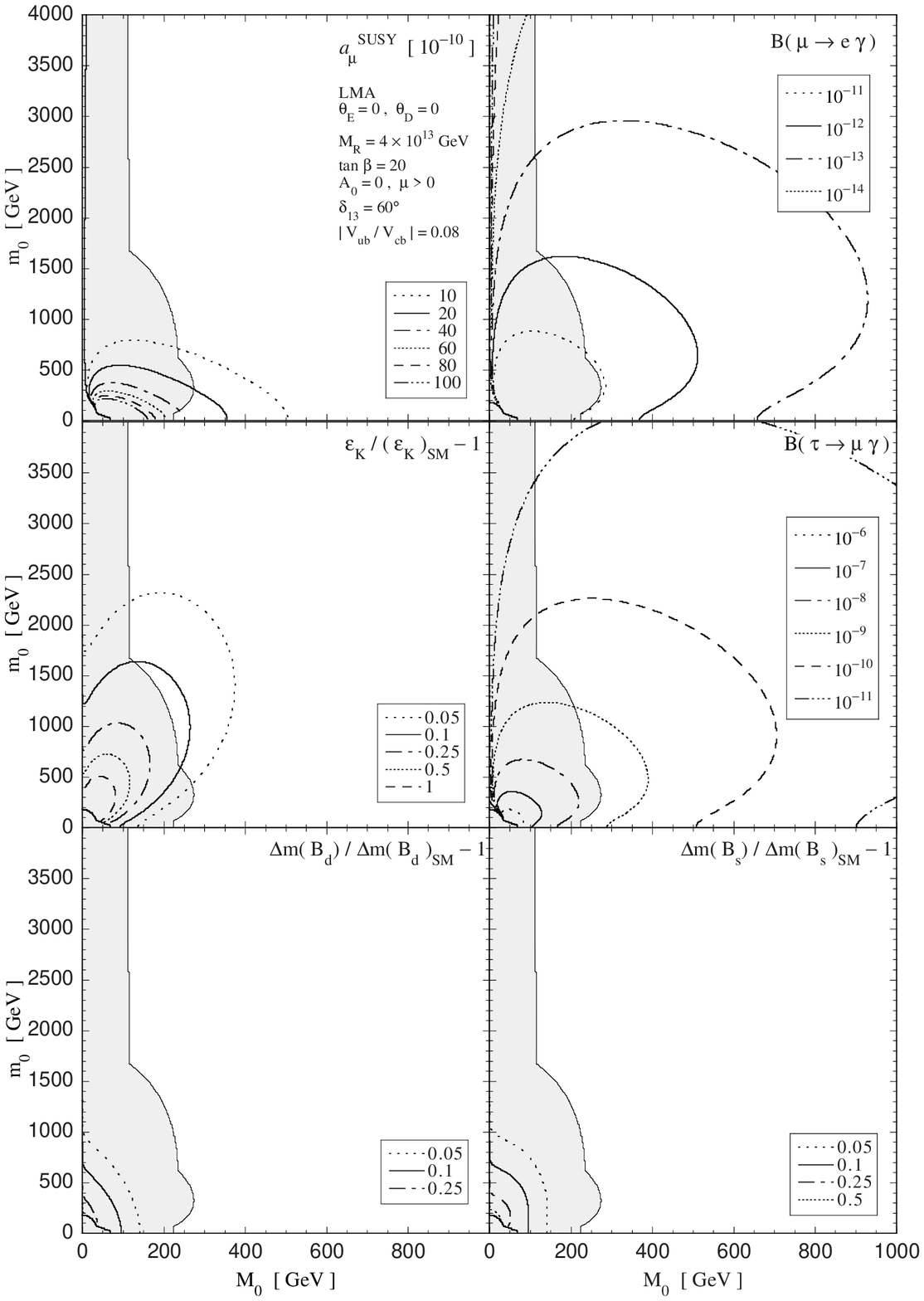}
\makebox[0em][r]{\raisebox{1ex}{(a)~~}}
}
\end{center}
\begin{center}
\makebox[0em]{
\def\epsfsize#1#2{#1}
\epsfbox{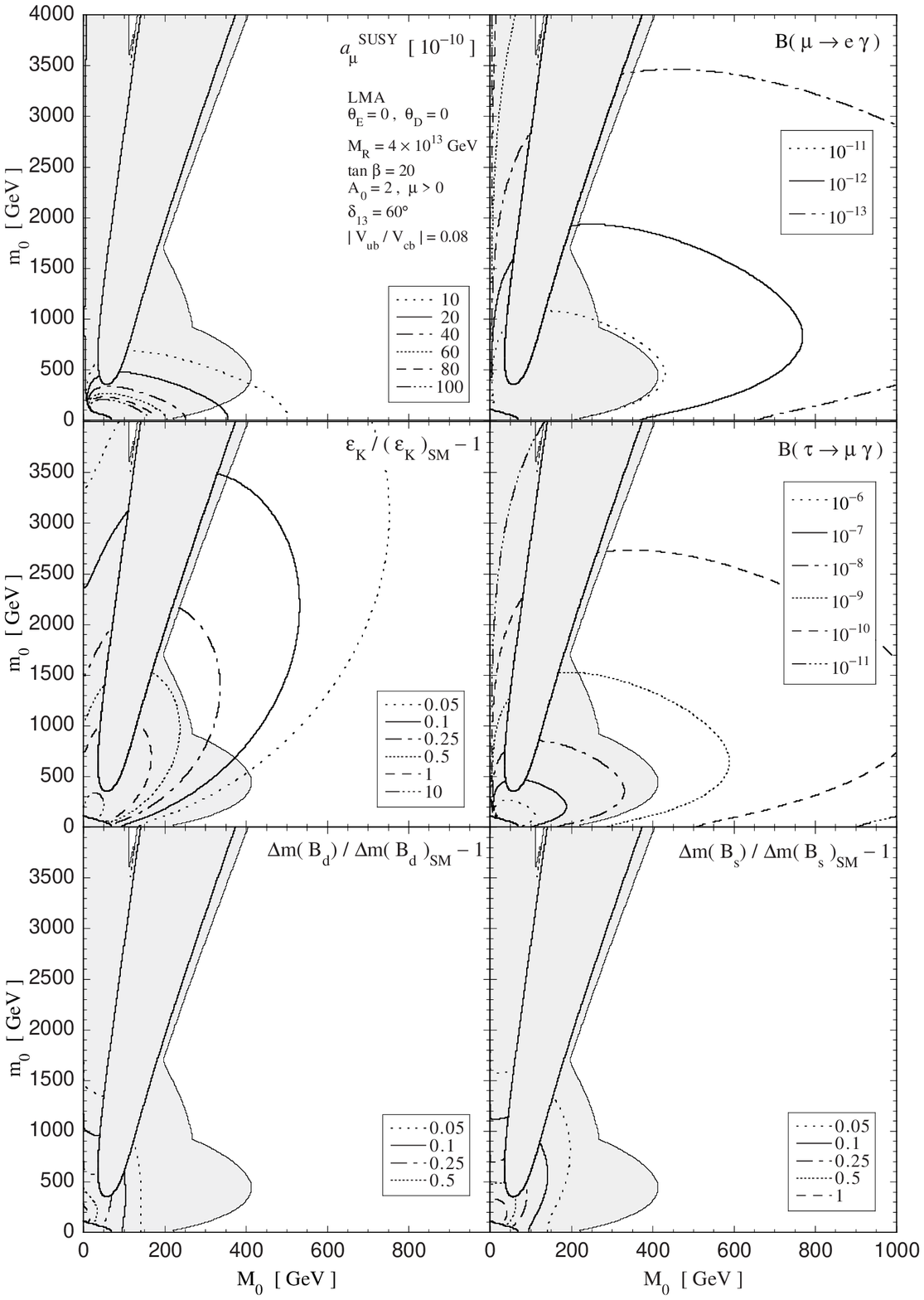}
\makebox[0em][r]{\raisebox{1ex}{(b)~~}}
}
\end{center}
\begin{center}
\makebox[0em]{
\def\epsfsize#1#2{#1}
\epsfbox{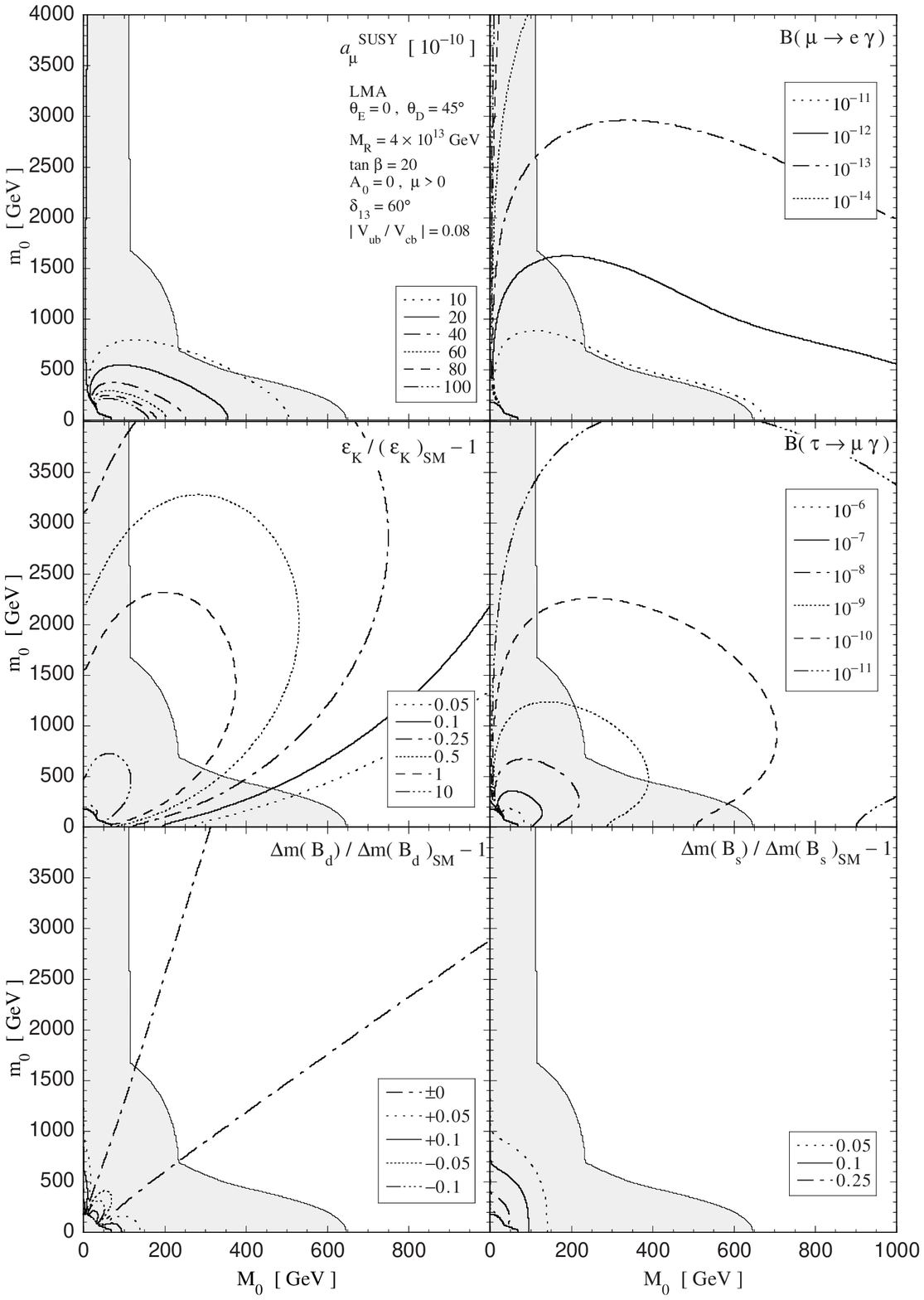}
\makebox[0em][r]{\raisebox{1ex}{(c)~~}}
}
\end{center}
\begin{center}
\makebox[0em]{
\def\epsfsize#1#2{#1}
\epsfbox{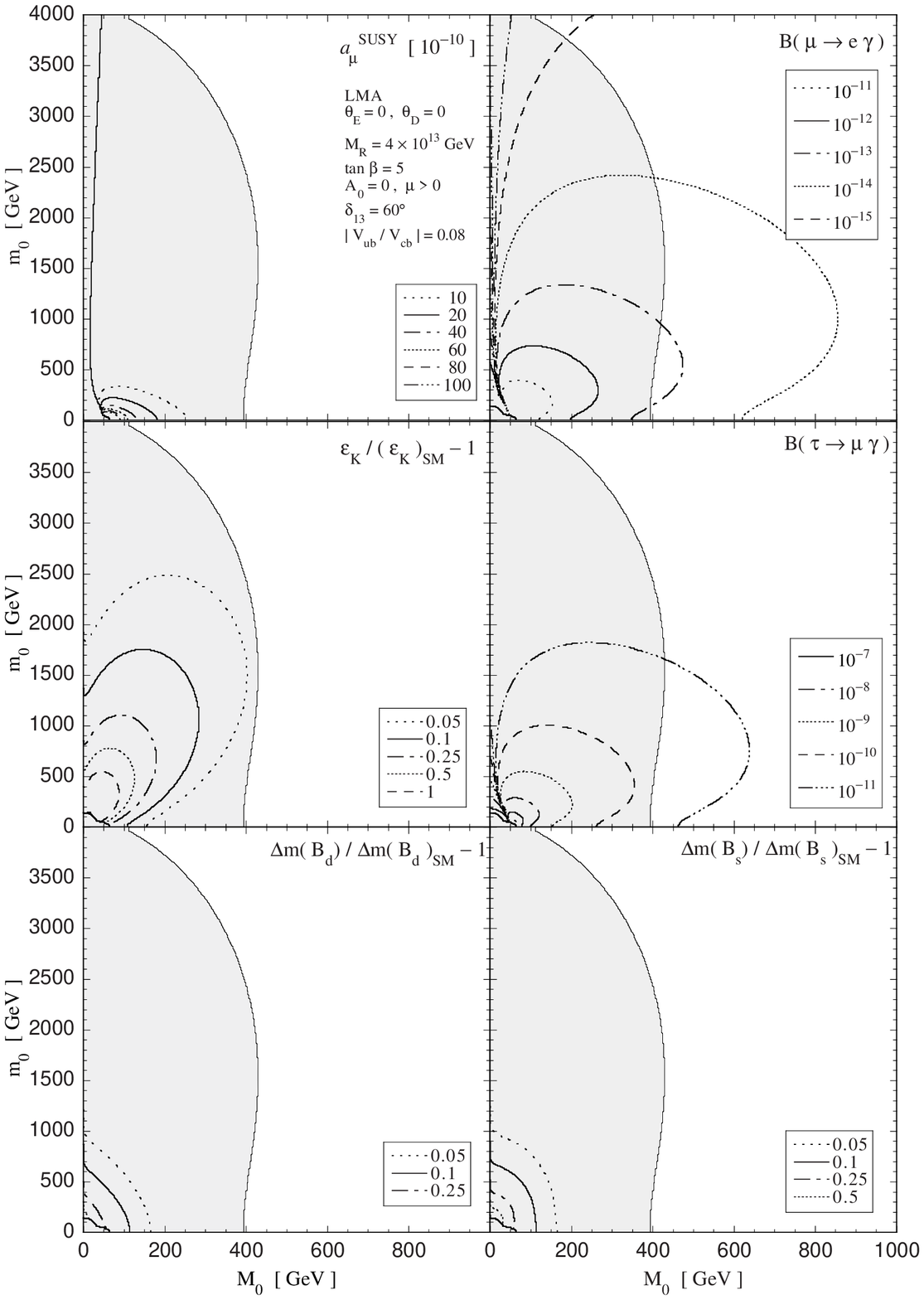}
\makebox[0em][r]{\raisebox{1ex}{(d)~~}}
}
\end{center}
\begin{center}
\makebox[0em]{
\def\epsfsize#1#2{#1}
\epsfbox{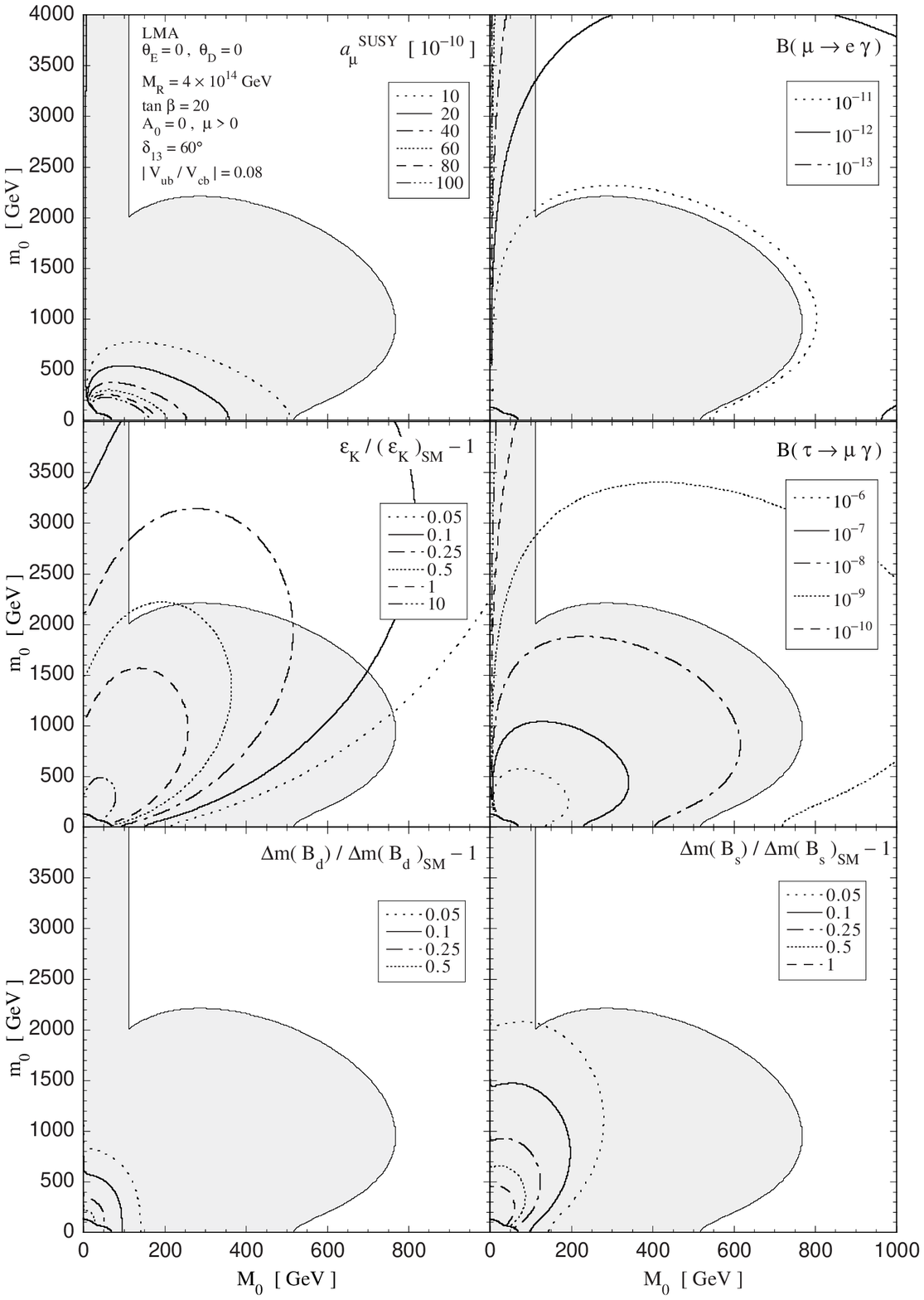}
\makebox[0em][r]{\raisebox{1ex}{(e)~~}}
}
\end{center}
\begin{center}
\makebox[0em]{
\def\epsfsize#1#2{#1}
\epsfbox{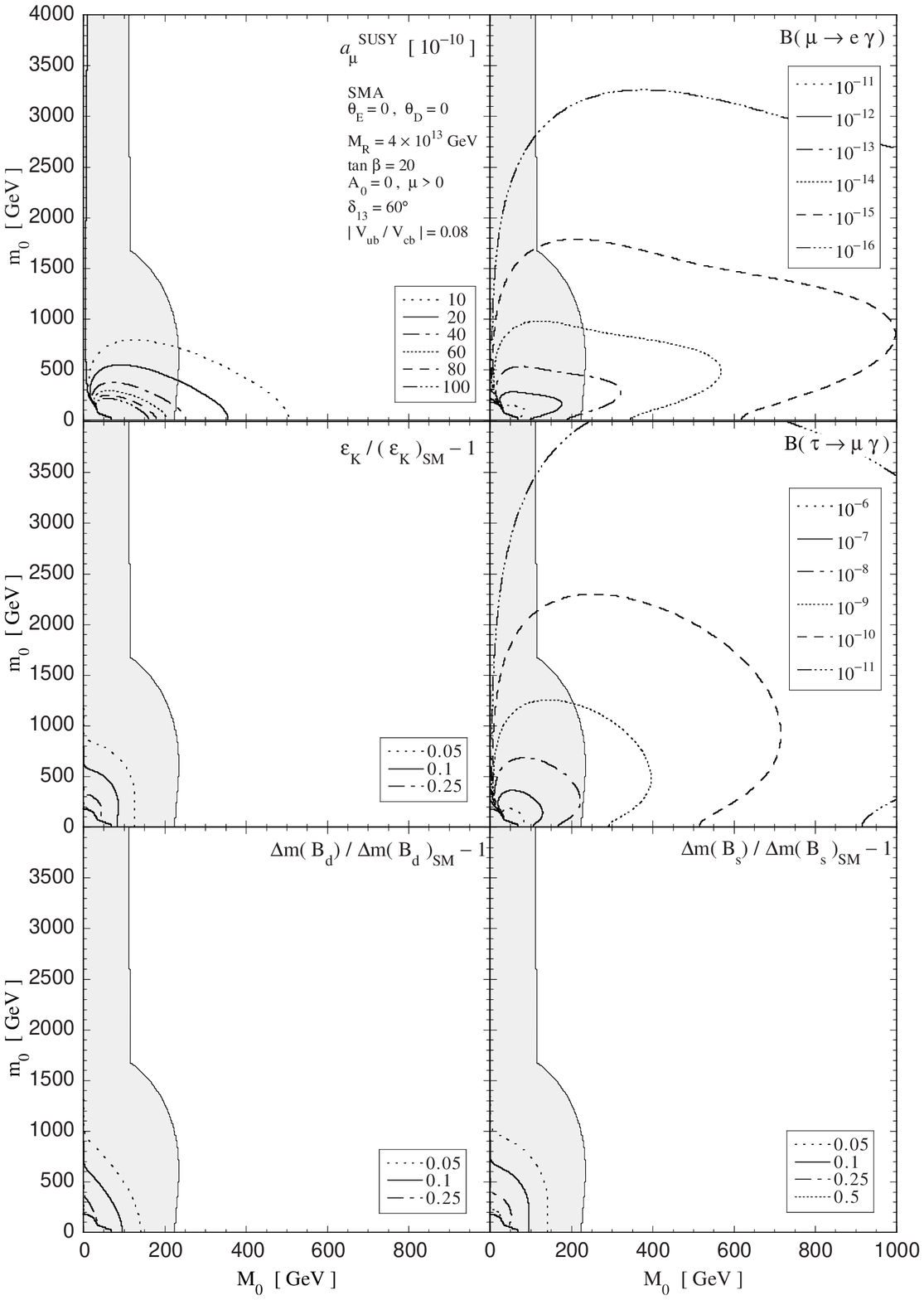}
\makebox[0em][r]{\raisebox{1ex}{(f)~~}}
}
\end{center}
\begin{center}
\makebox[0em]{
\def\epsfsize#1#2{#1}
\epsfbox{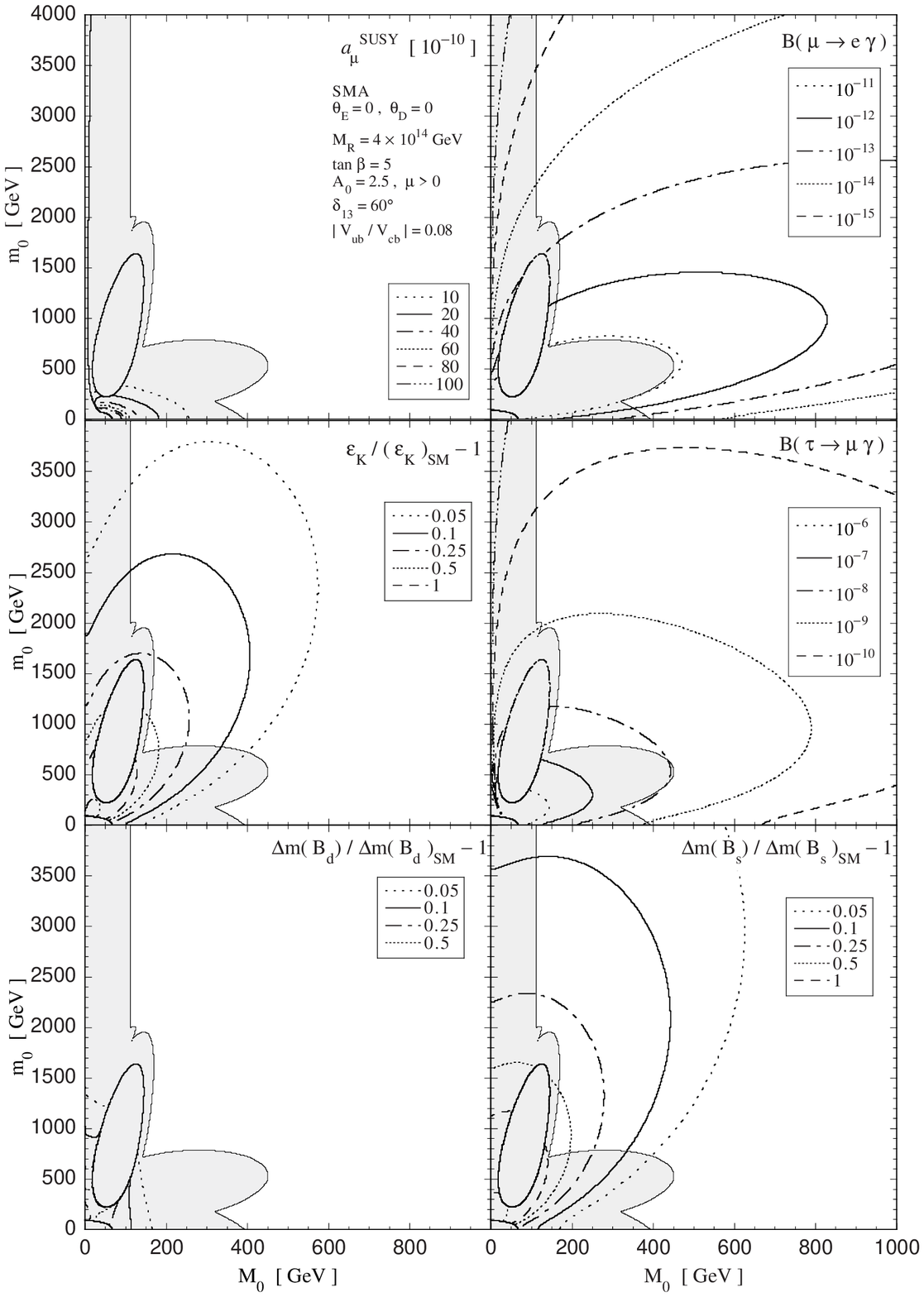}
\makebox[0em][r]{\raisebox{1ex}{(g)~~}}
}
\end{center}
\caption{
Contour plots of
\amu, \Bmeg, \Btmg, 
$\varepsilon_K/\varepsilon_K^{\rm SM}-1$,
$\Delta m_{B_d}/\Delta m_{B_d}^{\rm SM}-1$ and
$\Delta m_{B_s}/\Delta m_{B_s}^{\rm SM}-1$
on the $m_0$--$M_0$ plane for various choices of the parameters given in
Table~\protect\ref{tab:paramcontours}.
CKM parameters are fixed as $\delta_{13}=60^\circ$ and
$|V_{ub}/V_{cb}|=0.08$ and $\mu$ is taken as positive.
Shaded regions are excluded experimentally (see text).
In (b) and (g), the excluded regions correspond to negative stop mass
 squared are also shown.
}  
\label{fig:contours}
\end{figure}

In Fig.~\ref{fig:contours} we show contour plots of the SUSY
contribution to the muon anomalous magnetic moment \amu,
branching ratios of \meg\ and \tmg, and the
deviations from the SM values of \ek, \dmbd\ and \dmbs.
The input parameters used in each figure are given in
Table~\ref{tab:paramcontours}.
We also fix the sign of $\mu$ as $\mu>0$ in these figures.
$\mu<0$ region is disfavored because the SUSY contributions to \bsg\
decay amplitude interferes with the SM contribution constructively, so
that the branching ratio becomes too large in a large portion of the
parameter space.
Note that the constraints from \ek, \dmbd\ and \dmbs\ are not imposed
in Fig.~\ref{fig:contours} since these constraints depend on \dlt\ and
$|V_{ub}|$.
We show our result for each of these observables by taking a ratio to
the corresponding SM value, expecting that the most of the dependences on
the CKM parameters cancel.
In fact we have checked that the plots do not change when a different
value of \dlt\ is used.
The dependences of \amu, \Bmeg\ and \Btmg\ on the CKM parameters are
also small.

In Fig.~\ref{fig:contours}(a) we take the LMA case for the neutrino
masses and mixing, $M_R=4\times10^{13}$~GeV, $\theta_E=\theta_D=0$,
$\tan\beta=20$ and $A_0=0$ as a reference point.
Shaded regions are experimentally excluded region.
The constraints mainly come from the LEP~II Higgs boson search and the
upper bound on \Bmeg.
We see that there is a parameter region with
$20\times10^{-10}\lsim a_\mu^{\rm SUSY} \lsim 60\times10^{-10}$, which
is favored by the E821 result and in that region \Bmeg\ becomes larger
than $10^{-12}$.
In the allowed parameter region within the plotted range
$M_0<1$~TeV and $m_0<4$~TeV, \Bmeg\ varies $O(10^{-14})$ to
$O(10^{-11})$ and \Btmg\ varies $O(10^{-11})$ to $O(10^{-9})$.
Both branching ratios depend similarly on $M_0$ and $m_0$.
Also we see that the deviation of \ek\ from the SM value is about ten
percent at most and the deviations of \dmbd\ and \dmbs\ are small.

In Fig.~\ref{fig:contours}(b) plots for $A_0=2$ are given.
In this case the stop mass squared becomes negative in some parameter
region, which is shown in the figure.
The excluded region by the \Bmeg\ constraint is enlarged, due to the
enhancement of the branching ratio by the change in the left-right
mixing in the slepton mass matrices.
The deviation of \ek\ is also enhanced and there is an allowed parameter
region where \ek\ is enhanced by more than 25 percent.
It is noticeable that the allowed parameter region with a large
enhancement of \ek\ is different from the E821-favored region.
The region which corresponds to both a favorable \amu\
 and a large enhancement of \ek\ is
excluded by other constraint, such as \Bmeg.

Comparing Fig.~\ref{fig:contours}(a) and (c), we can see the dependence
on $\theta_D$.
Since $\theta_D$ affects the mixing between the first and the second
generations, the difference appears mainly for \Bmeg\ and \ek.
For a nonvanishing $\theta_D=45^\circ$ (Fig.~\ref{fig:contours}(c)),
\Bmeg\ is enhanced for $m_0\lsim700$~GeV and the excluded region is
enlarged.
Also the SUSY contribution to \ek\ can be larger
than the SM contribution in a part of the allowed parameter region as
shown in Fig.~\ref{fig:ek-thD}.
\amu\ does not depend on $\theta_D$ so that the parameter region with
$a_\mu^{\rm SUSY}\gsim10\times10^{-10}$ is excluded in the case (c).
The behavior of \Btmg\ and \dmbs\ are unchanged also.
Although \dmbd\ depends on $\theta_D$, the deviation is quite small in
either case.

The plots for $\tan\beta=5$ are given in Fig.~\ref{fig:contours}(d).
Since both \Bmeg\ and \Btmg\ are proportional to $\tan^2\beta$, possible
values are suppressed as $\text{B}(\mu\to e\,\gamma)\lsim 10^{-13}$ and 
$\text{B}(\tau\to \mu\,\gamma)\lsim 10^{-10}$.
\amu\ is proportional to $\tan\beta$ and is also suppressed.
The excluded region is larger than the $\tan\beta=20$ case because the
constraint from the Higgs mass bound is stronger for a smaller
$\tan\beta$.
The plots for \ek, \dmbd\ and \dmbs\ are the same as those in the case
(a) except that the excluded region is enlarged.

Fig.~\ref{fig:contours}(e) shows the case with a larger
$M_R=4\times10^{14}$~GeV.
Since the magnitude of the neutrino Yukawa coupling constants are
proportional to $\sqrt{M_R}$, the flavor mixings in $m_D^2$ and $m_L^2$
are enhanced for a larger $M_R$.
As a result we see that \Bmeg, \Btmg\ and \ek\ are significantly
enhanced in this case, compared to the $M_R=4\times10^{13}$~GeV case
(a).
Although the excluded region due to the constraint from \Bmeg\ is
enlarged, there is still an allowed parameter region where \ek\ is
enhanced more than fifty percent of the SM value.
Also \Btmg\ can be close to $O(10^{-8})$.
In the allowed parameter region, the deviations of \dmbd\ and \dmbs\ are
small.
\amu\ is unaffected by the change of $M_R$ and the E821-favored region is
excluded by the \Bmeg\ constraint.

Fig.~\ref{fig:contours}(f) shows the SMA case.
Other parameters are taken to be the same as those in the case (a).
In this case, the mixing between the first and the second generations is
suppressed compared to the LMA case.
Consequently \Bmeg\ is at most $O(10^{-13})$ in the allowed region and
the deviation of \ek\ is smaller.
\amu, \Btmg, \dmbd\ and \dmbs\ look the same as those in the case (a).

\begin{figure}
\begin{center}
\makebox[0em]{
\def\epsfsize#1#2{0.5#1}
\epsfbox{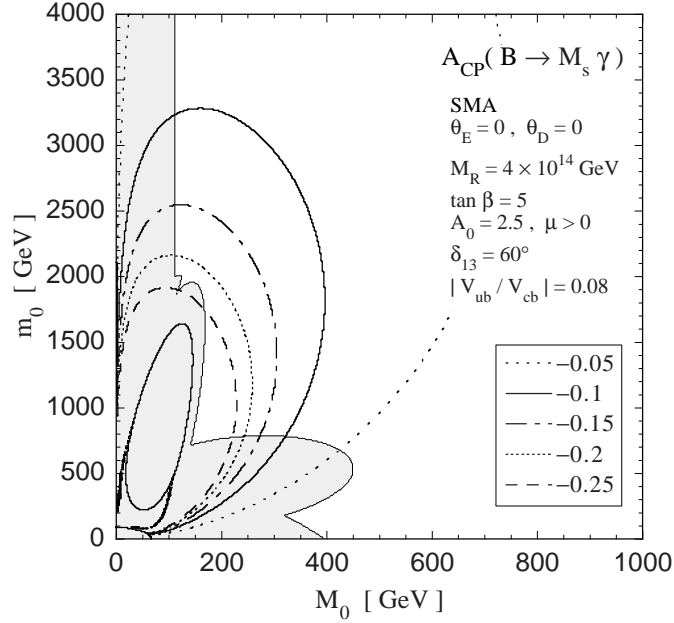}
}
\end{center}
\caption{
Contour plot of the time-dependent CP asymmetry of \Bsg\
decay with the same parameter set as Fig.~\protect\ref{fig:contours}(g).
}
\label{fig:Atbsg}
\end{figure}

In all the above cases the deviation of the \bb\ mixing from the SM
value is small.
Let us now show an example with a large enhancement of the \bsbs\ mixing
in Fig.~\ref{fig:contours}(g).
We see that \dmbs\ differs from the SM value by more than 50 percent in
a parameter region $m_0\gsim700$~GeV and $M_0\lsim200$~GeV.
In the same region \ek\ is also enhanced by a similar amount.
Note that this enhancement comes from the mixing in the right-handed
down-type squarks induced by the neutrino Yukawa coupling.
In such a case the time-dependent CP asymmetry of \Bsg\
decay is also enhanced.
Fig.~\ref{fig:Atbsg} shows \AtBsg\ with the same parameter set.
We see that this asymmetry can be larger than 25 percent in the
parameter region where \dmbs\ is enhanced.
In this case \Btmg\ can be close to $10^{-7}$.

In Table~\ref{tab:contourssummary} we summarize the possible SUSY
contributions to the observables given in Fig.~\ref{fig:contours}.
We can see that, except for the case (g), a large deviation from the SM
is possible only in \amu, \Bmeg\ and \ek.

\begin{figure}
\begin{center}
\makebox[0em]{
\def\epsfsize#1#2{#1}
\epsfbox{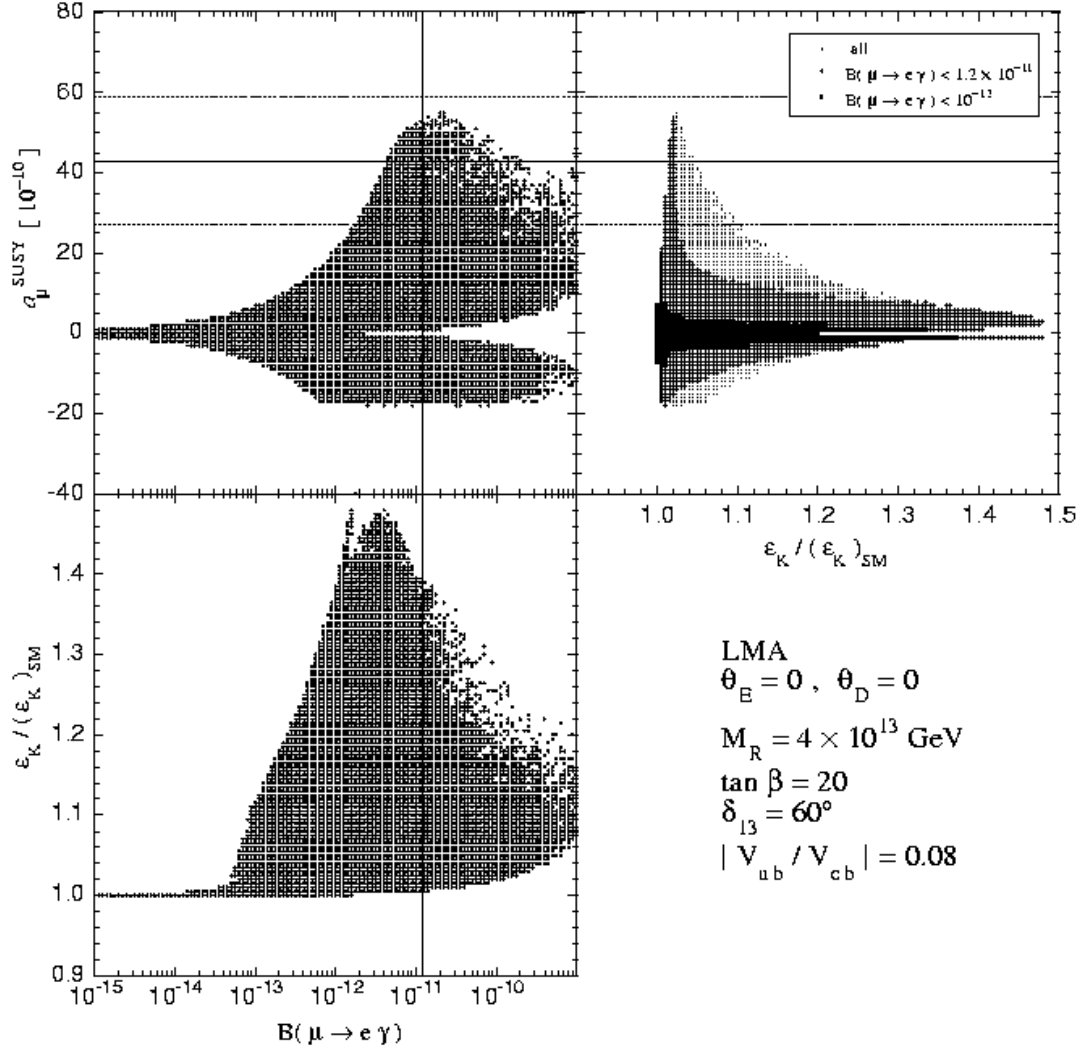}
}
\end{center}
\caption{
Correlation among \amu, \Bmeg\ and \ek.
The vertical line shows the experimental upper bound
$\text{B}(\mu\to e\gamma)< 1.2\times10^{-11}$ \protect\cite{Brooks:1999pu} and the 
horizontal solid and dotted lines show the E821-favored region 
$a_\mu^{\rm SUSY} = ( 43 \pm 16 )\times10^{-10}$ \protect\cite{E821}.
}
\label{fig:amu-meg-rek}
\end{figure}

Let us see the correlation among \amu, \Bmeg\ and \ek\ more closely.
Fig.~\ref{fig:amu-meg-rek} shows the correlation between \amu\ and
\Bmeg, \amu\ and \rek, and \rek\ and \Bmeg.
Here SUSY breaking parameters $m_0$, $M_0$ and $A_0$ are scanned within
the range $m_0,\,M_0<3$~TeV and $-5<A_0<5$.
Other parameters are taken to be the same as those in
Fig.~\ref{fig:contours}(a) and (b).
In the plot of the correlation between \amu\ and \Bmeg,
The $a_\mu^{\rm SUSY}<0$ branch corresponds to $\mu<0$ and
$a_\mu^{\rm SUSY}\lsim-20\times10^{-10}$ region is excluded by the
\Bbsg\ constraint.
Notice that the parameter region where \amu\ saturates the E821 result
is different from that with a large \rek.
As can be seen in the plot of \amu\ and \rek, when \ek\ is enhanced by
$\sim50$ percent, the magnitude of \amu\ is small.

\subsection{
  Allowed region of \dmbsd\ and the CP asymmetry of \BJKS
}

Finally, let us discuss the effect of varying the CKM parameters
$|V_{ub}/V_{cb}|$ and \dlt.
Within the SM, these parameters are determined by combining the
measurements of several observables: $b\to u\ell\bar{\nu}$ semileptonic
decays, \ek, \dmbd, \dmbsd\ and the time-dependent CP asymmetry of
\BJKS\ decay.
However, in the present case we have shown that there can be a
significant SUSY contributions to these observables, especially for \ek.
In such a case the allowed range of \dlt\ given by the measured value of
\ek\ is different from the SM case, and then this change affects the
other observables.
As for $|V_{ub}/V_{cb}|$, we have no change since the $b\to u\ell\bar{\nu}$
decay is dominated by the tree-level SM amplitude.

\begin{figure}
\begin{center}
\makebox[0em]{
\def\epsfsize#1#2{0.5#1}
\epsfbox{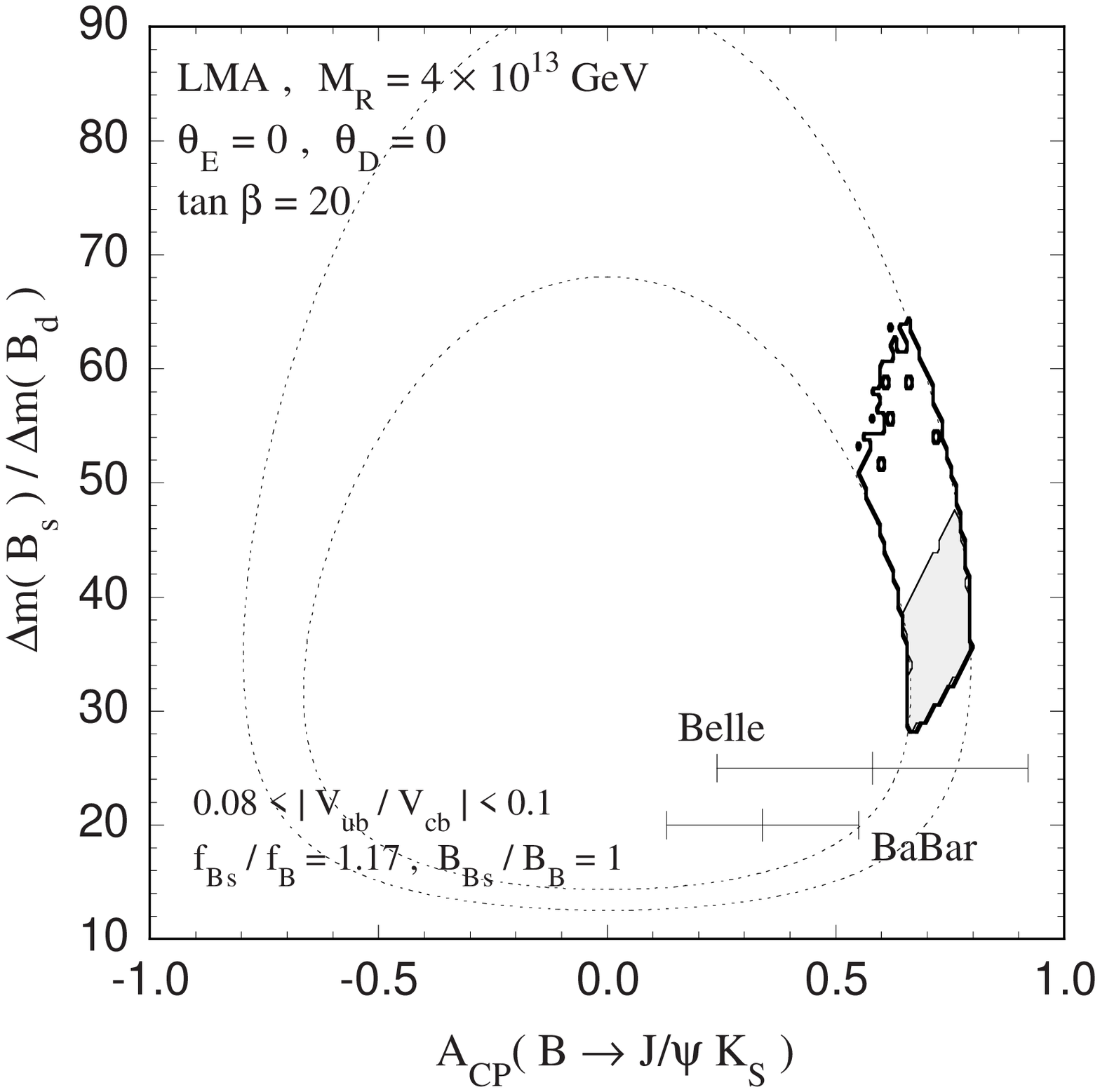}
\makebox[0em][r]{\raisebox{1ex}{(a)~~}}
\def\epsfsize#1#2{0.5#1}
\epsfbox{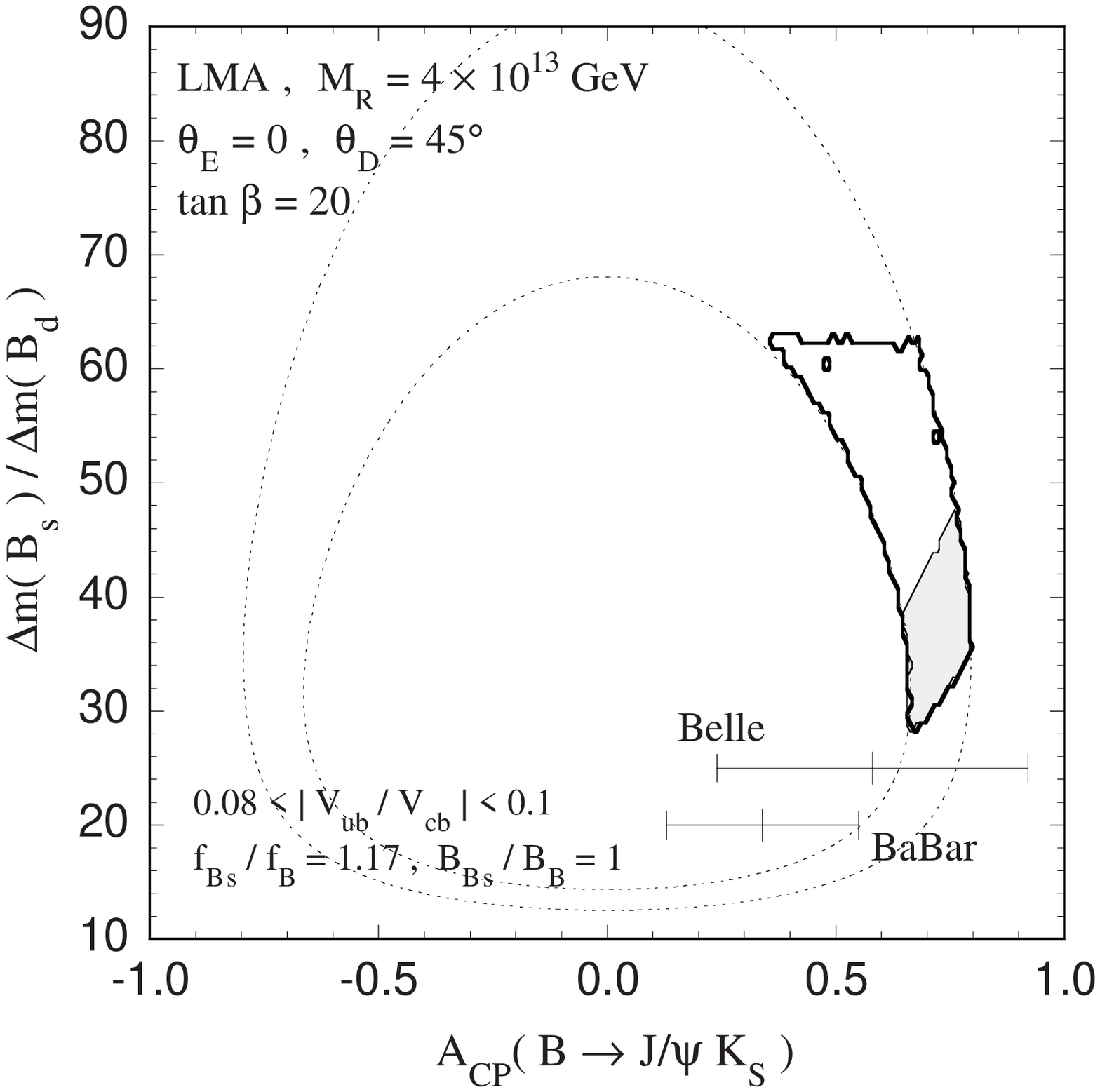}
\makebox[0em][r]{\raisebox{1ex}{(b)~~}}
}
\end{center}
\begin{center}
\makebox[0em]{
\def\epsfsize#1#2{0.5#1}
\epsfbox{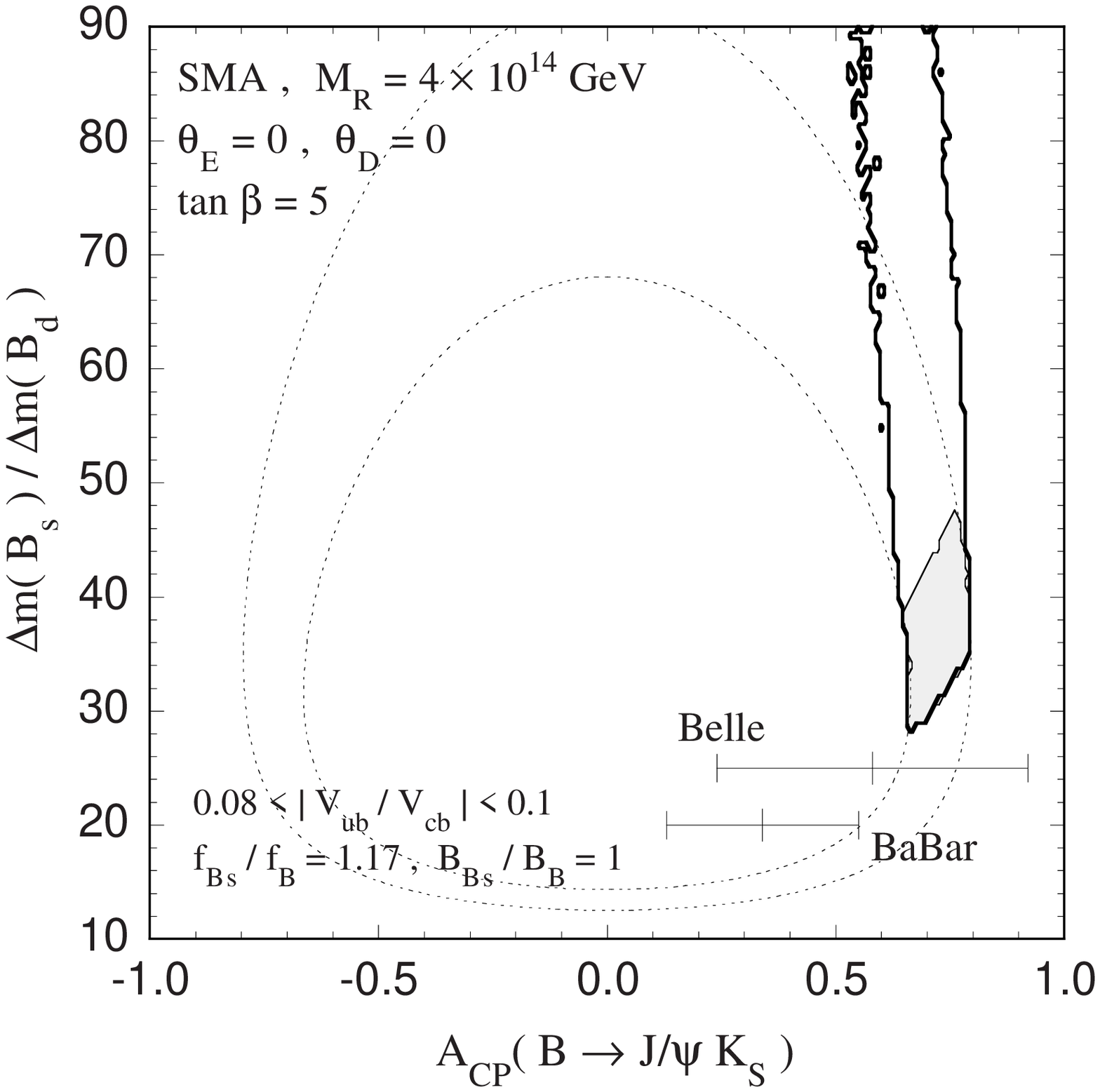}
\makebox[0em][r]{\raisebox{1ex}{(c)~~}}
}
\end{center}
\caption{
Allowed region in \dmbsd--$A_{CP}(B\to J/\psi\,K_S)$ plane.
\dlt\ and $|V_{ub}/V_{cb}|$ are varied and constraints from \ek, \dmbd\
and \dmbsd\ are imposed.
1 $\sigma$ ranges of the CP asymmetry from Belle and BaBar experiments
 are also shown \protect\cite{AtJKs}.
}
\label{fig:dmbsd-AtJKS}
\end{figure}

We show how this effect will be observed in Fig.~\ref{fig:dmbsd-AtJKS},
where the possible region in the space of \dmbsd\ and the time-dependent
CP asymmetry of \BJKS\ is presented.
In this figure, we vary $|V_{ub}/V_{cb}|$ and \dlt\ within the ranges
$0.08<|V_{ub}/V_{cb}|<0.1$ and $0<\delta_{13}<360^\circ$.
The dotted lines in each plot show the SM values of \dmbsd\
and \AtBJKS\ for the whole range of \dlt\ and 
$|V_{ub}/V_{cb}|=0.1$ (outer line) and 0.08 (inner line).
The shaded region is allowed in the SM case.
We impose the constraints from the measured values of
$\varepsilon_K=2.28\times10^{-3}$ and $\Delta m_{B_d}=0.482$~ps$^{-1}$
and from the lower limit of $\Delta m_{B_s}>14.3$~ps$^{-1}$.
In the calculation of \ek, \dmbd\ and \dmbs\ we fix the bag parameters
and the decay constant of the $B$ meson $f_{B_d,B_s}$ as given in
Table~\ref{tab:bagparam} \cite{lattice-parameters}.
When we impose the experimental constraints, we allow $\pm15$~\% and
$\pm40$~\% deviations for \ek\ and \dmbd, respectively, in order to
take theoretical uncertainties in the bag parameters into account.
Since this uncertainty is expected to be reduced in the ratio \dmbsd, we
use the lower limit of the ratio \dmbsd\ instead of \dmbs\ itself.

Fig.~\ref{fig:dmbsd-AtJKS}(a) shows the result with the same parameter
set as Fig.~\ref{fig:contours}(a) except that $A_0$ is scanned within
$-5<A_0<+5$ (see
Table~\ref{tab:paramcontours}).
As shown in Fig.~\ref{fig:contours}(a) and (b), the SUSY
contributions to \bdbd\ mixing and \bsbs\ mixing are quite small in this
case so that the allowed region lies between the dotted lines.
The difference of the allowed regions from the SM one comes from the
fact that the SUSY contribution to \ek\ can be as large as 50 percent of
the SM value, which can be seen in Fig.~\ref{fig:amu-meg-rek}.

In Fig.~\ref{fig:dmbsd-AtJKS}(b) we take $\theta_D=45^\circ$
as in Fig.~\ref{fig:contours}(c) and $A_0$ is scanned within
$-5<A_0<+5$.
In this case the enhancement of \ek\ is more significant compared to the
$\theta_D=0$ case.
Consequently a region with smaller \dlt\ is now allowed and hence a
smaller $A_{CP}(B\to J/\psi\,K_S)\sim 0.4$ is possible, compared to the SM
value $A_{CP}(B\to J/\psi\,K_S)\sim 0.7$.
At the same time \dmbsd\ can be as large as 60.

Fig.~\ref{fig:dmbsd-AtJKS}(c) is the case corresponding to
Fig.~\ref{fig:contours}(g) with $-5<A_0<+5$.
In this case the allowed region can be outside of the dotted circles,
since a large deviation of \dmbs\ is possible.
On the other hand, we see that the deviation of \AtBJKS\
from the SM value is small.

At present we only have lower bound for \dmbs\ and the CP asymmetry of
\BJKS\ and related modes is not precise enough \cite{AtJKs,AtJKs-CDF}.
In a few years we expect that the \dmbs\ will be measured at Tevatron
and the precision of the CP violating asymmetry will be improved to 10
percent level at Belle, BaBar and Tevatron experiments.
It is conceivable that the deviation shown in Fig.~\ref{fig:dmbsd-AtJKS}
will be clearly seen in these experiments.

\section{
  Conclusion and discussions
}
\label{sec:conclusion}

In this paper we have studied the FCNC and LFV processes as well as the
muon anomalous magnetic moment in the framework of SU(5) SUSY GUT with
right-handed neutrino motivated by the large mixing angle solutions for
the atmospheric  and solar neutrino anomalies.
In order to explain realistic mass relations for quarks and leptons,
we have taken into account effects of higher dimensional operators above
the GUT scale.
It has been shown that there appear new mixing angles in the
right-handed charged leptons and the right-handed down-type quarks due
to the higher dimensional operators.
We have calculated various low-energy observables by changing parameters
of the model, namely SUSY parameters, neutrino parameters (LMA or SMA,
and $M_R$) and the above new mixing angles.
We have shown that, within the current experimental bound of \Bmeg,
large SUSY contributions are possible either in the muon anomalous
magnetic moment or in \ek.
The parameter regions which have a large correction in one case is
different from that in the other case.
In the former case, the favorable value of the recent result of the BNL
E821 experiment can be accommodated.
In the latter case, the allowed region of the Kobayashi-Maskawa phase can be
different from the predictions within the SM, and therefore the
measurements of the CP asymmetry of \BJKS\ mode and \dmbs\ can
discriminate this case from the SM.
We also show that the \tmg\ branching ratio can be close to the current
experimental upper bound and the mixing-induced CP asymmetry of the
radiative $B$ decay can be enhanced in the case where the neutrino
parameters correspond to the small mixing angle MSW solution.

Finally there are several remarks.
\begin{itemize}
\item
In this paper we have neglected the constraint from the nucleon decay.
If we take the minimal model for the Higgs sector at the GUT scale, it
is likely that the nucleon decay experiments excludes most of the
parameter space even if the squark mass is multi-TeV \cite{pdecay1}.
It is known, however, that there are several ways to suppress the
nucleon decay without changing the flavor signals discussed here
\cite{pdecay2}.
\item
For LFV search, $\mu\to e\,e\,e$ decay
and $\mu-e$ conversion in a muonic atom
 are also promising experimentally \cite{review}.
The rates of these processes have simple relations with
\Bmeg\ if the photonic operator Eq.~(\ref{eq:mueg amplitudes})
 gives dominant contribution \cite{99HiNo,Czarnecki:1998iz}:
\begin{mathletters}
\begin{eqnarray}
  \frac{\text{B}(\mu^+\to e^+\,e^+\,e^-)}{\text{B}(\mu^+\to e^+\,\gamma)}
  &\approx&
  \frac{\alpha}{3\pi}
  \left[
    \log\frac{m_\mu^2}{m_e^2} - \frac{11}{4}
  \right]
  \approx 0.006,
\\
  \frac{\text{B}(\mu^-\,N\to e^-\,N)}{\text{B}(\mu^+\to e^+\,\gamma)}
  &\approx&
  \frac{B(A,Z)}{428},
\end{eqnarray}
\end{mathletters}
where $B(A,Z)$ represents the rate dependence on the mass number $A$ and
the atomic number $Z$ of the target nucleus: $B(A,Z)\approx1.1$ for
$^{27}$Al, $B(A,Z)\approx1.8$ for $^{48}$Ti and $B(A,Z)\approx1.25$ for
$^{208}$Pb.
These relations hold also in our case.
\item
We also calculated \teg\ branching ratio.
In all cases \Bteg\ is smaller by two or three
 orders of magnitude than \Btmg.
\item
As shown in Fig.~\ref{fig:contours}(a) and (e), the flavor mixing effect
due to the neutrino Yukawa coupling is enhanced (suppressed) for a large
(small) $M_R$ since the neutrino Yukawa coupling constants are
proportional to $\sqrt{M_R}$ for given neutrino masses.
When we take a small value of $M_R$, such as $M_R\lsim10^{10}$~GeV, the
contributions of ${y_\nu}$ in $m_L^2$ and $m_D^2$ given in
Eqs.~(\ref{eq:mass correction 3}) and (\ref{eq:mass correction 4}) are
suppressed and hence the SUSY contribution to \ek\ becomes smaller than
$\sim10$ percent.
Even in this case, however, there are contributions to the \meg\ decay
amplitudes independent of the magnitude of ${y_\nu}$, as shown in
Eq.~(\ref{eq:mueg-approximate-formulas}).
The terms proportional to $a_1^n$ dominate the amplitude for
$\theta_{D,\,E}=O(1)$ and the branching ratio can be as large as
the experimental upper bound in some parameter region.
\item
We have assumed 1-3 element of the MNS matrix to be vanishing.
However, present experimental upper bound is given as
$\sin^22\theta_{13}<0.1$ \cite{Apollonio:1999ae}.
When a nonvanishing $\theta_{13}$ is introduced, \meg\ and \ek\ are
generally enhanced since the loop diagrams including the
third-generation squarks/sleptons in the internal lines give large
contributions \cite{99HiNo}.
Consequently, the constraint from the upper bound of \Bmeg\ is
significant even in the SMA case.
In the allowed region, \amu\ and the SUSY contributions to \dmbd\ and
\dmbs\ are smaller than those in the $\theta_{13}=0$ case shown in
Fig.~\ref{fig:contours}.
We see that the large deviation of \dmbsd\ outside of the dotted lines
given in Fig.~\ref{fig:dmbsd-AtJKS}(c) disappears when we take
$\sin^22\theta_{13}\gsim0.001$, and the corresponding plot looks similar
to Fig.~\ref{fig:dmbsd-AtJKS}(b).
\item
Let us now discuss about the validity of the simplification imposed in
the mixing matrices $V_D$ and $V_E$.
We have numerically checked that, when we require $|(\kappa_d)_{ij}|<4$
for example, the mixing angles for the second-third and first-third
generation mixings are restricted to be smaller than $\sim15^\circ$ for
the $\tan\beta=20$ case.
In this case \Bmeg\ varies within the range which is several times
larger than those shown in Fig.~\ref{fig:meg-thDthE}.
In addition to the mixing angles, CP-violating complex phases of $O(1)$
can be introduced in $V_D$ and $V_E$.
It turns out that the SUSY contributions to \ek\ can be twice as large
as those given in Fig.~\ref{fig:ek-thD}.
These complex phases also contributes to the EDMs of the neutron ($d_n$)
and the electron ($d_e$).
We calculated EDMs and obtained that
 $|d_n|\lsim10^{-26}~e$~cm and
$|d_e|\lsim10^{-27}~e$~cm for $m_0=600$~GeV, $M_0=300$~GeV and $A_0=0$.
Thus the EDMs can be close to the present upper bounds
$|d_n|<6.3\times10^{-26}~e$~cm \cite{edmn} and
$|d_e|<4.0\times10^{-27}~e$~cm \cite{edme}.
\end{itemize}
As discussed above, if we relax
the simple assumptions for the mixing matrices $V_D$, $V_E$
and $V_{\rm MNS}$, typical patterns of the
deviation from the SM can be summarized in the following way.
(1) \Bmeg\ can be close to $10^{-11}$ and the deviation in
\dmbsd--\AtBJKS\ plane appears like Fig.~\ref{fig:dmbsd-AtJKS}(b).
In this case, \amu\ is quite small and the SUSY contribution does not
saturate the observed discrepancy of $a_\mu$.
(2) \amu\ is compatible with the E821 result and \Bmeg\ can be as large
as $10^{-11}$.
However, no deviation may be seen in \ek, \bdbd\ and \bsbs\ mixings in
this case.
From these
 observations we can
 conclude that it is important to search for new physics effects
 in the ongoing and near-future experiments, namely
 the BNL muon $g-2$ experiment, \meg\
 and $\mu-e$ conversion experiments \cite{psi,meco},
 $B$ physics experiments at $B$-factories and 
 Tevatron.
Combining results obtained in these experiments
 we  may be able to get some insights on
 interactions at the GUT or right-handed neutrino scales.

\section*{
Acknowledgments
}
The work of Y.~O.\ was supported in part by the Grant-in-Aid of the 
Ministry of Education, Culture, Sports, Science and Technology, 
Government of Japan (No.\ 09640381), Priority area ``Supersymmetry and 
Unified Theory of Elementary Particles'' (No.\ 707), and ``Physics of CP 
violation'' (No.\ 09246105).

\appendix

\section{
RGE and matching condition at the GUT scale
}
\label{sec:RGE}

In this appendix we show the detail of our numerical calculation
taking account of the effects of higher dimensional operators.
An outline of the calculation is as follows:
\begin{itemize}
\item
  We solve the RGEs for the gauge and Yukawa coupling constants between
  the EW scale and the GUT scale.
  The neutrino Yukawa coupling constants are calculated with
  Eq.~(\ref{eq:seesaw}) at the Majorana mass scale.
\item
  At the GUT scale, the coupling constants in the superpotential of the
  SU(5)RN SUSY GUT are determined from the Yukawa coupling constants for
  quarks and leptons using the matching condition explained in
  Subsection~\ref{sec:matching}.
  Then we solve the RGEs for these constants between the GUT scale and
  the Planck scale. 
\item
  At the Planck scale, we set the boundary conditions for the SUSY
  breaking parameters as Eq.~(\ref{eq:minimal-SUGRA}) and solve the RGEs
  for these parameters between the Planck scale to the EW scale.
\end{itemize}
The RGEs for the MSSM and MSSMRN are given for example in
 the reference \cite{91BeBoMaRi,99HiNo}
and we show the RGEs for the SU(5)RN SUSY GUT in Subsection
\ref{sec:RGE for SU(5)RN}.
In Subsection \ref{sec:matching} we explain the matching condition
at the GUT scale taking account of the higher dimensional terms in the
K\"ahler potential.

\subsection{
RGE for the SU(5) SUSY GUT with right-handed neutrino
}
\label{sec:RGE for SU(5)RN}

In this subsection we show one-loop RGEs for the SU(5)RN SUSY GUT.
In the derivation of RGEs, we only take account of diagrams to
which the higher dimensional operators are inserted at most one time.
In this approximation, the quadratic divergence does not appear in
the calculation.

The RGEs for the SU(5) gauge coupling constant $g_5$ and the gaugino
mass parameter $M_5$ are given by
\begin{mathletters}
\begin{eqnarray}
  (4\pi)^2M\frac{d}{d M} g_5 &=& b_5 g_5^3, 
\\
  (4\pi)^2M\frac{d}{d M} M_5 &=&  2 b_5 g_5^2 M_5,
\end{eqnarray}
\end{mathletters}
where $M$ is the renormalization scale.
The coefficient of the beta function is given as $b_5 = -3$ for the
minimal field contents.

The RGEs for the coupling constants
 in the superpotentials, Eqs.~(\ref{eq:SU5RN})
 and (\ref{eq:dim5 superpotential}) are represented
 as follows:
\begin{mathletters}
\begin{eqnarray}
(4\pi)^2M\frac{d}{d M}(\lambda_u)_{ij} &=& (\lambda_u)_{kj}(\Theta_T)^k_{~i}
                                    +(\lambda_u)_{ik}(\Theta_T)^k_{~j}
                                    +(\lambda_u)_{ij} \Theta_H, 
\\
(4\pi)^2M\frac{d}{d M}(\lambda_d)_{ij} &=& (\lambda_d)_{kj}
                                   (\Theta_{\overline{F}})^k_{~i}
                                    +(\lambda_d)_{ik}(\Theta_T)^k_{~j}
                                  +(\lambda_d)_{ij} \Theta_{\overline{H}}, 
\\
(4\pi)^2 M\frac{d}{d M}(\lambda_{\nu})_{ij}
                                 &=& (\lambda_\nu)_{kj}
                                      (\Theta_{\overline{N}})^k_{~i}
                                    +(\lambda_\nu)_{ik}
                                     (\Theta_{\overline{F}})^k_{~j}
                                    +(\lambda_\nu)_{ij} \Theta_H, 
\\
(4\pi)^2 M\frac{d}{d M}(\kappa_u^\pm)_{ij} &=&
                                      (\kappa_u^{\pm})_{kj}(\Theta_T)^k_{~i}
                                    +(\kappa_u^{\pm})_{ik}(\Theta_T)^k_{~j}
                                    +(\kappa_u^{\pm})_{ij}
                                     (\Theta_H+\Theta_{\Sigma}), 
\\
(4\pi)^2 M\frac{d}{d M}(\kappa_d)_{ij} &=& (\kappa_d)_{kj}
                                     (\Theta_{\overline{F}})^k_{~i}
                                    +(\kappa_d)_{ik}(\Theta_T)^k_{~j}
                                    +(\kappa_d)_{ij}
                                     (\Theta_{\overline{H}}+\Theta_{\Sigma}),
\\
(4\pi)^2 M\frac{d}{d M}(\overline{\kappa}_d)_{ij} &=&
                                     (\overline{\kappa}_d)_{kj}
                                     (\Theta_{\overline{F}})^k_{~i}
                                    +(\overline{\kappa}
                                      _d)_{ik}(\Theta_T)^k_{~j}
                                    +(\overline{\kappa}_d)_{ij}
                                     (\Theta_{\overline{H}}+\Theta_{\Sigma}),
\\
(4\pi)^2M\frac{d}{d M}(\kappa_{\nu})_{ij} &=& (\kappa_{\nu})_{kj}
                                     (\Theta_{\overline{N}})^k_{~i}
                                    +(\kappa_{\nu})_{ik}(\Theta_T)^k_{~j}
                                    +(\kappa_{\nu})_{ij}
                                     (\Theta_H+\Theta_{\Sigma}),
\end{eqnarray}
\end{mathletters}
where $\Theta$'s are given by
\begin{mathletters}
\begin{eqnarray}
(\Theta_T)^i_{~j} &=& 2(\lambda_d^{\dag})^{ik}(\lambda_d)_{kj}
                     +3(\lambda_u^{\dag})^{ik}(\lambda_u)_{kj}
                     -\frac{36}{5} g_5^2 \delta^i_{j}, 
\\
(\Theta_{\overline{F}})^i_{~j}
                  &=& 4(\lambda_d^*)^{ik}(\lambda_d^T)_{kj}
                     + (\lambda_{\nu}^{\dag})^{ik}(\lambda_{\nu})_{kj}
                     -\frac{24}{5} g_5^2 \delta^i_{j}, 
\\
(\Theta_{\overline{N}})^i_{~j} &=&
                           5(\lambda_{\nu}^*)^{ik}(\lambda_{\nu}^T)_{kj}, 
\\
\Theta_{\overline{H}}~~~ &=& 4 \text{Tr}\left(\lambda_d^{\dag}\lambda_d\right)
                           -\frac{24}{5} g_5^2, 
\\
\Theta_{H}~~~ &=& \frac{3}{2} \text{Tr}\left(\lambda_u^{\dag}\lambda_u\right)
                           +\text{Tr}\left(\lambda_{\nu}^{\dag}\lambda_{\nu}
                                     \right)
                           -\frac{24}{5}g_5^2,
\\
\Theta_{\Sigma}~~~&=& -10g_5^2.
\end{eqnarray}
\end{mathletters}

The RGEs for the SUSY breaking parameters in Eq.~(\ref{eq:SUSY-Breaking-SU5RN})
 are written as follows: 
\begin{mathletters}
\begin{eqnarray}
(4\pi)^2M\frac{d}{d M}(\widetilde{\lambda}_u)_{ij} &=&
                                (\widetilde{\lambda}_u)_{kj}(\Theta_T)^k_{~i}
                               +(\widetilde{\lambda}_u)_{ik}(\Theta_T)^k_{~j}
                               +(\widetilde{\lambda}_u)_{ij} \Theta_H
\nonumber\\&&
                          +2\left\{
                           (\lambda_u)_{kj}(\widetilde{\Theta}_T)^k_{~i}
                          +(\lambda_u)_{ik}(\widetilde{\Theta}_T)^k_{~j}
                          +(\lambda_u)_{ij} \widetilde{\Theta}_H \right
                          \}, 
\\
(4\pi)^2M\frac{d}{d M}(\widetilde{\lambda}_d)_{ij} &=&
                                (\widetilde{\lambda}_d)_{kj}
                                (\Theta_{\overline{F}})^k_{~i}
                               +(\widetilde{\lambda}_d)_{ik}
                                (\Theta_T)^k_{~j}
                               +(\widetilde{\lambda}_d)_{ij}
                                 \Theta_{\overline{H}}
\nonumber\\&&
                           +2\left\{
                            (\lambda_d)_{kj}
                            (\widetilde{\Theta}_{\overline{F}})^k_{~i}
                            +(\lambda_d)_{ik}
                                (\widetilde{\Theta}_T)^k_{~j}
                               +(\lambda_d)_{ij}
                                 \widetilde{\Theta}_{\overline{H}} 
                             \right\}, 
\\
(4\pi)^2M\frac{d}{d M}(\widetilde{\lambda}_{\nu})_{ij} &=&
                                (\widetilde{\lambda}_\nu)_{kj}
                                 (\Theta_{\overline{N}})^k_{~i}
                               +(\widetilde{\lambda}_\nu)_{ik}
                                                 (\Theta_{\overline{F}})^k_{~j}
                               +(\widetilde{\lambda}_\nu)_{ij} \Theta_H
\nonumber \\&&
                          +2\left\{(\lambda_\nu)_{kj}
                                (\widetilde{\Theta}_{\overline{N}})^k_{~i}
                          +(\lambda_\nu)_{ik}
                                (\widetilde{\Theta}_{\overline{F}})^k_{~j}
                          +(\lambda_\nu)_{ij} \widetilde{\Theta}_H \right\}, 
\\
(4\pi)^2M\frac{d}{d M}(\widetilde{\kappa}_u^{\pm})_{ij} &=&
                                     (\widetilde{\kappa}_u^{\pm})_{kj}
                                     (\Theta_T)^k_{~i}
                                    +(\widetilde{\kappa}_u^{\pm})_{ik}
                                     (\Theta_T)^k_{~j}
                                    +(\widetilde{\kappa}_u^{\pm})_{ij}
                                     (\Theta_{\Sigma}+\Theta_H)
\nonumber \\&&
                           +2\left\{(\kappa_u^{\pm})_{kj}
                                     (\widetilde{\Theta}_T)^k_{~i}
                                    +(\kappa_u^{\pm})_{ik}
                                     (\widetilde{\Theta}_T)^k_{~j}
                                    +(\kappa_u^{\pm})_{ij}
                                     (\widetilde{\Theta}_{\Sigma}
                                      +\widetilde{\Theta}_H) \right\}
\\
(4\pi)^2M\frac{d}{d M}(\widetilde{\kappa}_d)_{ij} &=&
                                     (\widetilde{\kappa}_d)_{kj}
                                     (\Theta_{\overline{F}})^k_{~i}
                                    +(\widetilde{\kappa}_d)_{ik}
                                     (\Theta_T)^k_{~j}
                                    +(\widetilde{\kappa}_d)_{ij}
                                     (\Theta_{\Sigma}+\Theta_{\overline{H}})
\nonumber \\&&
                               +2\left\{(\kappa_d)_{kj}
                                     (\widetilde{\Theta}_{\overline{F}})^k_{~i}
                                    +(\kappa_d)_{ik}
                                     (\widetilde{\Theta}_T)^k_{~j}
                                    +(\kappa_d)_{ij}
                                     (\widetilde{\Theta}_{\Sigma}
                                      +\widetilde{\Theta}_{\overline{H}}) \right\}, 
\\
(4\pi)^2M\frac{d}{d M}(\widetilde{\overline{\kappa}}_d)_{ij}
                                 &=&
                                     (\widetilde{\overline{\kappa}}_d)_{kj}
                                     (\Theta_{\overline{F}})^k_{~i}
                                    +(\widetilde{\overline{\kappa}}
                                      _d)_{ik}(\Theta_T)^k_{~j}
                                    +(\widetilde{\overline{\kappa}}_d)_{ij}
                                     (\Theta_{\Sigma}+\Theta_{\overline{H}})
\nonumber \\&&
                                +2\left\{(\overline{\kappa}_d)_{kj}
                                     (\widetilde{\Theta}_{\overline{F}})^k_{~i}
                                    +(\overline{\kappa}
                                      _d)_{ik}(\widetilde{\Theta}_T)^k_{~j}
                                    +(\overline{\kappa}_d)_{ij}
                                     (\widetilde{\Theta}_{\Sigma}
                                      +\widetilde{\Theta}_{\overline{H}})\right\}, 
\\
(4\pi)^2M\frac{d}{d M}(\widetilde{\kappa}_{\nu})_{ij}
                                 &=& (\widetilde{\kappa}_{\nu})_{kj}
                                     (\Theta_{\overline{N}})^k_{~i}
                                    +(\widetilde{\kappa}_{\nu})_{ik}
                                     (\Theta_{\overline{F}})^k_{~j}
                                    +(\widetilde{\kappa}_{\nu})_{ij}
                                     (\Theta_{\Sigma}+\Theta_H),
\nonumber \\&&
                               +2\left\{ (\kappa_{\nu})_{kj}
                                     (\widetilde{\Theta}_{\overline{N}})^k_{~i}
                                    +(\kappa_{\nu})_{ik}
                                     (\widetilde{\Theta}_{\overline{F}})^k_{~j}
                                    +(\kappa_{\nu})_{ij}
                                     (\widetilde{\Theta}_{\Sigma}
                                      +\widetilde{\Theta}_H) \right\},
\end{eqnarray}
\begin{eqnarray}
(4\pi)^2M\frac{d}{d M} (m_T^2)^i_{~j} &=& (\Theta_T)^i_{~k}(m_T^2)^k_{~j}
                                  +(m_T^2)^i_{~k}(\Theta_T)^k_{~j}
\nonumber\\ &&
               +2\left[3(\lambda_u^{\dag})^{ik}\left\{({m_T^2}^T)_k^{~l}
               +(m_H^2)\delta_k^{l}\right\}(\lambda_u)_{lj}
               \right.
\nonumber\\&&
\phantom{+2}
               \left.
               +2(\lambda_d^{\dag})^{ik}\left\{({m_{\overline{F}}^2}^T)_k^{~l}
               +(m_{\overline{H}}^2)\delta_k^{l}\right\}(\lambda_d)_{lj}
               \right.
\nonumber\\&&
\phantom{+2}
               \left.
               +3 (\widetilde{\lambda}_u^{\dag})^{ik}
                    (\widetilde{\lambda}_u)_{kj}
               +2 (\widetilde{\lambda}_d^{\dag})^{ik}
                    (\widetilde{\lambda}_d)_{kj} \right]
\nonumber\\&&
               +\frac{72}{5} g_5^2
                  \left\{(m_T^2)^i_{~j}-2|M_5|^2\delta^i_{j}\right\},
\\
(4\pi)^2M\frac{d}{d M} (m_{\overline{F}}^2)^i_{~j} &=&
                       (\Theta_{\overline{F}})^i_{~k}
                       (m_{\overline{F}}^2)^k_{~j}
                       +(m_{\overline{F}}^2)^i_{~k}
                       (\Theta_{\overline{F}})^k_{~j}
\nonumber\\&&
               +2\left[ 4(\lambda_d^*)^{ik}\left\{({m_T^2}^T)_k^{~l}
               +(m_{\overline{H}}^2)\delta_k^{l}\right\}(\lambda_d^T)_{lj}
                 \right.
\nonumber\\&&
\phantom{+2}
                 \left.
               +(\lambda_{\nu}^{\dag})^{ik}\left\{({m_N^2}^T)_k^{~l}
               +(m_H^2)\delta_k^{l}\right\}(\lambda_{\nu})_{lj}
                 \right.
\nonumber\\&&
\phantom{+2}
                 \left.
               +4 (\widetilde{\lambda}_d^*)^{ik}
                   (\widetilde{\lambda}_d^T)_{kj}
               +(\widetilde{\lambda}_{\nu}^{\dag})^{ik}
                 (\widetilde{\lambda}_{\nu})_{kj} \right]
\nonumber\\&&
               +\frac{48}{5} g_5^2 
                  \left\{(m_{\overline{F}}^2)^i_{~j}-2|M_5|^2\delta^i_{j}
                  \right\},
\\
(4\pi)^2M\frac{d}{d M} (m_{\overline{N}}^2)^i_{~j} &=&
                  (\Theta_{\overline{N}})^i_{~k}(m_{\overline{N}}^2)^k_{~j}
                      +(m_{\overline{N}})^2)^i_{~k}
                        (\Theta_{\overline{N}})^k_{~j}
\nonumber\\&&
               +2\left[ 
                5(\lambda_{\nu}^*)^{ik}\left\{({m_{\overline{F}}^2}^T)_k^{~l} 
               +(m_H^2)\delta_k^{l}\right\}(\lambda_{\nu}^T)_{lj}
               +5 (\widetilde{\lambda}_{\nu}^*)^{ik}
                   (\widetilde{\lambda}_{\nu}^T)_{kj} \right],\\
(4\pi)^2M\frac{d}{d M} m_H^2~~ &=& 2 \Theta_H m_H^2
               +2\left\{
                   6\text{Tr}\left(\lambda_u^{\dag} {m_T^2}^T
                   \lambda_u\right)
                  +\text{Tr}\left(\lambda_{\nu}^\dag
                    {m_{\overline{N}}^2}^T
                    \lambda_{\nu}\right) 
                 \right.
\nonumber\\&&
                 \left.
                  +\text{Tr}\left(\lambda_{\nu}^* 
                    {m_{\overline{F}}^2}^T
                    \lambda_{\nu}^T\right) 
               +3\text{Tr}\left(\widetilde{\lambda}_u^{\dag}
                                \widetilde{\lambda}_u\right)
               +\text{Tr}\left(\widetilde{\lambda}_{\nu}^{\dag}
                               \widetilde{\lambda}_{\nu}\right) \right\} 
\nonumber\\&&
               +\frac{48}{5}g_5^2\left(m^2_H -2|M_5|^2 \right),
\\
(4\pi)^2M\frac{d}{d M} (m_{\overline{H}}^2)~~ &=&
                2 \Theta_{\overline{H}}  m_{\overline{H}}^2
               +2
            \left\{ 4\text{Tr}\left(\lambda_d^*{m_T^2}^T \lambda_d^T\right)
               +4\text{Tr}\left(\lambda_d^{\dag} {m_{\overline{F}}^2}^T
                     \lambda_d\right)
            \right.
\nonumber\\&&
            \left.         
               +4\text{Tr}\left(\widetilde{\lambda}_d^*
                                \widetilde{\lambda}_d^T\right) 
            \right\}  
               +\frac{48}{5}g_5^2\left( m^2_{\overline{H}} -2|M_5|^2\right),
\end{eqnarray}
\end{mathletters}
where $\widetilde{\Theta}$'s are given by
\begin{mathletters}
\begin{eqnarray}
(\widetilde{\Theta}_T)^i_{~j} &=& 2(\lambda_d^{\dag})^{ik}
                                   (\widetilde{\lambda}_d)_{kj}
                                 +3(\lambda_u^{\dag})^{ik}
                                   (\widetilde{\lambda}_u)_{kj}
                                 -\frac{36}{5} g_5^2 M_5 \delta^i_{j}, 
\\
(\widetilde{\Theta}_{\overline{F}})^i_{~j}
                  &=& 4(\lambda_d^*)^{ik}(\widetilde{\lambda}_d^T)_{kj}
                     +(\lambda_{\nu}^{\dag})^{ik}
                       (\widetilde{\lambda}_{\nu})_{kj}
                     -\frac{24}{5} g_5^2 M_5 \delta^i_{j}, 
\\ 
(\widetilde{\Theta}_{\overline{N}})^i_{~j} &=& 5(\lambda_{\nu}^*)^{ik}
                                   (\widetilde{\lambda}_{\nu}^T)_{kj}, 
\\
\widetilde{\Theta}_{\overline{H}}~~~ &=&
                           4 \text{Tr}\left
                            (\lambda_d^{\dag}\widetilde{\lambda}_d\right)
                           -\frac{24}{5} g_5^2 M_5, 
\\
\widetilde{\Theta}_H~~~ &=& \frac{3}{2} \text{Tr}
                           \left(\lambda_u^{\dag}\widetilde{\lambda}_u\right)
                           +\text{Tr}\left(\lambda_{\nu}^{\dag}
                             \widetilde{\lambda}_{\nu}\right)
                           -\frac{24}{5}g_5^2 M_5, 
\\
\widetilde{\Theta}_{\Sigma}~~~ &=& -10 g_5^2 M_5.
\end{eqnarray}
\end{mathletters}

\subsection{
Matching conditions at the GUT scale
}
\label{sec:matching}

In this subsection we show the matching conditions between
 the SU(5)RN SUSY GUT and the MSSMRN at the GUT scale
 taking account of the dimension five terms in the K\"ahler potential.  
Although we include only the renormalizable terms in the K\"ahler potential
 at the Planck scale, 
 higher dimensional terms are induced by the renormalization
 effects between the Planck scale and the GUT scale, 
because we introduce the higher dimensional operators in the superpotential.
In order to simplify the treatment, we use logarithmic approximation for
 induced terms in the K\"ahler potential.
For other coupling constants, we explicitly solve the RGEs in the previous
 subsection.

Up to dimension five terms, the corrections for
 the K\"ahler potential in the present model are parameterized as follows:
\begin{eqnarray}
\label{eq:dim5 kahler}
\Delta{\cal K}_{\rm SU(5)RN}
           &=& \frac{1}{M_X}
               \left\{      (\overline{k}_T)^i_{~j}
                            (T_i^\dag)_{ab}(\Sigma)^b_{~c}(T^j)^{ca}
                          +(\overline{k}_{\overline{F}})^i_{~j}
                            (\overline{F}_i^\dag)^a
                                  (\Sigma)^b_{~a}(\overline{F}_j)_b 
               \right.
\nonumber\\&&
\phantom{\frac{1}{M_X}}
               \left.
                    + \overline{k}_H (H^\dag)_a(\Sigma)^a_{~b}(H)^b
                    + \overline{k}_{\overline{H}} (\overline{H}^\dag)^a
                                  (\Sigma)^b_{~a}(\overline{H})_b
              \right\}
              +\text{H.c.},
\end{eqnarray}
where we include SUSY breaking parts in the coupling constants
 using the spurion method as follows:
\begin{eqnarray}
\label{eq:dim5-kahler-coupling}
 \overline{k}_X &=& k_X +\hat{k}_X \theta^2
                           +\check{k}_X \overline{\theta}^2
                           +\widetilde{k}_X \theta^2\overline{\theta}^2
                           ~~~~~(X = T, \overline{F}, H, \overline{H}).
\end{eqnarray}
At the Planck scale, we assume all the components of
 Eq.~(\ref{eq:dim5-kahler-coupling}) are zero.
These coupling constants are induced from the renormalization
 between the Planck scale and the GUT scale.
In the logarithmic approximation, $k$'s at the GUT scale are given by,
\begin{mathletters}
\label{eq:mSUGRA-kahler}
\begin{eqnarray}
(k_T)^i_{~j} &\approx& -2\left\{
                 (\lambda_u^{\dag})^{ik}(\kappa_u^+)_{kj}
                 -5(\lambda_u^{\dag})^{ik}(\kappa_u^-)_{kj}
                 +(\lambda_d^{\dag})^{ik}(\kappa_d)_{kj}
                 +(\lambda_d^{\dag})^{ik}(\overline{\kappa}_d)_{kj}
                       \right\} t_G,
\\
(k_{\overline{F}})^i_{~j}
             &\approx& -2\left\{
                   (\lambda_d^*)(\kappa_d)_{jk}-4(\lambda_d^*)_{ik}
                   (\overline{\kappa}_d)_{jk}
                  -(\lambda_{\nu}^{\dag})^{ik}(\kappa_{\nu})_{kj}
                       \right\} t_G,
\\
k_H~~ &\approx& -2\left\{ 3\text{Tr}\left(\lambda_u^{\dag} \kappa_u\right)
                        -\text{Tr}
                         \left(\lambda_{\nu}^{\dag} \kappa_{\nu}\right)
                \right\} t_G,
\\
k_{\overline{H}}~~ &\approx& -2\left\{\text{Tr}
                                     \left(\lambda_d^{\dag} \kappa_d\right)
                     -4\text{Tr}
                             \left(\lambda_d^{\dag} \overline{\kappa}_d\right)
                             \right\} t_G.
\end{eqnarray}
\end{mathletters}
In the same approximation, $\hat{k}_X$, $\check{k}_X$ and $\widetilde{k}_X$
 at the GUT scale are proportional to $k_X$ as follows:
\begin{mathletters}
\label{eq:kahler-soft}
\begin{eqnarray}
 \hat{k}_X &\approx& m_0(A_0 +\Delta A_0) k_X, 
\\
 \check{k}_X &\approx& m_0 A_0 k_X, 
\\
 \widetilde{k}_X &\approx& m_0^2\left\{A_0^*(A_0 +\Delta A_0)+2
                                       \right\}
                                        k_X,
~~~~~(X = T, \overline{F}, H, \overline{H}).
\end{eqnarray}
\end{mathletters}
The dimension five terms in Eq.~(\ref{eq:dim5 kahler})
 modify the normalization of
 the K\"ahler potential after the SU(5) symmetry breaking. 
In order to obtain the correct normalization up to $O(\xi)$,
 we introduce the following new chiral superfields:
\begin{mathletters}
\begin{eqnarray}
({T^{i}}^{\prime})^{ab} &=& (T^i)^{ab}
+\frac{1}{M_X}
              \left\{(k_T)^i_{~j}+\theta^2 (\hat{k}_T)^i_{~j} \right\}
              \left\{(\Sigma)^a_{~c}(T^j)^{cb}
                         -(\Sigma)^b_{~c}(T^j)^{ca}
              \right\},
\\
({\overline{F}^{i}}^{\prime})_a &=& (\overline{F}^i)_a
+\frac{1}{M_X}
              \left\{(k_{\overline{F}})^i_{~j}
                     +\theta^2 (\hat{k}_{\overline{F}})^i_{~j}
              \right\}(\Sigma)^b_{~a}(\overline{F}^j)_b, 
\\
({H}^\prime)^a &=& (H)^a +\frac{1}{M_X}
                               \left( k_H +\theta^2 \hat{k}_H
                               \right)(\Sigma)^a_{~b}H^b
\\
({\overline{H}}^\prime)_a &=& (\overline{H})_a
                 +\frac{1}{M_X}
                               \left( k_{\overline{H}}
                                      +\theta^2 \hat{k}_{\overline{H}}
                               \right)
               (\Sigma)^b_{~a}(\overline{H})_b. 
\end{eqnarray}
\end{mathletters}
Substituting these superfields for Eqs.~(\ref{eq:SU5RN}) and 
(\ref{eq:dim5 superpotential}),
 we can define the coupling constants
 in terms of the new chiral superfields as follows:
\begin{mathletters}
\label{eq:dim5-redefinition}
\begin{eqnarray}
({\kappa_u^+}^\prime)_{ij} &=& (\kappa_u^+)_{ij}
 -\frac{1}{2}(\lambda_u)_{ik}(k_T)^k_{~j}
 -\frac{1}{2}(\lambda_u)_{kj}(k_T)^k_{~i}
 +(\lambda_u)_{ij} k_H, 
\\
({\kappa_u^-}^\prime)_{ij} &=& (\kappa_u^-)_{ij}
 +\frac{1}{2}(\lambda_u)_{ik}(k_T)^k_{~j}
 -\frac{1}{2}(\lambda_u)_{kj}(k_T)^k_{~i}, 
\\
({\kappa_d}^\prime)_{ij} &=& (\kappa_d)_{ij}
 -(\lambda_d)_{ik}(k_T)^k_{~j}
 -(\lambda_d)_{kj}(k_{\overline{F}})^k_{~i},   
\\
({\overline{\kappa}_d}^\prime)_{ij} &=& (\overline{\kappa}_d)_{ij}
 -(\lambda_d)_{ik}(k_T)^k_{~j}
 -(\lambda_d)_{ij} k_{\overline{H}}, \\
({\kappa_{\nu}}^\prime)_{ij} &=& (\kappa_{\nu})_{ij}
-(\lambda_{\nu})_{ik}(k_{\overline{F}})^k_{~j}
-(\lambda_{\nu})_{ij} k_H .
\end{eqnarray}
\end{mathletters}
The matching conditions for the Yukawa coupling matrices for quarks
 and leptons are written with the above new coupling constants as follows
\footnote{$\kappa$'s in Eqs.~(\ref{eq:yukawa-matching})
 of Sec.~\ref{sec:model} should be read as those with prime
 if we take account of the higher dimensional terms radiatively 
 induced in the K\"ahler potential.}:
\begin{mathletters}
\label{eq:yukawa-matching-kahler}
\begin{eqnarray}
(y_u)_{ij} &=& (\lambda_u)_{ij} +\xi\left\{
                       \frac{1}{2}({\kappa^+_u}^\prime)_{ij}
                      +\frac{5}{6}({\kappa^-_u}^\prime)_{ij} \right\}, 
\\
(y_d)_{ij} &=& (\lambda_d)_{ij} +\xi\left\{
                       \frac{1}{3}({\kappa_d}^\prime)_{ij}
                      -\frac{1}{2}({\overline{\kappa}_d}^\prime)_{ij}
               \right\}, 
\\
(y_e)_{ij} &=& (\lambda_d^T)_{ij} +\xi\left\{
                       -\frac{1}{2}({\kappa_d}^{\prime T})_{ij}
                       -\frac{1}{2}({\overline{\kappa}_d}^{\prime T})_{ij}
               \right\}, 
\\
(y_{\nu})_{ij} &=& (\lambda_{\nu})_{ij}-\frac{\xi}{2}
                    ({\kappa_{\nu}}^\prime)_{ij}. 
\end{eqnarray}
\end{mathletters}
Using Eqs.~(\ref{eq:mSUGRA-kahler})--(\ref{eq:yukawa-matching-kahler})
 we can relate $y$'s and $\lambda$'s and $\kappa$'s
 at the GUT scale.
The SUSY breaking parameters are also defined in terms of
 the new superfields as follows:
\begin{mathletters}
\label{eq:SUSYb-redefinition}
\begin{eqnarray}
(\widetilde{\kappa}_u^{+ \prime})_{ij} &=& (\widetilde{\kappa}_u^+)_{ij}
 -\frac{1}{2}(\widetilde{\lambda}_u)_{ik}(k_T)^k_{~j}
 -\frac{1}{2}(\widetilde{\lambda}_u)_{kj}(k_T)^k_{~i}
 +(\widetilde{\lambda}_u)_{ij} k_H 
\nonumber\\&&
\phantom{(\widetilde{\kappa}_u^{+ \prime})_{ij}}
 -\frac{1}{2}(\lambda_u)_{ik}(\hat{k}_T)^k_{~j}
 -\frac{1}{2}(\lambda_u)_{kj}(\hat{k}_T)^k_{~i}
 +(\lambda_u)_{ij} \hat{k}_H, 
\\
(\widetilde{\kappa}_u^{- \prime})_{ij} &=& (\widetilde{\kappa}_u^-)_{ij}
 -\frac{1}{2}(\widetilde{\lambda}_u)_{ik}(k_T)^k_{~j}
 +\frac{1}{2}(\widetilde{\lambda}_u)_{kj}(k_T)^k_{~i} 
\nonumber\\&&
\phantom{(\widetilde{\kappa}_u^{- \prime})_{ij}}
 -\frac{1}{2}(\lambda_u)_{ik}(\hat{k}_T)^k_{~j}
 +\frac{1}{2}(\lambda_u)_{kj}(\hat{k}_T)^k_{~i},
\\
({\widetilde{\kappa}_d}^{\prime})_{ij} &=& (\widetilde{\kappa}_d)_{ij}
 -(\widetilde{\lambda}_d)_{ik}(k_T)^k_{~j}-(\widetilde{\lambda}_d)_{kj}
(k_{\overline{F}})^k_{~i}
 -(\lambda_d)_{ik}(\hat{k}_T)^k_{~j}-(\lambda_d)_{kj}
(\hat{k}_{\overline{F}})^k_{~i}, 
\\
({\widetilde{\overline{\kappa}}_d}^{\prime})_{ij} &=&
 (\widetilde{\overline{\kappa}}_d)_{ij}
 -(\widetilde{\lambda}_d)_{ik}(k_T)^k_{~j}
 -(\widetilde{\lambda}_d)_{ij} k_{\overline{H}} 
\nonumber\\&&
\phantom{({\widetilde{\overline{\kappa}}_d}^{\prime})_{ij}}
 -(\lambda_d)_{ik}(\hat{k}_T)^k_{~j} -(\lambda_d)_{ij}
 \hat{k}_{\overline{H}}, 
\\
({\widetilde{\kappa}_{\nu}}^{\prime})_{ij} &=& (\widetilde{\kappa}_{\nu})_{ij}
-(\widetilde{\lambda}_{\nu})_{ik}(k_{\overline{F}})^k_{~j}
-(\widetilde{\lambda}_{\nu})_{ij} k_H
-(\lambda_{\nu})_{ik}(\hat{k}_{\overline{F}})^k_{~j}
-(\lambda_{\nu})_{ij} \hat{k}_H,
\end{eqnarray}
\begin{eqnarray}
({\widetilde{k}_T}^{\prime})^i_{~j} &=& (\widetilde{k}_T)^i_{~j}
 + (m_T^2)^i_{~k}(k_T)^k_{~j}, 
\\
({\widetilde{k}_{\overline{F}}}^{\prime})^i_{~j} &=&
 (\widetilde{k}_{\overline{F}})^i_{~j}
 + (m_{\overline{F}}^2)^i_{~k}(k_{\overline{F}})^k_{~j}, 
\\
\widetilde{k}_H^{\prime} &=& \widetilde{k}_H
 + m_H^2 k_H,
\\
\widetilde{k}_{\overline{H}}^{\prime} &=& \widetilde{k}_{\overline{H}}
 + m_{\overline{H}}^2 k_{\overline{H}}.
\end{eqnarray}
\end{mathletters}
The soft SUSY breaking parameters can be expressed using the above new coupling
 constants as follows:
\begin{mathletters}
\label{eq:soft-matching}
\begin{eqnarray}
(\widetilde{y}_u)_{ij} &=& (\widetilde{\lambda}_u)_{ij} +\xi
                      \left\{
                       \frac{1}{2}(\widetilde{\kappa}_u^{+ \prime})_{ij}
                      +\frac{5}{6}(\widetilde{\kappa}_u^{- \prime})_{ij} 
                      \right.
\nonumber\\&&
\phantom{(\widetilde{\lambda}_u)_{ij} +\xi}
                      \left.
                      +\frac{1}{6}(\lambda_u)_{ik}(\check{k}_T^{\dag})^k_{~j}
                      -\frac{2}{3}(\lambda_u)_{kj}(\check{k}_T^{\dag})^k_{~i}
                      +\frac{1}{2}(\lambda_u)_{ij} \check{k}_H^*
                      \right\}, 
\\
(\widetilde{y}_d)_{ij} &=& (\widetilde{\lambda}_d)_{ij} +\xi
                      \left\{
                       \frac{1}{3}({\widetilde{\kappa}_d}^{\prime})_{ij}
                      -\frac{1}{2}({\widetilde{\overline{\kappa}}_d}
                        ^{\prime})_{ij}
                      \right.
\nonumber\\&&
\phantom{(\widetilde{\lambda}_d)_{ij} +\xi}
                      \left.
                       -\frac{1}{6}(\lambda_d)_{ik}(\check{k}_T^{\dag})^k_{~j}
                       -\frac{1}{3}(\lambda_d)_{kj}
                             (\check{k}_{\overline{F}}^{\dag})^k_{~i}
                       +\frac{1}{2} (\lambda_d)_{ij}
                                     \check{k}_{\overline{H}}^*
                      \right\}, 
\\
(\widetilde{y}_e)_{ij} &=& (\widetilde{\lambda}_d^T)_{ij} +\xi
                      \left\{
                       -\frac{1}{2}({\widetilde{\kappa}_d}^{\prime T})_{ij}
                       -\frac{1}{2}({\widetilde{\overline{\kappa}}_d}
                          ^{\prime T})_{ij}
                      \right.
\nonumber\\&&
\phantom{(\widetilde{\lambda}_d^T)_{ij} +\xi}
                      \left.
                       +(\lambda_d^T)_{kj}(\check{k}_T^{\dag})^k_{~i}
                       +\frac{1}{2}(\lambda_d^T)_{ik}
                             (\check{k}_{\overline{F}}^{\dag})^k_{~j}
                       +\frac{1}{2} (\lambda_d^T)_{ij}
                                     \check{k}_{\overline{H}}^*
                      \right\}, 
\\
(\widetilde{y}_{\nu})_{ij} &=& (\widetilde{\lambda}_{\nu})_{ij}+\xi
                      \left\{
                       -\frac{1}{2}({\widetilde{\kappa}_{\nu}}^{\prime})_{ij}
                       +\frac{1}{2}(\lambda_{\nu})_{ik}
                         (\check{k}_{\overline{F}}^{\dag})^k_{~j}
                       +\frac{1}{2}(\lambda_{\nu})_{ij}
                          \check{k}_H^* 
                      \right\},
\end{eqnarray}
\begin{eqnarray}
(m_Q^2)^i_{~j} &=& (m_T^2)^i_{~j} + \frac{1}{6}\xi
                                   \left\{({\widetilde{k}_T}^{\prime})^i_{~j}
                                    +({\widetilde{k}_T}^{\prime \dag})^i_{~j}
                                   \right\}, 
\\
(m_U^2)_i^{~j} &=& (m_T^2)^j_{~i}
                                  - \frac{2}{3}\xi
                                   \left\{({\widetilde{k}_T}^{\prime})^j_{~i}
                                    +({\widetilde{k}_T}^{\prime \dag})^j_{~i}
                                   \right\}, 
\\
(m_E^2)_i^{~j} &=& (m_T^2)^j_{~i}
                                  + \xi
                                   \left\{({\widetilde{k}_T}^{\prime})^j_{~i}
                                    +({\widetilde{k}_T}^{\prime \dag})^j_{~i}
                                   \right\}, 
\\
(m_D^2)_i^{~j} &=& (m_{\overline{F}}^2)^j_{~i}
                 - \frac{1}{3}\xi
                  \left\{({\widetilde{k}_{\overline{F}}}^{\prime})^j_{~i}
                   + ({\widetilde{k}_{\overline{F}}}^{\prime \dag})^j_{~i}
                  \right\}, 
\\
(m_L^{2})^i_{~j} &=& (m_{\overline{F}}^2)^i_{~j}
                 + \frac{1}{2}\xi
                  \left\{({\widetilde{k}_{\overline{F}}}^{\prime})^i_{~j}
                   + ({\widetilde{k}_{\overline{F}}}^{\prime \dag})^i_{~j}
                  \right\}, 
\\
m_{H_1}^2~ &=& m_{\overline{H}}^2~~+\xi~{\widetilde{k}_{\overline{H}}}
               ^{\prime}, 
\\
m_{H_2}^2~ &=& m_H^2~~+\xi~{\widetilde{k}_H}^{\prime}.
\end{eqnarray}
\end{mathletters}
Using Eqs.~(\ref{eq:SUSYb-redefinition}) and (\ref{eq:soft-matching})
, we can express the soft SUSY breaking terms of the MSSMRN by the input
 parameters at the Planck scale, Eq.~(\ref{eq:minimal-SUGRA}).

\section{
Approximate expressions of the photon-penguin amplitudes for \meg\
process
}
\label{sec:approximation-mueg}

In this appendix we show the explicit forms of the functions
 which appear in the approximated expressions
 of the photon-penguin amplitudes given in
 Eq.~(\ref{eq:mueg-approximate-formulas}).
 We assume the following conditions to derive the expressions:
\begin{itemize}
\item The off-diagonal elements of the slepton mass matrices, $(m^2_E)_i^{~j}$
 and $(m^2_L)^i_{~j}$
 are given by Eqs.~(\ref{eq:mass correction 2})
 and (\ref{eq:mass correction 4}), and
 they  are diagonalized with good approximation
 in the basis where $(y_{\nu})_{ij}$ and $(y_{CR})_{ij}$
 in Eqs.~(\ref{eq:MSSM Yukawa}) and (\ref{eq:colored-Higgs-Yukawa})
 are diagonal.
\item In this basis, the left-right mixing mass of the slepton
 can be treated as perturbation
 to diagonalize the 6$\times$6 charged slepton mass matrix.
\item The eigenvalues of the slepton mass matrices are 
 almost degenerate and represented by $\overline{m}^2$.
\end{itemize}
With these conditions,
the SUSY contributions to the photon-penguin amplitudes correspond
 to Figs.~(\ref{fig:mueg1})-(\ref{fig:mueg3}) are expressed as
 Eq.~(\ref{eq:mueg-approximate-formulas}). 
In this formula, $a^n_2$, $a^c$ and $a^n_1$ are given by
\begin{mathletters}
\begin{eqnarray}
a^n_2 &=& -\frac{e}{32\pi^2}\tan\theta_W\sum^4_{A=1}
            (O_{N}^*)_{A1}\{(O_{N})_{A2}+\tan\theta_W(O_{N})_{A1}\}
            f^n_2\left(\frac{\overline{m}^2}{m_{\widetilde{\chi}^0_A}^2}\right)
            \left(\frac{m_W}{\overline{m}}\right)^2
            \left(\frac{m_0}{\overline{m}}\right)^4
\nonumber\\&&
\phantom{-\frac{e}{32\pi^2}\tan\theta_W\sum^4_{A=1}}
            \times
            \frac{m_0(A_e+\frac{3}{5}\Delta A)+\mu^*\tan\beta}{\overline{m}}
            (3+|A_0|^2)^2t_G(t_G+t_R), 
\\
a^c   &=& \frac{\sqrt{2}e}{32\pi^2\cos\beta}\sum^2_{A=1}
          (O_{CL}^*)_{A2}(O_{CR})_{A1}
          f^c\left(\frac{\overline{m}^2}{m_{\widetilde{\chi}^-_A}^2}\right)
          \left(\frac{m_W}{\overline{m}}\right)
          (3+|A_0|^2)(t_G+t_R),\\
a^n_1 &=& -\frac{e}{32\pi^2}\tan\theta_W\sum^4_{A=1}
            (O_{N}^*)_{A1}\{(O_{N})_{A2}+\tan\theta_W(O_{N})_{A1}\}
            f^n_1\left(\frac{\overline{m}^2}{m_{\widetilde{\chi}^0_A}^2}\right)
\nonumber\\&&
\phantom{-\frac{e}{32\pi^2}\tan\theta_W\sum^4_{A=1}}
            \times
            \left(\frac{m_W}{\overline{m}}\right)^2
            \left(\frac{m_0}{\overline{m}}\right)
            \frac{3}{5}\frac{m_s}{m_{\mu}}\Delta A,
\end{eqnarray}
\end{mathletters}
where $\theta_W$ is the Weinberg angle.
$m_{\widetilde{\chi}^0_A}$ and $m_{\widetilde{\chi}^-_A}$
 represent the masses of neutralinos and charginos, respectively
 and $O_N$, $O_{CR}$ and $O_{CL}$ are unitary matrices which
 are used to diagonalize the neutralino and chargino
 mass matrices, $M_{\widetilde{\chi}^0}$
 and $M_{\widetilde{\chi}^-}$ as follows:
\begin{mathletters}
\begin{eqnarray}
O_{N} M_{\widetilde{\chi}^0} O_{N}^{T}
&=& \text{diag}(m_{\widetilde{\chi}^{0}_{1}},
                                m_{\widetilde{\chi}^{0}_{2}},
                                m_{\widetilde{\chi}^{0}_{3}},
                                m_{\widetilde{\chi}^{0}_{4}}),
\\
M_{\widetilde{\chi}^0} &=&
 \left(
\begin{array}{cccc}
M_{1} & 0     & -m_{Z} s_{W}\cos\beta
      & m_{Z} s_{W}\sin\beta \\
0     & M_{2} & m_{Z} c_{W}\cos\beta
      & -m_{Z} c_{W}\sin\beta \\
        -m_{Z} s_{W}\cos\beta & m_{Z} c_{W}\cos\beta
      & 0 & -\mu \\
        m_{Z} s_{W}\sin\beta & -m_{Z} c_{W}\sin\beta
      & -\mu & 0
\end{array}
\right), 
\\
O_{CR} M_{\widetilde{\chi}^-}
O_{CL}^{\dag} &=& \text{diag}(m_{\tilde{\chi}^{-}_{1}},m_{\tilde{\chi}^{-}_{2}}),
\\
M_{\widetilde{\chi}^-} &=&
\left(
\begin{array}{cc}
M_{2} & \sqrt{2}m_{W}\cos\beta \\
\sqrt{2}m_{W}\sin\beta & \mu 
\end{array}
\right). \\
\end{eqnarray}
\end{mathletters}
where $m_Z$ and $m_W$ are the $Z$ boson mass and the $W$ boson
 mass, respectively 
 and $s_W = \sin \theta_W$ and $c_W = \cos \theta_W$. 
The mass functions $f^n_2$, $f^c$ and $f^n_1$ in the above formulas
 are given by
\begin{mathletters}
\begin{eqnarray}
f^n_2(x) &=& -\frac{x^{\frac{7}{2}}}{(1-x)^4}
             \{x^2+4x-5+2(2x+1)\ln (x)\}, 
\\
f^c(x) &=& -\frac{\sqrt{x}}{2x^3(1-x)^4}
             \{14x^3-25x^2+14x-3-2x^2(4x-1)\ln (x)\},
\\
f^n_1(x) &=& \frac{1}{x^4(1-x)^5}
             \{5x^4-37x^3+27x^2+13x-8+6x(7x-3)\ln (x)\}.
\end{eqnarray}
\end{mathletters}

\section{
FCNC effective couplings in MSSM
}
\label{sec:FCNC-effective-couplings}
In this appendix we present the explicit forms of FCNC effective coupling
 constants for \kk , \bdbd\ and \bsbs\ mixings.
For \kk\ mixing these coupling constants are defined
 in Eq.~(\ref{eq:epsilon_K}) of Sec.~\ref{sec:fcnc-lfv-processes}
 and for \bdbd\ and \bsbs\ mixings the coupling constants are given by 
 substitution of flavor indices.
In the MSSM box diagrams exchanging charged Higgs, neutralino,
 chargino and gluino can contribute to these coupling constants. 
 We first define the following neutralino, chargino and gluino vertices
 for quarks and squarks,
\begin{eqnarray}
\label{eq:ino-squark vertices}
{\cal L}  & \equiv & \sum^3_{i=1}\sum^4_{A=1}\sum^6_{X=1}
                      \left\{
                       \overline{d_{i}}
                        \left( N^{d L}_{iAX}P_{L}+N^{d R}_{iAX}P_{R} \right)
                       \widetilde{\chi}^{0}_{A}\widetilde{d}_{X} 
                      \right.
\nonumber\\&&
\phantom{\sum^3_{i=1}\sum^4_{A=1}\sum^6_{X=1}}
                      \left.
                       + \overline{u_{i}}
                          \left( N^{u L}_{iAX}P_{L}
                                 +N^{u R}_{iAX}P_{R} \right)
                       \widetilde{\chi}^{0}_{A}\widetilde{u}_{X}
                      \right\}
\nonumber \\&&
                       + \sum^3_{i=1}\sum^2_{A=1}\sum^6_{X=1}
                       \left\{
                         \overline{d_{i}}
                          \left( C^{d L}_{iAX}P_{L}+C^{d R}_{iAX}P_{R}\right)
                       \widetilde{\chi}^{-}_{A}\widetilde{u}_{X} 
                       \right.
\nonumber\\&&           
\phantom{\sum^3_{i=1}\sum^4_{A=1}\sum^6_{X=1}}
                      \left.
                       + \overline{u_{i}}
                          \left(C^{u L}_{iAX}P_{L}+C^{u R}_{iAX}P_{R}\right)
                       \widetilde{\chi}^{-}_{A}\widetilde{d}_{X} 
                      \right\}
\nonumber \\&&
                     + \sum^3_{i=1}\sum^6_{X=1}\sum^8_{a=1}
                      \left\{
                           \overline{d_i}
                            \left(\Gamma^{d L}_{iX}P_L+\Gamma^{d R}_{iX}P_R
                            \right)
                            \widetilde{G}^a T^a \widetilde{d}_X   
                      \right.
\nonumber\\&&  
\phantom{\sum^3_{i=1}\sum^6_{X=1}\sum^8_{a=1}}
                      \left.
                          + \overline{u_i}
                             \left(\Gamma^{u L}_{iX}P_L
                                  +\Gamma^{u R}_{iX}P_R\right)
                            \widetilde{G}^a T^a \widetilde{u}_X 
                      \right\} + {\rm H.c.},
\end{eqnarray}
where $P_L$ and $P_R$ are projection operators defined by
 $P_L=(1-\gamma_5)/2$ and $P_R=(1+\gamma_5)/2$ and 
$T^a$ is the generator of SU(3) gauge group.
The neutralino-squark coupling constants
 appear in the above formula are given by
\begin{mathletters}
\begin{eqnarray}
N^{d L}_{iAX} & = & -\sqrt{2} g_2
                     \left\{ 
                            \frac{1}{3}\tan\theta_{W}
                            (O_N)_{A1}(U_d^*)_{X i+3}
\phantom{\frac{{m_d}_i}{2 m_{W}\cos\beta}}
                     \right.
 \nonumber \\&&
                     \left.
\phantom{-\sqrt{2} g_2\frac{1}{3}}
                      +\frac{{m_d}_i}{2 m_{W}\cos\beta}
                       (O_N)_{A3}(U_d^*)_{Xi} \right\},
 \\
N^{d R}_{iAX} & = & -\sqrt{2} g_2
                      \left[
                       \left\{-\frac{1}{2}(O_N^*)_{A2}
                              +\frac{1}{6}\tan\theta_{W}(O_N^*)_{A1} 
                       \right\}
                        (U_{d}^*)_{Xi} 
\phantom{\frac{{m_d}_i}{2 m_{W}\cos\beta}}
                      \right.
\nonumber \\&&
                      \left.
\phantom{-\sqrt{2} g_2 -\frac{1}{2}}
                    +\frac{{m_d}_i}{2 m_{W}\cos\beta}
                       (O_N)_{A3}^{*}(U_{d}^*)_{Xi+3} 
                      \right], 
\\
N^{u L}_{iAX} & = & -\sqrt{2} g_2
                     \left\{
                            -\frac{2}{3}\tan\theta_{W}
                             (O_N)_{A1}(U_{u}^*)_{X i+3}
\phantom{\frac{{m_u}_i(V_{\rm CKM})^i_{~j}}{\sqrt{2}m_{W}\sin\beta}}   
                     \right.
\nonumber\\ &&
                     \left.
\phantom{-\sqrt{2} g_2 -\frac{2}{3}}
                    +\sum^3_{j=1}
                     \frac{{m_u}_i(V_{\rm CKM})^i_{~j}}{2 m_{W}\sin\beta}
                       (O_N)_{A3}(U_{u}^*)_{Xj} \right\},
\\
N^{u R}_{iAX} & = & -\sqrt{2} g_2
                     \left[
                       \left\{\frac{1}{2} (O_N^*)_{A2}
                              +\frac{1}{6}\tan\theta_{W}(O_N^*)_{A1} 
                       \right\}(U_{u}^*)_{Xi}
\phantom{\frac{(V_{\rm CKM}^{\dag})^i_{~j} {m_u}_j}{2 m_{W}\sin\beta}}
                     \right.
\nonumber \\&&
                     \left.
\phantom{-\sqrt{2} g_2 \frac{1}{2}}
                    +\sum^3_{j=1}
                     \frac{(V_{\rm CKM}^{\dag})^i_{~j} {m_u}_j}
                          {2 m_{W}\sin\beta}
                       (O_N^*)_{A3}(U_{u}^*)_{Xj+3} 
                     \right], 
\end{eqnarray}
\end{mathletters}
where $g_2$ is the SU(2) gauge coupling constant. 
$O_N$ is the diagonalization matrix for neutralino mass matrix defined in 
 Appendix \ref{sec:approximation-mueg}.
$U_d$ and $U_u$ are unitary matrices which appear in diagonalization of 
 the $6\times6$ squark mass matrices,
 $m^2_{\widetilde{d}}$ and $m^2_{\widetilde{u}}$ as follows:
\begin{mathletters}
\begin{eqnarray}
 U_{d}m_{\tilde{d}}^{2}U_{d}^{\dag} &=& \text{diag}(m_{\tilde{d}_{1}}^{2},
                                          m_{\tilde{d}_{2}}^{2},
                                          m_{\tilde{d}_{3}}^{2},
                                          m_{\tilde{d}_{4}}^{2},
                                          m_{\tilde{d}_{5}}^{2},
                                          m_{\tilde{d}_{6}}^{2}),
 \nonumber\\
m_{\tilde{d}}^{2} &=&
\left(
\begin{array}{cc}
\begin{array}{l}
m^{2}_{Q} + m_{d}^{\dag}m_{d} \\
+m_{Z}^{2}\cos 2\beta(-\frac{1}{2}+\frac{1}{3}\sin^{2}\theta_{W}){\bf 1}
\end{array}
 & 
\frac{v}{\sqrt{2}}\cos\beta( \widetilde{y}_d+y_d \mu^* \tan\beta)^{\dag} \\
\frac{v}{\sqrt{2}}\cos\beta( \widetilde{y}_d+y_d \mu^* \tan\beta) & 
\begin{array}{l}
m^{2}_{D}+m_{d}m_{d}^{\dag} \\
-\frac{1}{3}m_{Z}^{2}\cos 2\beta\sin^{2}\theta_{W}{\bf 1}
\end{array}
\end{array}
\right),\\
 U_{u}m_{\tilde{u}}^{2}U_{u}^{\dag} &=& \text{diag}(m_{\tilde{u}_{1}}^{2},
                                          m_{\tilde{u}_{2}}^{2},
                                          m_{\tilde{u}_{3}}^{2},
                                          m_{\tilde{u}_{4}}^{2},
                                          m_{\tilde{u}_{5}}^{2},
                                          m_{\tilde{u}_{6}}^{2}),
\nonumber\\
 m_{\tilde{u}}^{2} &=&
\left(
\begin{array}{cc}
\begin{array}{l}
m^{2}_{Q} + m_{u}^{\dag}m_{u} \\
+m_{z}^{2}\cos 2\beta(\frac{1}{2}-\frac{2}{3}\sin^{2}\theta_{W}){\bf 1} 
\end{array}
& 
-\frac{v}{\sqrt{2}}\sin\beta( \widetilde{y}_u+y_u \mu^* \cot\beta)^{\dag} \\
-\frac{v}{\sqrt{2}}\sin\beta( \widetilde{y}_u+y_u \mu^* \cot\beta) & 
\begin{array}{l}
m^{2}_{U}+m_{u}m_{u}^{\dag} \\
+\frac{2}{3}m_{z}^{2}\cos 2\beta\sin^{2}\theta_{W}{\bf 1}
\end{array}
\end{array}
\right),
\end{eqnarray}
\end{mathletters}
where generation indices are suppressed and
 the mass matrices for down-type and up-type quarks
 are given by $(m_d)_{ij}={m_d}_i\delta^i_{j}$ and
 $(m_u)_{ij}={m_u}_i(V_{\rm CKM})^i_{~j}$.
The chargino-squark coupling constants
 in Eq.~(\ref{eq:ino-squark vertices}) are given by
\begin{mathletters}
\begin{eqnarray}
C^{d L}_{iAX} & = & g_2\frac{{m_d}_i}{\sqrt{2}m_{W}\cos\beta}
                  (O_{CL}^*)_{A2}(U_{u}^*)_{Xi},
 \\
C^{d R}_{iAX} & = & -g_2\left\{(O_{CR}^*)_{A1}(U_{u}^*)_{Xi}
                        -\sum^3_{j=1}
                         \frac{(V_{\rm CKM}^{\dag})^i_{~j}{m_u}_j}
                         {\sqrt{2}m_W \sin\beta}
                         (O_{CR}^*)_{A2}(U_u^*)_{X j+3}\right\}, 
\\
C^{u L}_{iAX} & = & g_2\sum^3_{j=1}\frac{{m_u}_i(V_{\rm CKM})^i_{~j}}
                   {\sqrt{2}m_{W}\cos\beta}
                  (O_{CR})_{A2}(U_{d}^*)_{Xj},
 \\
C^{u R}_{iAX} & = & -g_2\left\{(O_{CL})_{A1}(U_{d}^*)_{Xi} 
                        -\frac{{m_d}_i}{\sqrt{2}m_W
                         \cos\beta}
                         (O_{CL})_{A2}(U_d^*)_{X i+3}\right\}, 
\end{eqnarray}
\end{mathletters}
where $O_{CL}$ and $O_{CR}$ are the diagonalization matrices for
 chargino mass matrix defined
 in Appendix \ref{sec:approximation-mueg}.
The gluino-squark coupling constants
 in Eq.~(\ref{eq:ino-squark vertices}) are given by
\begin{mathletters}
\begin{eqnarray}
\Gamma^{d L}_{iX} &=& -\sqrt{2} g_3 (U_d^{*})_{X i+3},
\\
\Gamma^{d R}_{iX} &=&  \sqrt{2} g_3 (U_d^{*})_{X i},
\\
\Gamma^{u L}_{iX} &=& -\sqrt{2} g_3 (U_u^{*})_{X i+3},
\\
\Gamma^{u R}_{iX} &=&  \sqrt{2} g_3 (U_u^{*})_{X i}.
\end{eqnarray}
\end{mathletters}
where $g_3$ is the SU(3) gauge coupling constant.

The SUSY contribution to the effective FCNC coupling constants is divided into
 six parts as follows:
\begin{eqnarray}
g &=& g(H^-) + g(H^- W) + g(\widetilde{\chi}^0) + g(\widetilde{\chi}^-)
          + g(\widetilde{G}) + g(\widetilde{G}\widetilde{\chi}^0),
\end{eqnarray}
where $g$ represents the effective coupling constants in
 Eq.~(\ref{eq:epsilon_K}) of Sec.~\ref{sec:fcnc-lfv-processes}
 and its generalization for \bb\ mixing.
In the following, coupling constants $g$ are associated with indices $i,j$.
\kk\ mixing corresponds to $i=1$, $j=2$ and \bdbd\ (\bsbs\ ) mixing
 corresponds to $i=1$, $j=3$ ($i=2$, $j=3$).
The contribution from box diagrams including 
 charged Higgs and up-type quark is given by
\begin{mathletters}
\begin{eqnarray}
g^V_R(H^-) &=& -\frac{\sqrt{2}G_F}{16\pi^2}\sum^3_{k,l=1}
               (V_{\rm CKM}^{\dag})^j_{~l}(V_{\rm CKM})^l_{~i}
               (V_{\rm CKM}^{\dag})^j_{~k}(V_{\rm CKM})^k_{~i} 
\nonumber\\&&
\phantom{-\frac{\sqrt{2}G_F}{16\pi^2}\sum^3_{k,l=1}}
               \times {m_d}_j^2 {m_d}_i^2
               \tan^4\beta ~d_2(m_{H^-}^2,m_{H^-}^2,{m_u}_k^2,{m_u}_l^2), 
\\
g^V_L(H^-) &=& -\frac{\sqrt{2}G_F}{16\pi^2}\sum^3_{k,l=1}
               (V_{\rm CKM}^{\dag})^j_{~l}(V_{\rm CKM})^l_{~i}
               (V_{\rm CKM}^{\dag})^j_{~k}(V_{\rm CKM})^k_{~i} 
\nonumber\\&&
\phantom{-\frac{\sqrt{2}G_F}{16\pi^2}\sum^3_{k,l=1}}
               \times {m_u}_l^2 {m_u}_k^2
               \cot^4\beta~d_2(m_{H^-}^2,m_{H^-}^2,{m_u}_k^2,{m_u}_l^2), 
\\
g^S_{RR}(H^-) &=& -\frac{\sqrt{2}G_F}{16\pi^2}\sum^3_{k,l=1}
               (V_{\rm CKM}^{\dag})^j_{~l}(V_{\rm CKM})^l_{~i}
               (V_{\rm CKM}^{\dag})^j_{~k}(V_{\rm CKM})^k_{~i} 
\nonumber\\&&
\phantom{-\frac{\sqrt{2}G_F}{16\pi^2}\sum^3_{k,l=1}}
               \times {m_d}_j^2 {m_u}_l^2 {m_u}_k^2
               ~d_0(m_{H^-}^2,m_{H^-}^2,{m_u}_k^2,{m_u}_l^2), 
\\
g^S_{LL}(H^-) &=& -\frac{\sqrt{2}G_F}{16\pi^2}\sum^3_{k,l=1}
               (V_{\rm CKM}^{\dag})^j_{~l}(V_{\rm CKM})^l_{~i}
               (V_{\rm CKM}^{\dag})^j_{~k}(V_{\rm CKM})^k_{~i} 
\nonumber\\&&
\phantom{-\frac{\sqrt{2}G_F}{16\pi^2}\sum^3_{k,l=1}}
               \times {m_u}_l^2 {m_u}_k^2 {m_d}_i^2
               ~d_0(m_{H^-}^2,m_{H^-}^2,{m_u}_k^2,{m_u}_l^2), 
\\
{g^S_{RR}}'(H^-) &=& {g^S_{LL}}'(H^-) = 0, 
\phantom{-\frac{\sqrt{2}G_F}{16\pi^2}\sum^3_{k,l=1}}
\\
g^S_{RL}(H^-) &=& -\frac{\sqrt{2}G_F}{16\pi^2}\sum^3_{k,l=1}
               (V_{\rm CKM}^{\dag})^j_{~l}(V_{\rm CKM})^l_{~i}
               (V_{\rm CKM}^{\dag})^j_{~k}(V_{\rm CKM})^k_{~i} 
\nonumber\\&&
\phantom{-\frac{\sqrt{2}G_F}{16\pi^2}\sum^3_{k,l=1}}
               \times {m_d}_j {m_u}_l^2 {m_u}_k^2 {m_d}_i
               ~d_0(m_{H^-}^2,m_{H^-}^2,{m_u}_k^2,{m_u}_l^2), 
\\
{g^S_{RL}}'(H^-) &=& \frac{\sqrt{2}G_F}{8\pi^2}\sum^3_{k,l=1}
               (V_{\rm CKM}^{\dag})^j_{~l}(V_{\rm CKM})^l_{~i}
               (V_{\rm CKM}^{\dag})^j_{~k}(V_{\rm CKM})^k_{~i} 
\nonumber\\&&
\phantom{-\frac{\sqrt{2}G_F}{16\pi^2}\sum^3_{k,l=1}}
               \times {m_d}_j {m_u}_k^2 {m_d}_i
               ~d_2(m_{H^-}^2,m_{H^-}^2,{m_u}_k^2,{m_u}_l^2).
\end{eqnarray}
\end{mathletters}
where $m_{H^-}$ is the charged Higgs mass.
The contribution from box diagrams including charged Higgs, W boson
 and up-type quark is given by
\begin{mathletters}
\begin{eqnarray}
g^V_R(H^- W) &=& 0, 
\phantom{-\frac{\sqrt{2}G_F}{8\pi^2}\sum^3_{k,l=1}}
\\
g^V_L(H^- W) &=& -\frac{\sqrt{2}G_F}{8\pi^2}\sum^3_{k,l=1}
                 (V_{\rm CKM}^\dag)^j_{~l}(V_{\rm CKM})^l_{~i} 
                 (V_{\rm CKM}^\dag)^j_{~k}(V_{\rm CKM})^k_{~i}
\nonumber\\&& 
\phantom{-\frac{\sqrt{2}G_F}{8\pi^2}\sum^3_{k,l=1}}
                 \times{m_u}_l^2 {m_u}_k^2 \cot^2\beta
                 \left\{
                 d_2(m_{H^-}^2, m_W^2, {m_u}_k^2, {m_u}_l^2)   
                 \right.
\nonumber\\&& 
\phantom{-\frac{\sqrt{2}G_F}{8\pi^2}\sum^3_{k,l=1}}
                 \left.
                 \phantom{\times{m_u}_l^2 {m_u}_k^2 \cot^2\beta}
                 -m_W^2 d_0(m_{H^-}^2, m_W^2, {m_u}_k^2, {m_u}_l^2)
                 \right\}, 
\\
g^S_{RR}(H^- W) &=& g^S_{LL}(H^- W) = {g^S_{RR}}'(H^- W)
 = {g^S_{LL}}'(H^- W) = 0,
\phantom{-\frac{\sqrt{2}G_F}{8\pi^2}\sum^3_{k,l=1}}
\\
g^S_{RL}(H^- W) &=& \frac{\sqrt{2}G_F}{2\pi^2}\sum^3_{k,l=1}
                 (V_{\rm CKM}^\dag)^j_{~l}(V_{\rm CKM})^l_{~i} 
                 (V_{\rm CKM}^\dag)^j_{~k}(V_{\rm CKM})^k_{~i}
\nonumber\\&&
                 \times{m_d}_j {m_d}_i m_W^2 \tan^2\beta 
                 \left\{
                 d_2(m_{H^-}^2, m_W^2, {m_u}_k^2, {m_u}_l^2)   
                 \phantom{\frac{{m_u}_k^2 {m_u}_l^2}{4 m_W^2}}\right.
\nonumber\\&&
                 \left.
                 \phantom{\times{m_d}_j {m_d}_i m_W^2 \tan^2\beta}
                 -\frac{{m_u}_k^2 {m_u}_l^2}{4 m_W^2}
                 d_0(m_{H^-}^2, m_W^2, {m_u}_k^2, {m_u}_l^2)
                 \right\},
\\
{g^S_{RL}}'(H^- W) &=& 0.
\end{eqnarray}
\end{mathletters}
The contribution from box diagrams including neutralino and 
 down-type squark
 is given by
\begin{mathletters}
\begin{eqnarray}
g^V_R(\widetilde{\chi}^0) &=&
          -\frac{\sqrt{2}}{128\pi^2G_F}\sum^{4}_{A,B=1}\sum^{6}_{X,Y=1}
          N^{dL}_{jBY}N^{dL*}_{iAY}
\nonumber\\&&
\phantom{-\frac{\sqrt{2}}{128\pi^2G_F}}
           \times \left\{
                 N^{dL}_{jAX}N^{dL*}_{iBX}
                  ~d_2(m^2_{\chi^0_A}, m^2_{\chi^0_B}, 
                       m^2_{\widetilde{d}_X}, m^2_{\widetilde{d}_Y})
\phantom{\frac{m_{\chi^0_A}m_{\chi^0_B}}{2}} 
           \right.
\nonumber\\&&
\phantom{-\frac{\sqrt{2}}{128\pi^2G_F}\sum^{4}_{A,B=1}\sum^{6}_{X,Y=1}}
           \left.
                 +N^{dL}_{jBX}N^{dL*}_{iAX}
                   \frac{m_{\chi^0_A}m_{\chi^0_B}}{2}
                   ~d_0(m^2_{\chi^0_A}, m^2_{\chi^0_B}, 
                        m^2_{\widetilde{d}_X}, m^2_{\widetilde{d}_Y})
           \right\}, 
\\
g^S_{RR}(\widetilde{\chi}^0) &=&
          \frac{\sqrt{2}}{128\pi^2G_F}\sum^{4}_{A,B=1}\sum^{6}_{X,Y=1}
          N^{dR}_{jBY}N^{dL*}_{iAY}N^{dR}_{jBX}N^{dL*}_{iAX} 
\nonumber\\&&
\phantom{\frac{\sqrt{2}}{128\pi^2G_F}\sum^{4}_{A,B=1}\sum^{6}_{X,Y=1}}
            \times m_{\widetilde{\chi}^0_A}m_{\widetilde{\chi}^0_B} 
            ~d_0(m^2_{\widetilde{\chi}^0_A}, m^2_{\widetilde{\chi}^0_B}, 
                m^2_{\widetilde{d}_X}, m^2_{\widetilde{d}_Y}), 
\\
{g^S_{RR}}'(\widetilde{\chi}^0) &=&
          \frac{\sqrt{2}}{128\pi^2G_F}\sum^{4}_{A,B=1}\sum^{6}_{X,Y=1}
          N^{dR}_{jBY}N^{dL*}_{iAY}
           \left( 
           N^{dR}_{jBX}N^{dL*}_{iAX}-N^{dR}_{jAX}N^{dL*}_{iBX} 
           \right)
\nonumber\\&&
\phantom{-\frac{\sqrt{2}}{128\pi^2G_F}\sum^{4}_{A,B=1}\sum^{6}_{X,Y=1}}
             \times m_{\widetilde{\chi}^0_A}m_{\widetilde{\chi}^0_B}
            ~d_0(m^2_{\widetilde{\chi}^0_A}, m^2_{\widetilde{\chi}^0_B}, 
                m^2_{\widetilde{d}_X}, m^2_{\widetilde{d}_Y}), 
\\
g^S_{RL}(\widetilde{\chi}^0) &=&
          \frac{\sqrt{2}}{64\pi^2G_F}\sum^{4}_{A,B=1}\sum^{6}_{X,Y=1}
          N^{dR}_{jBY}N^{dL*}_{iAY}
           \left( N^{dL}_{jBX}N^{dR*}_{iAX}+N^{dL}_{jAX}N^{dR*}_{iBX}
           \right) 
\nonumber\\&&
\phantom{\frac{\sqrt{2}}{64\pi^2G_F}\sum^{4}_{A,B=1}\sum^{6}_{X,Y=1}}
         \times ~d_2(m^2_{\widetilde{\chi}^0_A},
                                  m^2_{\widetilde{\chi}^0_B}, 
                m^2_{\widetilde{d}_X}, m^2_{\widetilde{d}_Y}), 
\\
{g^S_{RL}}'(\widetilde{\chi}^0) &=& 
          -\frac{\sqrt{2}}{128\pi^2G_F} \sum^{4}_{A,B=1}\sum^{6}_{X,Y=1}
          N^{dR}_{jBY}N^{dR*}_{iAY}
\nonumber\\&&
\phantom{-\frac{\sqrt{2}}{128\pi^2G_F}}
           \left\{
                 2N^{dL}_{jBX}N^{dL*}_{iAX}
                  ~d_2(m^2_{\widetilde{\chi}^0_A}, m^2_{\widetilde{\chi}^0_B}, 
                 m^2_{\widetilde{d}_X}, m^2_{\widetilde{d}_Y}) 
\phantom{\frac{1}{2}}
           \right.
\nonumber\\&&
\phantom{-\frac{\sqrt{2}}{128\pi^2G_F} \sum^{4}_{A,B=1}\sum^{6}_{X,Y=1}}
           \left.
                 +
                 N^{dL}_{jAX}N^{dL*}_{iBX}
                 m_{\widetilde{\chi}^0_A}m_{\widetilde{\chi}^0_B}
                 ~d_0(m^2_{\widetilde{\chi}^0_A}, m^2_{\widetilde{\chi}^0_B}, 
                      m^2_{\widetilde{d}_X}, m^2_{\widetilde{d}_Y})
            \right\},
\end{eqnarray}
\end{mathletters}
The contribution from box diagrams including chargino and 
 up-type squark
 is given by
\begin{mathletters}
\begin{eqnarray}
g^V_R(\widetilde{\chi}^-) &=& 
          -\frac{\sqrt{2}}{128\pi^2G_F}\sum^{2}_{A,B=1}\sum^{6}_{X,Y=1}
          C^{dL}_{jBY}C^{dL*}_{iAY}C^{dL}_{jAX}C^{dL*}_{iBX}
           ~d_2(m^2_{\chi^-_A}, m^2_{\chi^-_B},
                m^2_{\widetilde{u}_X}, m^2_{\widetilde{u}_Y}),
\\
g^S_{RR}(\widetilde{\chi}^-) &=& 0, 
\phantom{-\frac{\sqrt{2}}{128\pi^2G_F}\sum^{2}_{A,B=1}\sum^{6}_{X,Y=1}}
\\
{g^S_{RR}}'(\widetilde{\chi}^-) &=& 
          -\frac{\sqrt{2}}{128\pi^2G_F}\sum^{2}_{A,B=1}\sum^{6}_{X,Y=1}
          C^{dR}_{jBY}C^{dL*}_{iAY}C^{dR}_{jAX}C^{dL*}_{iBX} 
\nonumber\\&&
\phantom{-\frac{\sqrt{2}}{128\pi^2G_F}\sum^{2}_{A,B=1}\sum^{6}_{X,Y=1}}
            \times~m_{\widetilde{\chi}^-_A}m_{\widetilde{\chi}^-_B}
            ~d_0(m^2_{\widetilde{\chi}^-_A}, m^2_{\widetilde{\chi}^-_B}, 
                m^2_{\widetilde{u}_X}, m^2_{\widetilde{u}_Y}), 
\\
g^S_{RL}(\widetilde{\chi}^-) &=& \frac{\sqrt{2}}{64\pi^2G_F}
          \sum^{2}_{A,B=1}\sum^{6}_{X,Y=1}
          C^{dR}_{jBY}C^{dL*}_{iAY}C^{dL}_{jAX}C^{dR*}_{iBX}
            ~d_2(m^2_{\widetilde{\chi}^-_A}, m^2_{\widetilde{\chi}^-_B}, 
                m^2_{\widetilde{u}_X}, m^2_{\widetilde{u}_Y}), 
\\
{g^S_{RL}}'(\widetilde{\chi}^-) &=& -\frac{\sqrt{2}}{128\pi^2G_F}
          \sum^{2}_{A,B=1}\sum^{6}_{X,Y=1}
            C^{dL}_{jBY}C^{dL*}_{iAY}C^{dR}_{jAX}C^{dR*}_{iBX} 
\nonumber\\&&
\phantom{-\frac{\sqrt{2}}{128\pi^2G_F}\sum^{2}_{A,B=1}\sum^{3}_{X,Y=1}}
           \times
            m_{\widetilde{\chi}^-_A}m_{\widetilde{\chi}^-_B}
            ~d_0(m^2_{\widetilde{\chi}^-_A}, m^2_{\widetilde{\chi}^-_B}, 
                m^2_{\widetilde{u}_X}, m^2_{\widetilde{u}_Y}),
\end{eqnarray}
\end{mathletters}

The contribution from box diagrams including gluino and
 down-type squark
 is given by
\begin{mathletters}
\begin{eqnarray}
g^V_R(\widetilde{G}) &=&
           -\frac{\sqrt{2}}{128\pi^2G_F}\sum^6_{X,Y=1}
           \Gamma^{dL}_{jY}\Gamma^{dL*}_{iY}\Gamma^{dL}_{jX}\Gamma^{dL*}_{iX}
           \left\{
             \frac{11}{18}
              ~d_2(M_3^2, M_3^2, 
                  m^2_{\widetilde{d}_X}, m^2_{\widetilde{d}_Y})
           \right.
\nonumber\\&&
\phantom{-\frac{\sqrt{2}}{128\pi^2G_F}\sum^6_{X,Y=1}}
           \left.
             +\frac{1}{18} M_3^2
              ~d_0(M_3^2, M_3^2, 
                  m^2_{\widetilde{d}_X}, m^2_{\widetilde{d}_Y})
           \right\},
\\
g^S_{RR}(\widetilde{G}) &=& 
           -\frac{\sqrt{2}}{128\pi^2G_F}\sum^6_{X,Y=1}
           \frac{17}{36}
           \Gamma^{dR}_{jY}\Gamma^{dL*}_{iY}\Gamma^{dR}_{jX}\Gamma^{dL*}_{iX}
             M_3^2
              ~d_0(M_3^2, M_3^2, 
                  m^2_{\widetilde{d}_X}, m^2_{\widetilde{d}_Y}),
\\
{g^S_{RR}}'(\widetilde{G}) &=& 
           \frac{\sqrt{2}}{128\pi^2G_F}\sum^6_{X,Y=1}
           \frac{1}{12}
           \Gamma^{dR}_{jY}\Gamma^{dL*}_{iY}\Gamma^{dR}_{jX}\Gamma^{dL*}_{iX}
             M_3^2
              ~d_2(M_3^2, M_3^2, 
                  m^2_{\widetilde{d}_X}, m^2_{\widetilde{d}_Y}),
\\
g^S_{RL}(\widetilde{G}) &=& 
           -\frac{\sqrt{2}}{128\pi^2G_F}\sum^6_{X,Y=1}
           \left\{
            \left(
             -\frac{1}{3}\Gamma^{dR}_{jY}\Gamma^{dR*}_{iY}
               \Gamma^{dL}_{jX}\Gamma^{dL*}_{iX}
             -\frac{11}{18}\Gamma^{dL}_{jY}\Gamma^{dR*}_{iY}
               \Gamma^{dR}_{jX}\Gamma^{dL*}_{iX}
            \right) 
           \right.
\nonumber\\&&
\phantom{-\frac{\sqrt{2}}{128\pi^2G_F}\sum^6_{X,Y=1}-\frac{1}{3}}
           \left.
             \times~d_2(M_3^2, M_3^2, 
              m^2_{\widetilde{d}_X}, m^2_{\widetilde{d}_Y}) 
           \right.
\nonumber\\&&
\phantom{-\frac{\sqrt{2}}{128\pi^2G_F}\sum^6_{X,Y=1}}
           \left.
         +\frac{7}{12}\Gamma^{dL}_{jY}\Gamma^{dL*}_{iY}
           \Gamma^{dR}_{jX}\Gamma^{dR*}_{iX}
              M_3^2~d_0(M_3^2, M_3^2, 
                  m^2_{\widetilde{d}_X}, m^2_{\widetilde{d}_Y})
           \right\},
\\
{g^S_{RL}}'(\widetilde{G}) &=& 
           -\frac{\sqrt{2}}{128\pi^2G_F}\sum^6_{X,Y=1}
            \left\{
             \left(
              -\frac{5}{6}\Gamma^{dL}_{jY}\Gamma^{dR*}_{iY}
                \Gamma^{dR}_{jX}\Gamma^{dL*}_{iX}
             +\frac{5}{9}\Gamma^{dR}_{jY}\Gamma^{dR*}_{iY}
               \Gamma^{dL}_{jX}\Gamma^{dL*}_{iX}
              \right) 
             \right.
\nonumber\\&&
\phantom{-\frac{\sqrt{2}}{128\pi^2G_F}\sum^6_{X,Y=1}-\frac{5}{6}}
             \left.
                 \times~d_2(M_3^2, M_3^2, 
                  m^2_{\widetilde{d}_X}, m^2_{\widetilde{d}_Y}) 
             \right.
\nonumber\\&&
\phantom{-\frac{\sqrt{2}}{128\pi^2G_F}\sum^6_{X,Y=1}}
             \left.
             +\frac{1}{36}\Gamma^{dL}_{jY}\Gamma^{dL*}_{iY}
               \Gamma^{dR}_{jX}\Gamma^{dR*}_{iX}
              M_3^2~d_0(M_3^2, M_3^2, 
                  m^2_{\widetilde{d}_X}, m^2_{\widetilde{d}_Y})
             \right\},
\end{eqnarray}
\end{mathletters}
The contribution from box diagrams including neutralino, gluino
 and down-type squark is given by
\begin{mathletters}
\begin{eqnarray}
g^V_R(\widetilde{G}\widetilde{\chi}^0) &=&
          -\frac{\sqrt{2}}{128\pi^2G_F}
          \sum^{4}_{A=1}\sum^{6}_{X,Y=1}
          \left\{
            \frac{2}{3}N^{dL}_{jAY}\Gamma^{dL*}_{iY}
             \Gamma^{dL}_{jX}N^{dL*}_{iAX}
             ~d_2(m^2_{\widetilde{\chi}^0_A}, M_3^2, 
                  m^2_{\widetilde{d}_X}, m^2_{\widetilde{d}_Y}) 
        \right.
\nonumber\\&&
        \left.
        \phantom{-\frac{\sqrt{2}}{256\pi^2G_F}
          \sum^{4}_{A=1}\sum^{6}_{X,Y=1}}
            +\frac{1}{6}\left(
             N^{dL}_{jAY}\Gamma^{dL*}_{iY}
              N^{dL}_{jAX}\Gamma^{dL*}_{iX}
            +\Gamma^{dL}_{jY}N^{dL*}_{iAY}
              \Gamma^{dL}_{jX}N^{dL*}_{iAX} \right) 
        \right.
\nonumber\\&&
        \phantom{-\frac{\sqrt{2}}{256\pi^2G_F}
          \sum^{4}_{A=1}\sum^6_{X,Y=1} }
        \left.
        \phantom{+\frac{1}{6}}
             \times m_{\widetilde{\chi}^0_A} M_3 
             ~d_0(m^2_{\widetilde{\chi}^0_A}, M_3^2, 
                  m^2_{\widetilde{d}_X}, m^2_{\widetilde{d}_Y})\right\},
\\
g^S_{RR}(\widetilde{G}\widetilde{\chi}^0) &=&
          -\frac{\sqrt{2}}{128\pi^2G_F}
          \sum^{4}_{A=1}\sum^{6}_{X,Y=1}\left\{
             N^{dR}_{jAY}\Gamma^{dL*}_{iY}
             \Gamma^{dR}_{jX}N^{dL*}_{iAX} 
\phantom{-\frac{1}{3}}\right.
\nonumber\\&&
\phantom{-\frac{\sqrt{2}}{128\pi^2G_F}
          \sum^{4}_{A=1}\sum^{6}_{X,Y=1}}
            \left.
             -\frac{1}{3}\left(
             N^{dR}_{jAY}\Gamma^{dL*}_{iY}
              N^{dR}_{jAX}\Gamma^{dL*}_{iX}
            +\Gamma^{dR}_{jY}N^{dL*}_{iAY}
              \Gamma^{dR}_{jX}N^{dL*}_{iAX}
              \right)\right\} 
\nonumber\\&&
\phantom{-\frac{\sqrt{2}}{128\pi^2G_F}
          \sum^{4}_{A=1}\sum^{6}_{X,Y=1}-\frac{1}{3}}
             \times m_{\widetilde{\chi}^0_A} M_3 
             ~d_0(m^2_{\widetilde{\chi}^0_A}, M_3^2, 
                  m^2_{\widetilde{d}_X}, m^2_{\widetilde{d}_Y}),
\\
{g^S_{RR}}'(\widetilde{G}\widetilde{\chi}^0) &=& 
             \frac{\sqrt{2}}{128\pi^2G_F}
             \sum^{4}_{A=1}\sum^{6}_{X,Y=1}
             \frac{1}{3}\left(
             N^{dR}_{jAY}\Gamma^{dL*}_{iY}\Gamma^{dR}_{jX}N^{dL*}_{iAX} 
          \right.
\nonumber\\&&
\phantom{\frac{\sqrt{2}}{128\pi^2G_F}
         \sum^{4}_{A=1}\sum^{6}_{X,Y=1}\frac{1}{3}}
          \left.
            +N^{dR}_{jAY}\Gamma^{dL*}_{iY}
               N^{dR}_{jAX}\Gamma^{dL*}_{iX}
            +\Gamma^{dR}_{jY}N^{dL*}_{iAY}
                \Gamma^{dR}_{jX}N^{dL*}_{iAX} \right) 
\nonumber\\&&
\phantom{\frac{\sqrt{2}}{128\pi^2G_F}
         \sum^{4}_{A=1}\sum^{6}_{X,Y=1}\frac{1}{3}+}
             \times m_{\widetilde{\chi}^0_A} M_3
             ~d_0(m^2_{\widetilde{\chi}^0_A}, M_3^2, 
                  m^2_{\widetilde{d}_X}, m^2_{\widetilde{d}_Y}),
\\
g^S_{RL}(\widetilde{G}\widetilde{\chi}^0) &=&
          -\frac{\sqrt{2}}{128\pi^2G_F}
          \sum^{4}_{A=1}\sum^{6}_{X,Y=1}\left[
            \left\{\frac{1}{3}
             (N^{dL}_{jAY}\Gamma^{dR*}_{iY}
              +\Gamma^{dL}_{jY}N^{dR*}_{iAY})
             (\Gamma^{dR}_{jX}N^{dL*}_{iAX}
              +N^{dR}_{jAX}\Gamma^{dL*}_{iX})
            \right. 
         \right.
\nonumber\\&&
         \phantom{-\frac{\sqrt{2}}{256\pi^2G_F}
          \sum^{4}_{A=1}\sum^{6}_{X,Y=1}}
         \left.
                \left.
\phantom{\frac{1}{3}}
                +(N^{dR}_{jAY}\Gamma^{dR*}_{iY}N^{dL}_{jAX}\Gamma^{dL*}_{iX}
                +\Gamma^{dR}_{jY}N^{dR*}_{iAY}\Gamma^{dL}_{jX}N^{dL*}_{iAX})
                \right\}
         \right.
\nonumber\\&&
\phantom{-\frac{\sqrt{2}}{128\pi^2G_F}
\sum^{4}_{A=1}\sum^{6}_{X,Y=1}}
         \left.
\phantom{\frac{1}{3}+2}
                  \times ~d_2(m^2_{\widetilde{\chi}^0_A}, M_3^2, 
                  m^2_{\widetilde{d}_X}, m^2_{\widetilde{d}_Y}) 
         \right.
\nonumber\\&&
\phantom{-\frac{\sqrt{2}}{128\pi^2G_F}
\sum^{4}_{A=1}\sum^{6}_{X,Y=1}}
         \left.
            +\frac{1}{2}\left(
             N^{dL}_{jAY}\Gamma^{dL*}_{iY}
              \Gamma^{dR}_{jX}N^{dR*}_{iAX}
            +\Gamma^{dL}_{jY}N^{dL*}_{iAY}
              N^{dR}_{jAX}\Gamma^{dR*}_{iX} \right) 
         \right.
\nonumber\\&&
\phantom{-\frac{\sqrt{2}}{128\pi^2G_F}
\sum^{4}_{A=1}\sum^{6}_{X,Y=1}}
         \left.
\phantom{+\frac{1}{3}}
             \times m_{\widetilde{\chi}^0_A} M_3 
             ~d_0(m^2_{\widetilde{\chi}^0_A}, M_3^2, 
                  m^2_{\widetilde{d}_X}, m^2_{\widetilde{d}_Y})\right],
\\
{g^S_{RL}}'(\widetilde{G}\widetilde{\chi}^0) &=&
          \frac{\sqrt{2}}{128\pi^2G_F}
          \sum^{4}_{A=1}\sum^{6}_{X,Y=1}\left[
            \left\{
             (N^{dL}_{jAY}\Gamma^{dR*}_{iY}
              +\Gamma^{dL}_{jY}N^{dR*}_{iAY})
             (N^{dR}_{jAX}\Gamma^{dL*}_{iX}
              +\Gamma^{dR}_{jX}N^{dL*}_{iAX})
            \phantom{\frac{1}{3}}\right. 
          \right.
\nonumber\\&&
\phantom{\frac{\sqrt{2}}{128\pi^2G_F}
          \sum^{4}_{A=1}\sum^{6}_{X,Y=1}}
          \left.
            \left.
            +\frac{1}{3}
             \left(N^{dR}_{jAY}\Gamma^{dR*}_{iY}N^{dL}_{jAX}\Gamma^{dL*}_{iX}
              +\Gamma^{dR}_{jY}N^{dR*}_{iAY}\Gamma^{dL}_{jX}N^{dL*}_{iAX}
              \right)
             \right\} 
          \right.
\nonumber\\&&
\phantom{\frac{\sqrt{2}}{128\pi^2G_F}
          \sum^{4}_{A=1}\sum^{6}_{X,Y=1}}
          \left.
\phantom{+\frac{1}{3}}
                \times ~d_2(m^2_{\widetilde{\chi}^0_A}, M_3^2, 
                  m^2_{\widetilde{d}_X}, m^2_{\widetilde{d}_Y}) 
          \right.
\nonumber\\&&
\phantom{\frac{\sqrt{2}}{128\pi^2G_F}
          \sum^{4}_{A=1}\sum^{6}_{X,Y=1}}
          \left.
            +\frac{1}{6}\left(
             N^{dL}_{jAY}\Gamma^{dL*}_{iY}
              \Gamma^{dR}_{jX}N^{dR*}_{iAX}
            +\Gamma^{dL}_{jY}N^{dL*}_{iAY}
              N^{dR}_{jAX}\Gamma^{dR*}_{iX} \right)
          \right.
\nonumber\\&&
\phantom{\frac{\sqrt{2}}{128\pi^2G_F}
          \sum^{4}_{A=1}\sum^{6}_{X,Y=1}}
          \left.
\phantom{+\frac{1}{6}}
             \times M_3 m_{\widetilde{\chi}^0_A}
             ~d_0(m^2_{\widetilde{\chi}^0_A}, M_3^2, 
                  m^2_{\widetilde{d}_X}, m^2_{\widetilde{d}_Y})
          \right],
\end{eqnarray}
\end{mathletters}
The neutralino, chargino, gluino and neutralino-gluino contributions
 to $g^V_L$, $g^S_{LL}$ and ${g^S_{LL}}'$ are
 obtained by replacing the suffix $R$ with $L$ and $L$ with $R$ in the
 corresponding formulas for $g^V_R$, $g^S_{RR}$ and ${g^S_{RR}}'$,
 respectively. 
The mass functions which appear in the above formulas are defined as follows:
\begin{mathletters}
\begin{eqnarray}
d_{0}(x,y,z,w) & = &    \frac{x\ln(x)}{(y-x)(z-x)(w-x)} 
                       +\frac{y\ln(y)}{(x-y)(z-y)(w-y)} \nonumber \\
               &   &   +\frac{z\ln(z)}{(x-z)(y-z)(w-z)}
                       +\frac{w\ln(w)}{(x-w)(y-w)(z-w)},
 \\
d_{2}(x,y,z,w) & = &   \frac{1}{4} \left\{ \frac{x^{2}\ln(x)}{(y-x)(z-x)(w-x)} 
                       +\frac{y^{2}\ln(y)}{(x-y)(z-y)(w-y)}
                       \right.
 \nonumber \\&&
                       \left.
                       +\frac{z^{2}\ln(z)}{(x-z)(y-z)(w-z)}
                       +\frac{w^{2}\ln(w)}{(x-w)(y-w)(z-w)} \right\}.
\end{eqnarray}
\end{mathletters}


\begin{table}
\begin{tabular}{cccccc}
  $m_t^{\rm pole}$
& $m_b^{\rm pole}$
& $m_s^{\rm\overline{MS}}(2\,\text{GeV})$
& $V_{cb}$
& $|V_{ub}/V_{cb}|$
& \dlt
\\
\hline
  175 GeV
& 4.8 GeV
& 120 MeV
& 0.04
& 0.08
& $60^\circ$
\\
\end{tabular}
\caption{
Input parameters used in the numerical calculation.
$|V_{ub}/V_{cb}|$ and \dlt\ are varied in
Fig.~\ref{fig:dmbsd-AtJKS}.
}
\label{tab:fixedparameters}   
\end{table}

\begin{table}
\begin{tabular}{ccc}
& LMA & SMA
\\
\hline
$\sin^22\theta_{\rm sun}$ & $1$ & $5.5\times10^{-3}$
\\
$\sin^22\theta_{\rm atm}$ & $1$ & $1$
\\
$\theta_{13}$ & $0$ & $0$
\\
$\Delta m^2_{12}$ (eV$^2$) & $1.8\times10^{-5}$ & $5.0\times10^{-6}$
\\
$\Delta m^2_{23}$ (eV$^2$) & $3.5\times10^{-3}$ & $3.5\times10^{-3}$
\\
$m_{\nu_{1}}$ (eV) & $4.0\times10^{-3}$ & $2.2\times10^{-3}$
\end{tabular}
\caption{
Parameters for the neutrino sector.
LMA (SMA) corresponds to the large (small) mixing angle MSW solution for
the solar neutrino anomaly.
}
\label{tab:nuparam}
\end{table}

\begin{table}
\begin{tabular}{cccccccc}
            & (a)              & (b)              & (c)              & (d)              & (e)              & (f)              & (g)
\\                                                                                                                            
\hline                                                                                                                        
neutrino    & LMA              & LMA              & LMA              & LMA              & LMA              & SMA              & SMA
\\                                                                                                                            
$M_R$ (GeV) & $4\times10^{13}$ & $4\times10^{13}$ & $4\times10^{13}$ & $4\times10^{13}$ & $4\times10^{14}$ & $4\times10^{13}$ & $4\times10^{14}$
\\                                                                                                                            
$\theta_D$  & 0                & 0                & 45$^\circ$       & 0                & 0                & 0                & 0
\\                                                                                                                            
$\tan\beta$ & 20               & 20               & 20               & 5                & 20               & 20               & 5
\\                                                                                                                            
$A_0$       & 0                & 2                &  0               &  0               &  0               &  0               & 2.5
\end{tabular}
\caption{Parameters for Fig.~\protect\ref{fig:contours}.}
\label{tab:paramcontours}
\end{table}

\begin{table}
\begin{tabular}{cccccccc}
  $f_{B_d}$
& $f_{B_s}/f_{B_d}$
& $B_B$
& $(B_B)_{RL}^{S}$
& ${(B_B)_{RL}^{S}}'$
& $B_K$
& $(B_K)_{RL}^{S}$
& ${(B_K)_{RL}^{S}}'$
\\
\hline
  210 MeV
& 1.17
& 0.8
& 0.8
& 0.8
& 0.69
& 1.03
& 0.73
\end{tabular}
\caption{
Decay constants and bag parameters for \bb\ and \kk\ mixing matrix
elements used in the numerical calculation \protect\cite{lattice-parameters}.
}
\label{tab:bagparam}  
\end{table}

\begin{table}
\begin{tabular}{cccccccc}
                                             & (a)            & (b)            & (c)            & (d)             & (e)            & (f)             & (g)
\\                                                                                                                                                   
\hline                                                                                                                                               
\amu\ [$10^{-10}$]                           & $\lsim50$      & $\lsim50$      & $\lsim10$      & --              & --             & $\lsim50$       & --
\\                                                                                                                                                   
\Bmeg                                        & $\surd$        & $\surd$        & $\surd$        & $\lsim10^{-13}$ & $\surd$        & $\lsim10^{-13}$ & $\surd$
\\                                                                                                                                                   
\Btmg                                        & $\lsim10^{-9}$ & $\lsim10^{-9}$ & $\lsim10^{-9}$ & $\lsim10^{-10}$ & $\lsim10^{-9}$ & $\lsim10^{-8}$  & $\lsim10^{-7}$
\\                                                                                                                                                   
$\varepsilon_K/(\varepsilon_K)_{\rm SM}-1$   & $\lsim0.1$     & $\lsim0.5$     & $\lsim1$       & $\lsim0.05$     & $\lsim0.5$     & --              & $\lsim0.5$
\\                                                                                                                                                   
$\Delta m_{B_d}/(\Delta m_{B_d})_{\rm SM}-1$ & --             & --             & --             & --              & --             & --              & --
\\                                                                                                                                                   
$\Delta m_{B_s}/(\Delta m_{B_s})_{\rm SM}-1$ & --             & --             & --             & --              & $\lsim0.05$    & --              & $\lsim1$
\end{tabular}
\caption{
Summary of the SUSY contributions to the observables in
Fig.~\protect\ref{fig:contours}.
``$\surd$'' shows that some parameter region is excluded by the \meg\
constraint and hence the branching ratio can be just below the present
upper bound.
``--'' means that the SUSY contribution is negligible.
}
\label{tab:contourssummary}
\end{table}


\begin{references}
%
\bibitem{E821}
   Muon $g-2$ Collaboration, H.N.~Brown {\it et al.}, hep-ex/0102017.
\bibitem{SUSY-Flavor}
  J.~Ellis and D.V.~Nanopoulos, \Journal{\plb}{110}{1982}{44};
  R.~Barbieri and R.~Gatto, \Journal{\plb}{110}{1982}{211};
  T.~Inami and C.S.~Lim, \Journal{\npb}{207}{1982}{533}.
\bibitem{gluino-FCNC}
  M.J.~Duncan, \Journal{\npb}{221}{1983}{285};
  J.F.~Donoghue, H.P.~Nilles and D.~Wyler, \Journal{\plb}{128}{1983}{55};
  A.~Bouquet, J.~Kaplan and C.A.~Savoy, \Journal{\plb}{148}{1984}{69}.
\bibitem{86HaKoRa}
  L.J.~Hall, V.A.~Kostelecky and S.~Raby, \Journal{\npb}{267}{1986}{415}.
\bibitem{Krimoto-Cho}
  T.~Kurimoto, \Journal{\prd}{39}{1989}{3447};
  G.C.~Branco, G.C.~Cho, Y.~Kizukuri and N.~Oshimo,
  \Journal{\plb}{337}{1994}{316}.
\bibitem{91BeBoMaRi}
  S.~Bertolini, F.~Borzumati, A.~Masiero and G.~Ridolfi,
  \Journal{\npb}{353}{1991}{590}.
\bibitem{Goto}
  T.~Goto, T.~Nihei and Y.~Okada, \Journal{\prd}{53}{1996}{5233};
  Erratum-\Journal{\ibid}{54}{1996}{5904}; 
  T.~Goto, Y.~Okada and Y.~Shimizu, \Journal{\prd}{58}{1998}{094006}.
\bibitem{99GoOkSh}
  T.~Goto, Y.~Okada and Y.~Shimizu, KEK-Preprint-99-72, KEK-TH-611,
 hep-ph/9908499.
%
\bibitem{GUT-classic}
F.~Gabbiani and A.~Masiero, \Journal{\npb}{322}{1989}{235}:
J.S.~Hagelin, S.~Kelley and T.~Tanaka, \Journal{\npb}{415}{1994}{293}.
%
\bibitem{LFV}
  R.~Barbieri and L.J.~Hall, \Journal{\plb}{338}{1994}{212};
  R.~Barbieri, L.~Hall and A.~Strumia, \Journal{\npb}{445}{1995}{219}.
%
\bibitem{FCNC-GUT}
  R.~Barbieri, L.~Hall and A.~Strumia, \Journal{\npb}{449}{1995}{437};
  N.~G.~Deshpande, B.~Dutta and S.~Oh, \Journal{\prl}{77}{1996}{4499}.
%
\bibitem{GUT}
  P.~Ciafaloni, A.~Romanino and A.~Strumia, \Journal{\npb}{458}{1996}{3}; 
  N.~Arkani-Hamed, H.~Cheng and L.J.~Hall, \Journal{\prd}{53}{1996}{413};
  T.V.~Duong, B.~Dutta and E.~Keith \Journal{\plb}{378}{1996}{128}; 
  M.E.~G\'omez and H.~Goldberg, \Journal{\prd}{53}{1996}{5244};
  N.G.~Deshpande, B.~Dutta and E.~Keith \Journal{\prd}{54}{1996}{730};
  J.~Hisano, T.~Moroi, K.~Tobe and M.~Yamaguchi,
  \Journal{\plb}{391}{1997}{341}, Erratum-\Journal{\ibid}{397}{1997}{357};
  J.~Hisano, D.~Nomura, Y.~Okada, Y.~Shimizu and M.~Tanaka,
  \Journal{\prd}{58}{1998}{116010};
  G.~Barenboim, K.~Huitu and M.~Raidal, \Journal{\prd}{63}{2001}{055006}.
%
\bibitem{neutrino-exp}
  Super-Kamiokande Collaboration, Y.~Fukuda {\it et al.}, 
  \Journal{\prl}{81}{1998}{1562};
  E.~Kearns, talk at the 30th International Conference on High Energy
  Physics (ICHEP 2000), Osaka, 2000.
\bibitem{seesaw}
  T.~Yanagida,
  in {\it Proceedings of the Workshop on Unified Theory and Baryon Number
    of the Universe}, Tsukuba, Japan, 1979, edited by O.~Sawada and
  A.~Sugamoto;
  M.~Gell-Mann, P.~Ramond and R.~Slansky,
  in {\it Supergravity}, 1979, edited by P.~van~Nieuwenhuizen and
  D.~Freedman (North-Holland, Amsterdam).
%
\bibitem{ynu-LFV}
 F.~Borzumati and A.~Masiero, \Journal{\prl}{57}{1986}{961};
 J.~Hisano, T.~Moroi, K.~Tobe, M.~Yamaguchi and T.~Yanagida,
   \Journal{\plb}{357}{1995}{579};
 J.~Hisano, T.~Moroi, K.~Tobe and M.~Yamaguchi,
   \Journal{\prd}{53}{1996}{2442};
 J.A.~Casas and A.~Ibarra, hep-ph/0103065.
\bibitem{98HiNoYa}
  J.~Hisano and D.~Nomura, T.~Yanagida
  \Journal{\plb}{437}{1998}{351}.
\bibitem{99HiNo}
  J.~Hisano and D.~Nomura, \Journal{\prd}{59}{1999}{116005}.
%
\bibitem{ynu-LFV-GUT}
  M.E.~G\'omez, G.K.~Leontaris, S.~Lola and J.D.~Vargados,
   \Journal{\plb}{459}{1999}{116009};
  W.~Buchmuller, D.~Delepine and F.~Vissani,
   \Journal{\plb}{459}{1999}{171};
  J.~Ellis, M.E.~G\'omez, G.K.~Leontaris,
  S.~Lola and D.V.~Nanopoulos,
   \Journal{\epjc}{14}{2000}{319};
  W.~Buchmuller, D.~Delepine, L.T.~Handoko
   \Journal{\npb}{576}{2000}{445};
  J.L.~Feng, Y.~Nir and Y.~Shadmi, \Journal{\prd}{61}{2000}{113005};
  J.~Sato, K.~Tobe, T.~Yanagida, \Journal{\plb}{498}{2001}{189};
  J.~Sato and K.~Tobe, hep-ph/0012333.
%
\bibitem{01BaGoOkOk}
  S.~Baek, T.~Goto, Y.~Okada and K.~Okumura,
  \Journal{\prd}{63}{2001}{051701(R)};
  talk at the 30th International Conference on High-Energy
  Physics (ICHEP 2000), Osaka, 2000, hep-ph/0009196.
%
\bibitem{00Moroi}
  T.~Moroi, \Journal{\jh}{03}{2000}{019}; \Journal{\plb}{493}{2000}{366}.
%
\bibitem{g-2new}
  A.~Czarnecki and W.J.~Marciano, hep-ph/0102122;
  L.~Everett, G.L.~Kane, S.~Rigolin and L.-T.~Wang, hep-ph/0102145;
  J.L.~Feng and K.~Matchev, hep-ph/0102146;
  E.A.~Baltz, P.~Gondolo, hep-ph/0102147;
  U.~Chattopadhyay and P.~Nath, hep-ph/0102157;
  S.~Komine, T.~Moroi and M.~Yamaguchi, hep-ph/0102204;
  J.~Hisano and K.~Tobe, hep-ph/0102315;
  T.~Ibrahim, U.~Chattopadhyay and P.~Nath, hep-ph/0102324;
  J.~Ellis, D.V.~Nanopoulos and K.A.~Olive, hep-ph/0102331;
  K.~Choi, K.~Hwang, S.K.~Kang, K.Y.~Lee and W.Y.~Song, hep-ph/0103048;
  S.~Baek, P.~Ko and H.S.~Lee, hep-ph/0103218;
  D.F.~Carvalho, J.~Ellis, M.E.~G\'{o}mez and S.~Lola, hep-ph/0103256;
  H.~Baer, C.~Bal\'{a}zs, J.~Ferrandis and X.~Tata, hep-ph/0103280.
%
\bibitem{62MaNaSa}
  Z.~Maki, M.~Nakagawa, S.~Sakata, \Journal{\ptp}{28}{1962}{870}.
%
\bibitem{Apollonio:1999ae}
  CHOOZ Collaboration, M.~Apollonio {\it et al.}, 
  \Journal{\plb}{466}{1999}{466}.
%
\bibitem{g-2}
  P.~Fayet,
    in Unification of the Fundamental Particles Interactions, edited by
    S.~Ferrara, J.~Ellis and P.~van~Nieuwenhuizen (Plenum, New York, 1980)
    p.~587;
  J.A.~Grifols and A.~Mendez, \Journal{\prd}{26}{1982}{1809};
  J.~Ellis, H.S.~Hagelin and D.V.~Nanopoulos,
    \Journal{\plb}{116}{1982}{283};
  R.~Barbieri and L.~Maiani, \Journal{\plb}{117}{1982}{203};
  D.A.~Kosower, L.M.~Krauss, and N.~Sakai, \Journal{\plb}{133}{1983}{305};
  T.C.~Yuan, R.~Arnowitt, A.H.~Chamseddine, and P.~Nath,
  \Journal{\zpc}{26}{1984}{407};
  T.~Moroi, \Journal{\prd}{53}{1996}{6565}.
%
\bibitem{tanb-enhancement}
  J.~Lopez, D.V.Nanopoulos and X.~Wang, \Journal{\prd}{49}{1994}{366};
  U.~Chattopadhyay and P.~Nath, \Journal{\prd}{53}{1996}{1648}.
%
\bibitem{polarization}
  Y.~Okada, K.~Okumura and Y.~Shimizu, \Journal{\prd}{61}{2000}{094001}.
%
\bibitem{00KiOk}
  R.~Kitano and Y.~Okada, {KEK-preprint-2000-131, KEK-TH-732, hep-ph/0012040},
 to be published in \prd.
%
\bibitem{atbsg}
  D.~Atwood, M.~Gronau and A.~Soni, \Journal{\prl}{79}{1997}{185};
  C.~Chua, X.~He and W.~Hou, \Journal{\prd}{60}{1999}{014003}.
%
\bibitem{QCD-BBKK}
  A.J.~Buras, M.~Jamin and P.H.~Weisz, \Journal{\npb}{347}{1990}{491};
  I.I.~Bigi and F.~Gabbiani, \Journal{\npb}{352}{1991}{309};
  S.~Herrlich and U.~Nierste, \Journal{\npb}{419}{1994}{292};
  S.~Herrlich and U.~Nierste, \Journal{\npb}{476}{1996}{27};
  J.~Urban, F.~Krauss, U.~Jentschura and G.~Soff,
    \Journal{\npb}{523}{1998}{40}.
%
\bibitem{susy-search}
  CDF Collaboration, \Journal{\prd}{56}{1997}{1357};
  D0 Collaboration, \Journal{\prl}{75}{1995}{618};
  ALEPH Collaboration, \Journal{\plb}{499}{2001}{67}.
%
\bibitem{higgs-search}
  ALEPH Collaboration, \Journal{\plb}{499}{2001}{53};
  L3 Collaboration, \Journal{\plb}{503}{2001}{21};
  DELPHI Collaboration, \Journal{\plb}{499}{2001}{23};
  OPAL Collaboration, \Journal{\plb}{499}{2001}{38}.
%
\bibitem{bsg-exp}
  CLEO Collaboration, T.~Coan, talk at
  the 30th International Conference on High-Energy Physics (ICHEP 2000),
  Osaka, 2000;
  Belle Collaboration, KEK-preprint-2001-3, BELLE-preprint-2001-2
, hep-ex/0103042. 
%
\bibitem{CK-PDG}
  L.~Chau and W.~Keung,
  \Journal{\prl}{\bf 53}{1984}{1802}.
%
\bibitem{Brooks:1999pu}
  MEGA Collaboration, M.L.~Brooks {\it et al.},  
  \Journal{\prl}{83}{1999}{1521}.
%
\bibitem{AtJKs}
  Belle Collaboration, \Journal{\prl}{86}{2001}{2509};
  BaBar Collaboration, \Journal{\prl}{86}{2001}{2515}.
%
\bibitem{lattice-parameters}
  N.~Yamada, S.~Hashimoto, K.~Ishikawa, H.~Matsufuru and T.~Onogi,
  \Journal{\np Proc.\ Suppl.\ }{83}{2000}{340};
  JLQCD Collaboration, S.~Aoki {\it et al.},  
  \Journal{\prd}{60}{1999}{034511};
  R.~Gupta, T.~Bhattacharya and S.~Sharpe,
  \Journal{\prd}{55}{1997}{4036};
  M.~Ciuchini {\it et al.},
  \Journal{\jh}{10}{1998}{008};
  CP-PACS Collaboration, A.~Ali Khan {\it et al.},  
  \Journal{\np Proc.\ Suppl.\ }{83}{2000}{265}.
%
\bibitem{AtJKs-CDF}
  CDF Collaboration, \Journal{\prd}{61}{2000}{072005}.
%
\bibitem{pdecay1}
  T.~Goto and T.~Nihei,
  \Journal{\prd}{59}{1999}{115009};
  K.S.~Babu and M.J.~Strassler, hep-ph/9808447;
  T.~Goto and T.~Nihei, in
  {\it Proceedings of the KIAS-CTP International Symposium on
       Supersymmetry, Supergravity and Superstring}, Seoul, Korea, 1999,
       edited by J.E.~Kim and C.~Lee.
%
\bibitem{pdecay2}
  J.~Hisano, H.~Murayama and T.~Yanagida,
  \Journal{\plb}{291}{1992}{263};
  K.S.~Babu and S.M.~Barr,
  \Journal{\prd}{48}{1993}{5354};
  J.~Hisano, T.~Moroi, K.~Tobe and T.~Yanagida,
  \Journal{\plb}{342}{1995}{138};
  K.S.~Babu and S.M.~Barr,
  \Journal{\prd}{51}{1995}{2463}.
%
\bibitem{review}
  Y.~Kuno and Y.~Okada, \Journal{\rmp}{73}{2001}{151}.
%
\bibitem{Czarnecki:1998iz}
  A.~Czarnecki, W.J.~Marciano and K.~Melnikov, in
 {\it Proceedings of Workshop on Physics at the First Muon Collider and
 at the Front End of the Muon Collider}, Fermilab, 1997,
 edited by S.H.~Geer and R.~Raja, AIP Conf. Proc. No. 435 (AIP, New York), p.409.
%
\bibitem{edmn}
  P.G.~Harris \etal, \Journal{\prl}{82}{1999}{904}.
%
\bibitem{edme}
  E.D.~Commins \etal, \Journal{\pra}{50}{1994}{2960}.
%
\bibitem{psi}
L.M.~Barkov \etal, ``Search for the decay $\mu^+ \to e^+ \gamma$
 down to $10^{-14}$ branching ratio'', Research Proposal to Paul Scherrer
 Institut (1999).
%
\bibitem{meco}
MECO Collaboration, M.~Bachman \etal, ``A Search for $\mu^-N \to e^-N$ with
 sensitivity below $10^{-16}$'', experimental proposal E940 to Brookhaven
 National Laboratory, AGS (1997).
%
\end{references}
\end{document}